\documentclass[12pt]{iopart}
\usepackage{color}
\usepackage{graphicx}
\usepackage{epsf}
\usepackage{graphicx,epsfig}
\def\cs2{c_{s}^{2}}

 \def\al{\alpha}
 \def\b{\beta}
 
 \def\de{\delta}
 
 \def\ep{\varepsilon}

 \def\df{\delta\phi}
 \def\p{\partial}

 \def\epr{\eta^{'}}
 \def\eps{\eta^{''}}

 \def\be   {\begin{equation}}   \def\ee   {\end{equation}}
 \def\ba   {\begin{array}}      \def\ea   {\end{array}}
 \def\bea  {\begin{eqnarray}}   \def\eea  {\end{eqnarray}}
 \def\bean {\begin{eqnarray*}}  \def\eean {\end{eqnarray*}}
\begin{document}

\title{Anisotropic Trispectrum of Curvature Perturbations Induced by Primordial Non-Abelian Vector Fields}

\vspace{0.8cm}

\author{Nicola Bartolo$^{1,2}$, Emanuela Dimastrogiovanni$^{1,2}$, Sabino Matarrese$^{1,2}$ and 
Antonio Riotto$^{2,3}$}
\vspace{0.4cm}
\address{$^1$ Dipartimento di Fisica ``G. Galilei'', Universit\`{a} degli Studi di 
Padova, \\ via Marzolo 8, I-35131 Padova, Italy} 
\address{$^2$ INFN, Sezione di Padova, via Marzolo 8, I-35131 Padova, Italy}
\address{$^3$ CERN, Theory Division, CH-1211 Geneva 23, Switzerland}
\eads{\mailto{nicola.bartolo@pd.infn.it}, \mailto{dimastro@pd.infn.it}, 
\mailto{sabino.matarrese@pd.infn.it} and \mailto{riotto@mail.cern.ch}}

\date{\today}
\vspace{1cm}
\begin{abstract} Motivated by the interest in models of the early universe where statistical isotropy is broken and can be revealed in cosmological observations, we consider an $SU(2)$ theory of gauge interactions in a single scalar field inflationary scenario. We calculate the trispectrum of curvature perturbations, as a natural follow up to a previous paper of ours, where we studied the bispectrum in the same kind of models. The choice of a non-Abelian set-up turns out to be very convenient: on one hand, gauge boson self-interactions can be very interesting being responsible for extra non-trivial terms (naturally absent in the Abelian case) appearing in the cosmological correlation functions; on the other hand, its results can be easily reduced to the $U(1)$ case. As expected from the presence of the vector bosons, preferred spatial directions arise and the 
trispectrum reveals anisotropic signatures. We evaluate its amplitude $\tau_{NL}$, which receives contributions both from scalar and vector fields, and verify that, in a large subset of its parameter space, the latter contributions can be larger than the former. We carry out a shape analysis of the trispectrum; in particular we discuss, with some examples, how the anisotropy parameters appearing in the analytic expression of the trispectrum can modulate its profile and we show that the amplitude of the anisotropic part of the trispectrum can be of the same order of magnitude as the isotropic part.\\

DFPD 09-A-18 / CERN-PH-TH/2009-179
\end{abstract}

\maketitle

\section{Introduction}

The Cosmic Microwave Background (CMB) radiation is a window on the early universe physics \cite{Lyth:1998xn} and represents a major object of the {\it WMAP} 
\cite{http://map.gsfc.nasa.gov/,Spergel:2006hy,Komatsu:2008hk} and {\it Planck} mission \cite{http://planck.esa.int/} 
searches and of other future cosmological investigations.\\ 
A powerful feature of the CMB is its degree of non-Gaussianity \cite{Bartolo:2004if,Komatsu:2009kd}. A signal is considered Gaussian if the information it carries is completely encoded in the two-point correlation function, all higher order connected correlators being zero. Present measurements have already allowed to put interesting limits on the level of non-Gaussianity of 
the CMB, and the recently launched Planck satellite and future experiments are expected to provide either tighter constraints 
or possibly a detection of a primordial non-Gaussian signal. Many of the existing inflationary models are in excellent agreement with the CMB observations but differ from each other in terms of predicted amount of non-Gaussianity \cite{Acqua,Maldacena:2002vr,Bmulti,SL1,SL2,Lyth:2001nq,Lyth:2002my,Bartolo:2003jx,Ali,Chen:2006nt,ghostinfl}. If we want to be prepared for an accurate comparison with the data in the near future, it is therefore crucial to understand what the different models forecast in terms of non-Gaussianity. This translates into calculating three, four and higher-order correlation functions for the cosmological fluctuations. \\

In this paper, we focus on the trispectrum of curvature perturbations \cite{Okamoto:2002ik,Kogo:2006kh,Barttrisp}. This has been calculated in several models, such as multifield slow-roll inflation \cite{Seery:2006vu,SL3,Byrnes,Seery:2008ax}, in the curvaton mechanism \cite{SVW}, in theories with non-canonical kinetic terms both in single field \cite{Huang:2006eha,Arroja:2008ga,Arroja:2009pd,Chen:2009bc} and in the multifield case \cite{Gao:2009gd,Mizuno:2009mv}. 
~\footnote{At present there are no constraints on the non-linearity parameters characterizing the trispectrum from CMB analyses~\cite{Kunzetal,Komatsuthesis}. 
The only constraint available on the 
parameter $g_{\rm NL}$ of  ``local'' cubic non-linearities has been obtained from Large-Scale-Strucuture (LSS) data yielding~$|g_{\rm NL}| < 10^6$, and future CMB 
and LSS observations could probe $|g_{\rm NL}| \geq 10^4$~\cite{DS}.}
\\
We consider here another class of models, specifically those including primordial vector fields \cite{Dimopoulos:2006ms,Ack,Pullen:2007tu,Golovnev:2008cf,Dimopoulos:2008rf,Yokoyama:2008xw,Dimopoulos:2008yv,Karciauskas:2008bc,ValenzuelaToledo:2009af,Golovnev:2009ks}. The interest in these models finds its motivations, besides in the search for a strong non-Gaussian signature, also in the study of statistical anisotropy. In fact, 
several anomalies in the CMB data suggest signatures of statistical anisotropy breaking 
\cite{Eriksen:2003db,Hansen:2004vq,Jaffe:2005pw,Eriksen:2007pc,Groeneboom:2008fz}. This unexpected feature could, among other ways, be explained with the existence of preferred spatial directions defined by some primordial vector fields, in models where they can somehow leave their imprints on the cosmological fluctuations. \\
Recent works where two and three point functions of the curvature perturbation are computed in the presence of primordial vector fields include Refs.~\cite{Dimopoulos:2008yv,Karciauskas:2008bc,ValenzuelaToledo:2009af}, which consider a $U(1)$ gauge theory and different specific models (vector inflation, vector curvaton, time-dependent gauge coupling models); Ref.~\cite{Yokoyama:2008xw}, which works in the Abelian case and for a model of hybrid inflation; a previous paper of us, \cite{Bartolo:2009pa}, where an $SU(2)$ gauge theory is explored. The interest in extending the existing work to a non-Abelian case, was for us mostly driven by the possibility of new contributions to the correlation functions arising from self-interactions of the gauge fields in the Lagrangian. These new contributions are naturally absent in the Abelian case. For the bispectrum, we proved that they are potentially comparable and, in some case, even larger than the Abelian contributions. The same motivation leads us to study, in this paper, the trispectrum for the same kind of theories. 
In fact one of the reasons why the study of the trispectrum is important is that in most cases the three and four-point correlation 
functions are correlated to each other, being underlined by the same coupling constant. 
\\ 
        
Throughout the paper, we will employ the $\delta$N formula \cite{deltaN1,deltaN2,deltaN3,deltaN4}, in order to express the fluctuations in the e-folding number in terms of the fluctuations of all of the quantum fields on the initial temporal slice. In addition to that, the Schwinger-Keldysh formalism \cite{in-in1,in-in2,in-in3} will be used for calculating the non-Abelian contributions.\\

The organization of the paper goes as follows: in Section 2 we introduce and briefly discuss the Lagrangian of the model; in Section 3 we list all the terms contributing to the trispectrum of the curvature perturbation, classifying them in three separate categories, scalar, vector and mixed (depending whether they originate from scalar, vector or both fields); in Section 4 we calculate the mixed terms; in Section 5 we are faced with the vector terms, particularly with the non-Abelian contributions from the gauge field trispectrum, represented by vector-exchange and point-interaction diagrams; in Section 6 we provide an estimate of the $\tau_{NL}$ parameter; in Section 7 we study the shape of the trispectrum; in Section 8 we 
discuss how the anisotropic features modulate the trispectrum and show that the amplitude of the anisotropic part can be of the same order of magnitude as the isotropic one; in Section 9 we extend our calculations to models of gauge interactions where the kinetic term is multiplied by a function of time; in Section 10 we present our conclusions. The Appendices provide more details about some lengthy calculations and expressions shortened in the rest of the paper for the mixed contributions (Appendix A), for vector-exchange (Appendix B) and point-interation diagrams (Appendix C).

\section{The model}

Several models have been proposed which include primordial vector fields. We want to consider a scenario of a scalar-field driven inflation, where the inflaton field coexists with an $SU(2)$ gauge multiplet. The latter can play a non-negligible role in determining the amount of curvature perturbations through mechanisms that will be later discussed in the paper. A possible choice for the Lagrangian is

\bea\label{ac}\fl
S=\int d^{4}x \sqrt{-g}\left[\frac{m_{P}^{2}R}{2}-\frac{f^{2}(\phi)}{4}g^{\mu\al}g^{\nu\b}\sum_{a=1,2,3}F_{\mu\nu}^{a}F_{\al\b}^{a}-\left(\frac{m^2_{0}+\xi R}{2}\right)g^{\mu\nu}\sum_{a=1,2,3}B_{\mu}^{a}B_{\nu}^{a}+L_{\phi}\right],\nonumber\\
\eea
$L_{\phi}$ being the scalar field Lagrangian and $F_{\mu\nu}^{a}\equiv\p_{\mu}B^{a}_{\nu}-\p_{\nu}B^{a}_{\mu}+g_{c}\ep^{abc}B^{b}_{\mu}B^{c}_{\nu}$. The physical fields are defined as $A_{\mu}^{a}\equiv\left(B_{0}^{a},\vec{B}^{a}/a(t)\right)$ and $\xi$ is a numerical factor. The kinetic term for the gauge fields is multiplied by a generic function $f$ that can be viewed as a function of time.\\
The gauge field quantum fluctuations are expanded as follows
\bea
\fl
\delta A_{i}^{a}(\vec{x},\eta)=\int \frac{d^{3}q}{(2\pi)^{3}}e^{i\vec{q}\cdot\vec{x}}\sum_{\lambda=L,R,l}\Big[e^{\lambda}_{i}(\hat{q})a_{\vec{q}}^{a,\lambda}\delta A_{\lambda}^{a}(q,\eta)+e^{*\lambda}_{i}(-\hat{q})\left(a_{-\vec{q}}^{a,\lambda}\right)^{\dagger}\delta A_{\lambda}^{*a}(q,\eta)\Big],
\eea
where $a$ and $a^{\dagger}$ are creation and annihilation operators and the index $\lambda$ runs over left, right and longitudinal polarizations.\\
Depending on the functional form of $f$, on the value of the bare mass $m_{0}$ and of $\xi$, different models arise. In the special case $f=1$, it was proven in \cite{Dimopoulos:2006ms} that, for a small gauge field mass $m_{0}\ll H$ and choosing $\xi=1/6$, the transverse components of the gauge field perturbations satisfy equations of motion similar to the one of a minimally coupled light scalar field which, therefore, has a an almost scale-invariant spectrum on large scales. In this case, the gauge fields acquire an effective mass $M^{2}\equiv m_{0}^{2}-2 H^2$ (it is correct to use the approximation $R\simeq -12H^2$ during inflation). The reader can refer to Appendix A of \cite{Bartolo:2009pa} for all the details of the equations of motion for the background and for the gauge field perturbations and their solutions. On the other hand, the longitudinal wave function in this model has been shown to behave like a ghost, its kinetic energy being negative, and it is still debated whether or not it has a well-defined quantum field theory, also because it is governed by an equation of motion suffering from singularities. In spite of this, a regular solution was proposed in \cite{Dimopoulos:2008yv}. The physical validity of this model was questioned in \cite{Himmetoglu:2008zp,Himmetoglu:2008hx,Koivisto:2009sd,Himmetoglu:2009qi}. However these instabilities do not represent an issue in some models with a varying kinetic function where $f$ (and eventually the effective mass) is instead not a constant \cite{Yokoyama:2008xw,Dimopoulos:2009am,Dimopoulos:2009vu}. \\
In view of these different possibilities, we will employ for the longitudinal mode the parametrization in terms of the transverse mode $\delta B^{||}=n(x)\delta B^{T}$ we introduced in \cite{Bartolo:2009pa}, motivated by the intent of keeping our calculations as general as possible, in spite of having to make a specific choice for the Lagrangian. Our calculations can indeed be easily specified for different models. We are going to first consider the Lagrangian with $m_{0}\ll H$ and $\xi=1/6$, keeping $f=1$. In the last section of the paper, we show how to extend our calculations to $f(t)$ models.

\section{Trispectrum from $\delta$N formula}

We want to calculate the trispectrum of the curvature perturbation  
\bea\label{quartic,bisp}
\langle\zeta_{\vec{k}_{1}}\zeta_{\vec{k}_{2}}\zeta_{\vec{k}_{3}}\zeta_{\vec{k}_{4}} \rangle =(2 \pi)^3\delta^{3}(\vec{k}_{1}+\vec{k}_{2}+\vec{k}_{3}+\vec{k}_{4})T_{\zeta}(\vec{k}_{1},\vec{k}_{2},\vec{k}_{3},\vec{k}_{4})
\eea
using the $\delta N$ expansion. In the presence of primordial vector fields \cite{Dimopoulos:2008yv,Bartolo:2009pa}
\bea\label{expansion}\fl
\zeta(\vec{x},t)&=& N_{\phi}\df+N^{\mu}_{a}\delta A_{\mu}^{a}+\frac{1}{2}N_{\phi\phi}\left(\df\right)^2+\frac{1}{2}N^{\mu\nu}_{ab}\delta A_{\mu}^{a}\delta A_{\nu}^{b}+N_{\phi a}^{\mu}\df\delta A_{\mu}^{a}\nonumber\\\fl&+&\frac{1}{3!}N_{\phi\phi\phi}(\df)^3+\frac{1}{3!}N_{abc}^{\mu\nu\lambda}\de A_{\mu}^{a} \de A_{\nu}^{b} \de A_{\lambda}^{c}+\frac{1}{2}N_{\phi\phi a}^{\mu}(\df)^2\de A_{\mu}^{a}+\frac{1}{2}N_{\phi ab}^{\mu\nu}\df\de A_{\mu}^{a}\de A_{\nu}^{b}\nonumber\\\fl&+&\frac{1}{3!}N_{\phi\phi\phi\phi}(\df)^4+\frac{1}{3!}N_{abcd}^{\mu\nu\lambda\eta}\de A_{\mu}^{a}\de A_{\nu}^{b}\de A_{\lambda}^{c}\de A_{\eta}^{d}+...,
\eea
where $N$ is the number of e-foldings between an initial time $t^{*}$, which is often conveniently fixed at the horizon crossing era of the given wave number, and the final time $t$, labeling the slice of observation for $\zeta$. The partial derivatives of $N$ have been thus defined
\bea\label{NNN}
N_{\phi}\equiv \left(\frac{\p N}{\p \phi}\right)_{t^{*}},\,\,\,\,\,
N^{\mu}_{a}\equiv\left(\frac{\p N}{\p A^{a}_{\mu}}\right)_{t^{*}},\,\,\,\,\,N_{\phi a}^{\mu}\equiv\left(\frac{\p^{2} N}{\p \phi\p A^{a}_{\mu}}\right)_{t^{*}}
\eea
and so on for higher order derivatives.\\
Once Eq.(\ref{expansion}) is plugged in Eq.~(\ref{quartic,bisp}), several terms will arise. It is convenient to collect them in three categories: (purely) scalar (S), mixed (M) and vector (V)

\bea
T_{\zeta}\equiv T_{\zeta}^{S}+T_{\zeta}^{M}+T_{\zeta}^{V}.
\eea
We anticipate here that $T_{\zeta}^{M}$ and $T_{\zeta}^{V}$ can be written as the sum of an Abelian, $T_{\zeta\,(A)}^{M,V}$, and a non-Abelian, $T_{\zeta\,(NA)}^{M,V}$ , contributions. This will help stress what the origin of the different terms is and help derive the non-Abelian limit of our results.\\

Calculations of the purely scalar contributions already exist in the literature, whereas the mixed and the vector terms have not been computed so far. We report here the tree-level result \cite{SL3,Byrnes}

\bea\label{scalar}
T_{\zeta}^{S}&=&N_{\phi}^{4}T_{\phi}(\vec{k}_{1},\vec{k}_{2},\vec{k}_{3},\vec{k}_{4})\nonumber\\&+&N_{\phi}^{3}N_{\phi\phi}\left[P_{\phi}(k_{1})B_{\phi}(|\vec{k}_{1}+\vec{k}_{2}|,k_{3},k_{4})+perms.\right]\nonumber\\&+&N_{\phi}^{2}N_{\phi\phi}^{2}\left[P_{\phi}(k_{1})P_{\phi}(k_{2})P_{\phi}(|\vec{k}_{1}+\vec{k}_{3}|)+perms.\right]\nonumber\\&+&N_{\phi}^{3}N_{\phi\phi\phi}\left[P_{\phi}({k}_{1})P_{\phi}(k_{2})P_{\phi}(k_{3})+perms.\right].
\eea
We defined

\bea\fl
\langle\df_{\vec{k}_{1}}\df_{\vec{k}_{2}} \rangle=(2 \pi)^3\delta^{(3)}(\vec{k}_{1}+\vec{k}_{2})P_{\phi}(k),\\\fl
\langle\df_{\vec{k}_{1}}\df_{\vec{k}_{2}} \df_{\vec{k}_{3}}\rangle=(2 \pi)^3\delta^{(3)}(\vec{k}_{1}+\vec{k}_{2}+\vec{k}_{3})B_{\phi}(k_{1},k_{2},k_{3}),\\\fl
\langle\df_{\vec{k}_{1}}\df_{\vec{k}_{2}} \df_{\vec{k}_{3}}\df_{\vec{k}_{4}}\rangle=(2 \pi)^3\delta^{(3)}(\vec{k}_{1}+\vec{k}_{2}+\vec{k}_{3}+\vec{k}_{4})T_{\phi}(\vec{k}_{1},\vec{k}_{2},\vec{k}_{3},\vec{k}_{4}),
\eea
where, in single-field slow-roll inflation, to leading order we have: $P_{\phi}=H^{2}_{*}/(2k^3)$ ($H_{*}$ being the Hubble parameter evalutated at horizon exit $k=aH$); from \cite{SL2,Maldacena:2002vr} we learn the result
\bea
B_{\phi}\simeq\frac{\sqrt{\epsilon}H^{4}_{*}M(k_{i})}{m_{P}},
\eea 
$M$ being a function of the momenta moduli $k_{i}$ of dimension~$({\rm mass})^{-6}$; finally from \cite{Seery:2006vu,Seery:2008ax} we have 
\bea
T_{\phi}\simeq H^{6}_{*}\tilde{M}(\vec{k}_{i}),
\eea
where $\tilde{M}$ is a function of dimension~$({\rm mass})^{-11}$.\\
The new contributions from vectors will be evaluated in the next sections.

\section{Calculation of mixed contributions}

We list the total (connected) $T_{\zeta}$ contribution from the mixed (scalar-vector) terms

\bea\label{M}\fl
T_{\zeta}^{M}&=&N_{\phi}^{2}N_{\phi a}^{\mu}N_{\phi b}^{\nu}\left[P_{\mu\nu}^{ab}(\vec{k}_{1}+\vec{k}_{3})P_{\phi}^{}({k}_{1})P_{\phi}^{}({k}_{2})+perms.\right]\nonumber\\\fl&+&N_{a}^{\mu}N_{b}^{\nu}N_{\phi c}^{\rho}N_{\phi d}^{\sigma}\left[P_{\mu\rho}^{ac}(\vec{k}_{1})P_{\nu\sigma}^{bd}(\vec{k}_{2})P_{\phi}^{}(|\vec{k}_{1}+\vec{k}_{3}|)+perms.\right]\nonumber\\\fl&+&N_{\phi}^{2}N_{a}^{\mu}N_{\phi\phi b}^{\nu}\left[P_{\phi}^{}({k}_{1})P_{\phi}^{}({k}_{2})P_{\mu\nu}^{ab}(\vec{k}_{3})+perms.\right]\nonumber\\\fl&+&N_{\phi}^{}N_{a}^{\mu}N_{b}^{\nu}N_{\phi cd}^{\rho\sigma}\left[P_{\mu\rho}^{ac}(\vec{k}_{1})P_{\nu\sigma}^{bd}(\vec{k}_{2})P_{\phi}^{}({k}_{3})+perms.\right]\nonumber\\\fl&+&N_{\phi\phi}N_{\phi}N_{\phi a}^{\mu}N_{b}^{\nu}\left[P_{\phi}(k_{2})P_{\phi}(|\vec{k}_{1}+\vec{k}_{2}|)P_{\mu\nu}^{ab}(\vec{k}_{4})+perms.\right]\nonumber\\\fl&+&N_{ab}^{\mu\nu}N_{c}^{\rho}N_{\phi d}^{\sigma}N_{\phi}\left[P_{ac}^{\mu\rho}(\vec{k}_{2})P_{bd}^{\nu\sigma}(\vec{k}_{1}+\vec{k}_{2})P_{\phi}(k_{4})+perms.\right]\nonumber\\\fl&+&N_{\phi}^{2}N_{a}^{\mu}N_{\phi b}^{\nu}\left[P_{\mu\nu}^{ab}(\vec{k}_{3})B_{\phi}^{}(k_{1},k_{2},|\vec{k}_{3}+\vec{k}_{4}|)+perms.\right]\nonumber\\\fl&+&N_{a}^{\mu}N_{b}^{\nu}N_{\phi}N_{\phi c}^{\rho}\left[P_{\phi}^{}({k}_{3})B_{\mu\nu\rho}^{abc}(\vec{k}_{1},\vec{k}_{2},\vec{k}_{3}+\vec{k}_{4})+perms.\right]\nonumber\\\fl&+&[higher\,\,order\,\,terms],
\eea
where we defined as usual
\bea\label{exp-mixed}
\langle\de A_{\mu,\vec{k}_{1}}^{a}\de A_{\nu,\vec{k}_{2}}^{b} \rangle=(2 \pi)^3\delta^{(3)}(\vec{k}_{1}+\vec{k}_{2})P_{\mu\nu}^{ab}(\vec{k}),\\
\langle\de A_{\mu,\vec{k}_{1}}^{a}\de A_{\nu,\vec{k}_{2}}^{b} \de A_{\rho,\vec{k}_{3}}^{c}\rangle=(2 \pi)^3\delta^{(3)}(\vec{k}_{1}+\vec{k}_{2}+\vec{k}_{3})B_{\mu\nu\rho}^{abc}(\vec{k}_{1},\vec{k}_{2},\vec{k}_{3}).
\eea
In Eq.~(\ref{M}), we avoided including terms that involve mixed scalar-vector bispectra, the reason being that in the chosen Lagrangian no direct coupling exists between scalar (inflaton) and vector degrees of freedom. Before proceeding with the calculations, one more remark should be made which will hold for the rest of the paper: the partial derivatives of the e-folding number w.r.t. the temporal components of the gauge fields will be set to zero, the reason being that the $A_{0}^{a}$ are equal to zero (see Appendix A of \cite{Bartolo:2009pa}).\\
The power spectrum of the gauge fields was previously calculated in \cite{Bartolo:2009pa} and it reads
\bea\label{ps}
P_{ij}^{ab}(\vec{k})\equiv T^{even}_{ij}(\vec{k})P_{+}^{ab}+i T^{odd}_{ij}(\vec{k})P_{-}^{ab}+T^{long}_{ij}(\vec{k})P_{long}^{ab}
\eea 
where
\bea
T^{even}_{ij}(\vec{k})&=&\delta_{ij}-\hat{k}_{i}\hat{k}_{j},\\
T^{odd}_{ij}(\vec{k})&=&\epsilon_{ijk}\hat{k}_{k},\\
T^{long}_{ij}(\vec{k})&=&\hat{k}_{i}\hat{k}_{j}.
\eea
and
\bea
P_{R}^{ab}&\equiv& \delta_{ab}\delta A_{R}^{a}(k,t^{*})\delta A_{R}^{b*}(k,t^{*}),\\
P_{L}^{ab}&\equiv& \delta_{ab}\delta A_{L}^{a}(k,t^{*})\delta A_{L}^{b*}(k,t^{*}),\\
P_{long}^{ab}&\equiv& \delta_{ab}\delta A_{long}^{a}(k,t^{*})\delta A_{long}^{b*}(k,t^{*}),
\eea
$P_{\pm}^{ab}\equiv (1/2)(P_{R}^{ab}\pm P_{L}^{ab})$, $\delta A^{a}_{R,L,long}$ indicating the eigenfunctions for the gauge fields. The total power spectrum of $\zeta$ then becomes \cite{Bartolo:2009pa}

\bea\label{power-zeta}
P_{\zeta}(\vec{k})=P^{iso}(k)\left[1+g^{ab}\left(\hat{k}\cdot\vec{N}_{a}\right)\left(\hat{k}\cdot\vec{N}_{b}\right)+is^{ab}\hat{k}\cdot\left(\vec{N}_{a}\times\vec{N}_{b}\right)\right],
\eea
where $g^{ab}\equiv (P^{ab}_{long}-P^{ab}_{+})/[N^{2}_{\phi}P_{\phi}+\left(\vec{N}_{c}\cdot\vec{N}_{d}\right)P^{cd}_{+}]$ 
and $s^{ab}\equiv P^{ab}_{-}/[N^{2}_{\phi}P_{\phi}+\left(\vec{N}_{c}\cdot\vec{N}_{d}\right)P^{cd}_{+}]$
and the isotropic part of the power spectrum is 

\bea\label{piso}
P^{iso}(k)\equiv N^{2}_{\phi}P_{\phi}(k)+\left(\vec{N}_{c}\cdot\vec{N}_{d}\right)P^{cd}_{+}.
\eea

\noindent As to the vector modes bispectrum (in the last line of Eq.~(\ref{M})), this was also introduced and computed in \cite{Bartolo:2009pa} using the Schwinger-Keldysh formalism (see Eq.~(50) therein) and it turns out to be

\bea\label{bs}\fl
B_{ijk}^{abc}(\vec{k}_{1},\vec{k}_{2},\vec{k}_{3})=g_{c}^{2}H^{2}_{*}\left(\frac{x^{*}}{k}\right)^3\ep^{a^{'}ac}\ep^{a^{'}be}A^{e}_{l}\sum_{\alpha,\beta,\gamma}\left(\int dx\right)_{\alpha\beta\gamma}T^{\alpha}_{ip}T^{\beta}_{jp}T_{kl}^{\gamma}+perms.+c.c.
\eea
We define $x\equiv -k\eta$, where $k\equiv k_{1}+k_{2}+k_{3}+k_{4}$. As a reminder, $\left(\int dx\right)_{\alpha\beta\gamma}$ indicates the time integral over the wavefunctions of the internal legs multiplied by the wavefunctions of the external legs of the diagrams, the indices $\alpha,\beta$ and $\gamma$ runnig over longitudinal and transverse polarization states only. In the model we adopted, no violation of parity occurs, therefore $P^{ab}_{-}=0$ and $P_{+}^{ab}=P^{ab}_{R}$. \\
The next step is to calculate the time integrals in Eq~(\ref{bs}). It is well-known that, for the Lagrangian~(\ref{ac}) with $f=1$ and an
effective mass $M^2=m_0^2-2 H^2\simeq - 2 H^2$ the transverse mode behaves as a light scalar field during inflation \cite{Dimopoulos:2006ms}
\bea\label{T}
\delta B^{T}=-\frac{\sqrt{\pi x}}{2\sqrt{k}}\left[J_{3/2}(x)+iJ_{-3/2}(x)\right].
\eea
Throughout the paper, we find it convenient to adopt the parametrization introduced in \cite{Bartolo:2009pa} for the wavefunctions of the longitudinal mode in terms of the transverse mode, i.e.
\bea\label{L}
\delta B^{||}=n(x)\delta B^{T}\, ,
\eea
where $n$ is un unknown function of $x\equiv -k\eta$. Like in \cite{Bartolo:2009pa}, the choice of this parametrization is motivated by the convenience of a more general approach when handling longitudinal modes, in view of their still debated nature and of the possibility of making our analysis adaptable to more than one physical model.\\
When performing the time integrals, we should remind ourselves that, for the kind of wavefunctions we are dealing with, the major contribution comes from the region of integration around the horizon $H\simeq k/a$~\cite{Maldacena:2002vr}. Also, let us assume that $n(x)$ is a smooth function: this is reasonable, based on requirement that the longitudinal mode asymptotically scales as the transverse mode at late times so as to produce a scale invariant spectrum. For example this happens in the models with the transverse mode given by (\ref{T}) with $m_0\simeq 0$~ \cite{Dimopoulos:2006ms} , or in models with varying kinetic function and mass \cite{Dimopoulos:2009am,Dimopoulos:2009vu}. \footnote{The assumption that $m_{0}\simeq 0$ is crucial if by ``late time'' we mean a epoch within the end of inflation and well-after horizon crossing, since in the opposite case, i.e. for non-negligible $m_{0}$, the solution (\ref{T}) for the transverse mode would no longer apply and it would become harder to produce a scale-invariant power spectrum.} It then appears to be a good approximation evaluating $n(x)$ at the horizon crossing time $x^{*}$ and taking it out of the integrals in Eq.~(\ref{bs}), which leads to
\bea
\fl
\left(\int dx\right)_{EEE}&=&-\frac{1}{24k^{3}k_{1}^{2}k_{2}^{2}k_{3}^{2}x^{*5}}\left[A_{EEE}+\left(B_{EEE}\cos x^{*}+C_{EEE}\sin x^{*}\right)E_{i}x^{*}\right],\nonumber\\\fl
\left(\int dx\right)_{EEl}&=&\left(\int dx\right)_{ElE}=\left(\int dx\right)_{lEE}=n^{2}(x^{*})\left(\int dx\right)_{EEE},\nonumber\\\fl
\left(\int dx\right)_{llE}&=&\left(\int dx\right)_{lEl}=\left(\int dx\right)_{Ell}=n^{4}(x^{*})\left(\int dx\right)_{EEE},\nonumber\\\fl
\left(\int dx\right)_{lll}&=&n^{6}(x^{*})\left(\int dx\right)_{EEE}.
\eea
The letter `E' is an abbreviation for `even', whereas `l' stands for `longitudinal' and the definition of the functions $A_{EEE}$, $B_{EEE}$ and $C_{EEE}$ can be found in Appendix C of \cite{Bartolo:2009pa}. \\
The final expression for line $8$ of Eq.~(\ref{M}) becomes
\bea\label{final-M}\fl
T^{M}|_{line\,8}&=&N_{\phi}P_{\phi}(k_{3})g_{c}^{2}H^{2}_{*}\left(\frac{x^{*}}{k}\right)^3\left(\int dx \right)_{EEE}\big[{I}_{EEE}^{(8)}+n^{2}(x^{*})\left({I}_{EEl}^{(8)}+{I}_{ElE}^{(8)}+{I}_{lEE}^{(8)}\right)\nonumber\\\fl&+&n^{4}(x^{*})\left({I}_{llE}^{(8)}+{I}_{lEl}^{(8)}+{I}_{Ell}^{(8)}\right)+n^{6}(x^{*}){I}_{lll}^{(8)}\big]
\eea
where the expressions of the anisotropy coefficients ${I}_{\alpha\beta\gamma}^{(8)}$ can be found in Appendix A (Eqs.~(\ref{anisotropic-coefficients1}) through (\ref{anisotropic-coefficients2})). This is the only non-Abelian contributions to $T_{\zeta}^{M}$, the remaining terms are also present in the Abelian limit $g_{c}\rightarrow 0$.\\

In some vector field scenarios, the mixed scalar-vector derivatives $N_{\phi a...}^{\mu ...}$ vanish, so Eq.~(\ref{M}) does not contribute to the trispectrum. This can certainly happen in those models where a direct coupling between scalar and vector fields is missing, but the latter condition is not sufficient for concluding that the mixed derivatives are null. As an example, it is useful to refer to \cite{Vernizzi:2006ve}, which, among other things, includes an analytic study for the case of a set of slowly rolling fields with a separable quadratic potential. The number of e-folding is written as a sum of integrals over the different fields, to be evaluated between their values at an initial (generally set at around horizon crossing) and a final times. For each field, the value at the final time depends on the total field configuration at the initial time, so the mixed derivative of $N$ can in principle be non-zero. Anyway, if the final time approaches the end of inflation, it is reasonable to assume that, by then, the fields have stabilized to their equilibrium value and no longer carry the memory of their evolution. If this happens, the sum of integrals which defines $N$ becomes independent of the final field configuration and its mixed derivatives can therefore be shown to be zero. It turns out that we are allowed the same kind of analytic study, if we work with the Lagrangian in Eq.~(\ref{ac}) and if we introduce some slow-roll assumptions for the vector fields (see \cite{Bartolo:2009pa} for details, in particular Appendix A for a discussion about what these assumptions imply and Appendix B for the actual calculation of $N$ and its derivatives).

\section{Calculation of purely vector contributions}

Let us now turn to the vector trispectrum contributions: they are quite similar to the scalar ones (Eq.~(\ref{scalar}))

\bea\label{vector}
T_{\zeta}^{V}&=&N_{a}^{\mu}N_{b}^{\nu}N_{c}^{\rho}N_{d}^{\sigma}T_{\mu\nu\rho\sigma}^{abcd}(\vec{k}_{1},\vec{k}_{2},\vec{k}_{3},\vec{k}_{4})\nonumber\\&+&N_{a}^{\mu}N_{b}^{\nu}N_{c}^{\rho}N_{de}^{\sigma\delta}\left[P_{\mu\sigma}^{ad}(\vec{k}_{1})B_{\nu\rho\delta}^{bce}(\vec{k}_{1}+\vec{k}_{2},\vec{k}_{3},\vec{k}_{4})+perms.\right]\nonumber\\&+&N_{a}^{\mu}N_{b}^{\nu}N_{cd}^{\rho\sigma}N_{ef}^{\delta\eta}\left[P_{\mu\rho}^{ac}(\vec{k}_{1})P_{\nu\delta}^{be}(\vec{k}_{2})P_{\sigma\eta}^{df}(\vec{k}_{1}+\vec{k}_{3})+perms.\right]\nonumber\\&+&N_{a}^{\mu}N_{b}^{\nu}N_{c}^{\rho}N_{def}^{\sigma\delta\eta}\left[P_{\mu\sigma}^{ad}(\vec{k}_{1})P_{\nu\delta}^{be}(\vec{k}_{2})P_{\rho\eta}^{cf}(\vec{k}_{3})+perms.\right]
\eea
where
\bea\fl
\langle\de A_{\mu,\vec{k}_{1}}^{a}\de A_{\nu,\vec{k}_{2}}^{b} \de A_{\rho,\vec{k}_{3}}^{c}\de A_{\sigma,\vec{k}_{4}}^{d}\rangle=(2 \pi)^3\delta^{(3)}(\vec{k}_{1}+\vec{k}_{2}+\vec{k}_{3}+\vec{k}_{4})T_{\mu\nu\rho\sigma}^{abcd}(\vec{k}_{1},\vec{k}_{2},\vec{k}_{3},\vec{k}_{4})\, .
\eea

\noindent The power spectra and the bispectrum appearing in Eq.~(\ref{vector}) were provided in the previous section (Eqs.(\ref{ps}) and (\ref{bs})). All we are left with is then calculating the trispectrum of the gauge fields, i.e. $T_{\mu\nu\rho\sigma}^{abcd}$ in the first line of Eq.~(\ref{vector}). However, in the next pages, we are also going to rewrite the other lines of the expression above for completeness and for the computation of the non-Gaussianity parameter 
$\tau_{NL}$ in Sec. 6. Like in \cite{Bartolo:2009pa}, the terms that are proportional to $N_{0}^{a}$ will not be taken into account since $A^{a}_{0}=0$ for all $a=1,2,3$.

\subsection{Abelian terms in $T_{\zeta}^{V}$}

They include the third and fourth lines of Eq.(\ref{vector}). Let us rewrite them in such a way that their anisotropy coefficients are easily readable.\\
Let us first introduce the following quantity (we use the same notation as in \cite{Bartolo:2009pa})
\bea\fl
\label{defMs}
M_{k}^{c}(\vec{k})\equiv N_{a}^{i}P_{ik}^{ac}(\vec{k})=P^{ac}_{+}(k)\left[\delta_{ik}{N}_{a}^{i}+p^{ac}(k)\hat{k}_{k}\left(\hat{k}\cdot\vec{N}_{a}\right)+iq^{ac}(k)\left(\hat{k}\times\vec{N}_{a}\right)_{k}\right] 
\eea
where $p^{ac}(k)\equiv \left(P^{ac}_{long}-P^{ac}_{+}\right)/(P^{ac}_{+})$, $q^{ac}(k) \equiv P^{ac}_{-}/P^{ac}_{+}$ and $\vec{N}_{a}\equiv (N_{a}^{1},N_{a}^{2},N_{a}^{3})$.\\
Also, let us define
\bea\fl
L_{ce}^{jl}(\vec{k})&\equiv& N_{cd}^{ji}P_{ik}^{df}(\vec{k})N_{fe}^{kl}\nonumber\\\fl&=&P_{+}^{df}(\vec{k})\big[\vec{N}_{cd}^{j}\cdot\vec{N}_{ef}^{l}+p^{df}(k)\left(\hat{k}\cdot\vec{N}_{cd}^{j}\right)\left(\hat{k}\cdot\vec{N}_{ef}^{l}\right)+iq^{df}(k)\hat{k}\cdot\vec{N}^{j}_{cd}\times \vec{N}_{ef}^{l}\big]
\eea
where $\vec{N}_{cd}^{j}\equiv (N_{cd}^{j1},N_{cd}^{j2},N_{cd}^{j3})$.\\
Using the equations above, the third and the fourth lines of (\ref{vector}) become
\bea\label{VV}
T_{\zeta(A)}^{V}= M_{i}^{c}L_{ce}^{ij}M_{j}^{e}+M_{i}^{f}M_{j}^{e}M_{k}^{d}N_{fed}^{ijk}
\eea
where the subscript $A$ always stands for ``Abelian'' and $N_{fed}^{ijk}$ is the third derivative of $N$ w.r.t. the vector fields. Notice that the anisotropy coefficients $p^{ac}$ and $q^{ac}$ become zero respectively for theories in which transverse and longitudinal modes evolve exactly in the same way and $P_{-}^{ab}=0$.\\ 
In the situation where the vectors $\vec{N}_{a}$ are all aligned along a unique spatial direction, some more considerations can be made about the level of statistical anisotropy in Eq.~(\ref{VV}). In particular, for theories where $q^{ab}=0$ applies, $T_{\zeta(A)}^{V}$ isotropizes if the four wave vectors $\vec{k}_{i}$ lie in a plane perpendicular to the direction of the gauge vectors; otherwise, for theories where $p^{ab}= 0$, $T_{\zeta(A)}^{V}$  will appear isotropic when considering wave vectors parallel to the gauge vectors.

\subsection{Non-Abelian terms in $T_{\zeta}^{V}$}

The non-Abelian contributions to $T_{\zeta}^{V}$ are given by the first and second line of Eq.(\ref{vector}).\\
The bispectrum $B_{\nu\rho\delta}^{bce}$ was reported in Eq.(\ref{bs}). If we define $\tilde{N}^{k}_{c}\equiv M_{i}^{d}N^{ik}_{dc}$ (where $M_{i}^{d}$ 
was introduced in Eq.~(\ref{defMs})), then the second line of (\ref{vector}) can be rewritten as
\bea
{T_{\zeta}^{V}}|_{line\,\,2}=N_{a}^{i}N_{b}^{j}\tilde{N}^{k}_{c}B_{ijk}^{abc}(\vec{k}_{1}+\vec{k}_{2},\vec{k}_{3},\vec{k}_{4})
\eea
which is formally similar to Eq.(50) of \cite{Bartolo:2009pa}, with one of the $N_{a}^{i}$ replaced by $\tilde{N}_{a}^{i}$. The result is therefore given by
\bea\label{final-V}\fl
{T_{\zeta}^{V}}|_{line\,\,2}&=&g_{c}^{2}H^{2}_{*}{\left(\frac{x^{*}}{k}\right)}^{3} \left(\int dx\right)_{EEE}\big[\tilde{I}_{EEE}+n^{2}(x^{*})\left(\tilde{I}_{EEl}+\tilde{I}_{ElE}+\tilde{I}_{lEE}\right)\nonumber\\\fl&+&n^{4}(x^{*})\left(\tilde{I}_{llE}+\tilde{I}_{lEl}+\tilde{I}_{Ell}\right)+n^{6}(x^{*})\tilde{I}_{lll}\big]+perms.
\eea
where the $\tilde{I}_{\alpha\beta\gamma}$ are provided in Appendix A after Eqs.~(\ref{anisotropic-coefficients1})-(\ref{anisotropic-coefficients2}). It is interesting to notice that, in those models where the longitudinal and the transverse modes have the same time-evolution (i.e. $n(x)=1$) and parity is preserved (i.e. $P_{-}^{ab}=0$), the contribution from Eq.~(\ref{final-V}) becomes isotropic

\bea\label{anisotropo}\fl
{T_{\zeta}^{V}}|_{line\,\,2}&=&g_{c}^{2}H^{2}_{*}{\left(\frac{x^{*}}{k}\right)}^{3}\ep^{a^{'}ab}\ep^{a^{'}ce}\vec{N}_{a}\cdot\vec{N}_{c}^{}\tilde{\vec{N}}_{b}\cdot\vec{A}_{}^{e}\\\fl&\times&\left[-\frac{1}{24k^{3}k_{1}^{2}k_{2}^{2}k_{3}^{2}x^{*5}}\left(A_{EEE}+\left(B_{EEE}\cos x^{*}+C_{EEE}\sin x^{*}\right)E_{i}x^{*}\right)+perms\right]\nonumber.
\eea
Another observation worth to be made is that the $\tilde{I}_{\alpha\beta\gamma}$ become all equal to zero in the case where the three gauge vectors point in the same spatial direction. In order to prove this, one should employ, besides their definitions in (\ref{anisotropic-coefficients1})-(\ref{anisotropic-coefficients2}), also the explicit expressions for the derivatives of $N$ (which can be found in \cite{Dimopoulos:2008yv} or also in \cite{Bartolo:2009pa} and that will be used in Section~6 in the evaluation of $\tau_{NL}$).\\

We are now left with one last term to be analysed, i.e. the first line of Eq.(\ref{vector}). Like for the other non-Abelian contributions, this will be calculated using the Schwinger-Keldysh formula

\be\label{compact}
\langle\Omega|\Theta(t)|\Omega\rangle=\left\langle 0\left|\left[\bar{T}\left(e^{i {\int}^{t}_{0}H_{I}(t')dt'}\right)\right]\Theta_{I}(t)\left[T \left(e^{-i {\int}^{t}_{0}H_{I}(t')dt'}\right)\right]\right|0\right\rangle,
\ee  
where $|\Omega\rangle$ and $|0\rangle$ are the vacua respectively of the interaction and of the free theories. Both time-ordering, $T$, and anti-time-ordering operators, $\bar{T}$, are needed to account for the fact that the quantity on the left-hand side is not a scattering amplitude between an initial and a final state, it is instead an expectation value of a given cosmological observable at a given time. The subscript $I$ reminds the reader that we are working in interaction picture, so all the fields can be treated as free and thus expanded in terms of creation and annihilation operators in the usual way.\\
To leading order in the $SU(2)$ coupling  and in the perturbative expansion of the gauge fields, there are two types of diagrams, the vector-exchange and the point interaction diagrams (see Fig.~1), which arise respectively from third and fourth-order interactions in the Hamiltonian

\bea\label{third}
H_{int}^{(3)}&=&g_{c} \ep^{abc}g^{ik}g^{jl}\p_{i}\de B^{a}_{j}\de B^{b}_{k}\de B^{c}_{l}\\\label{fourth}
H_{int}^{(4)}&=&g_{c}^2 \ep^{eab}\ep^{ecd}g^{ij}g^{kl}\de B^{a}_{i}\de B^{b}_{k}\de B^{c}_{j}\de B^{d}_{l}.
\eea
Both of these contributions to $T_{\zeta}^{V}$ are expected to be of order $\sim g_{c}^{2}H_{*}^{4}$.\\

\begin{figure}\centering
 \includegraphics[width=0.3\textwidth]{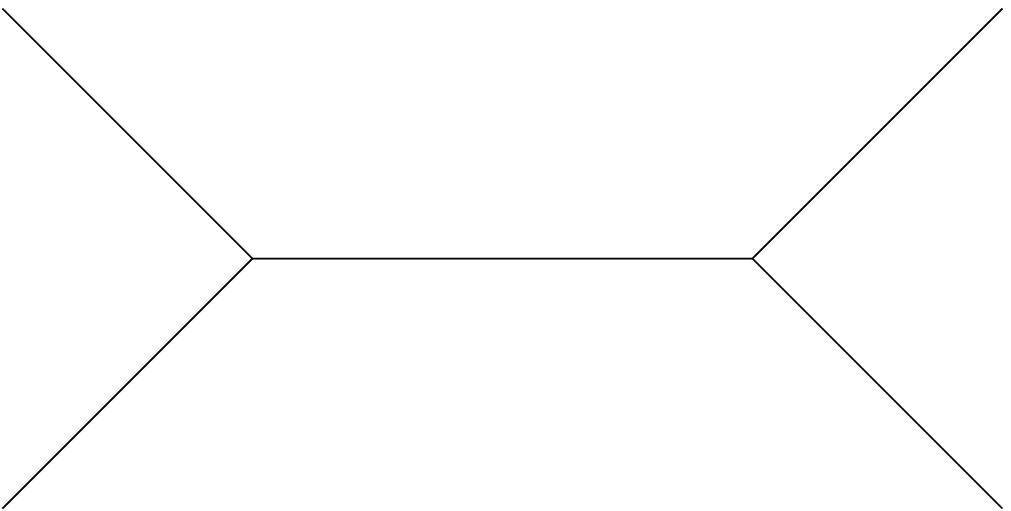}
\hspace{0.17\textwidth}
 \includegraphics[width=0.3\textwidth]{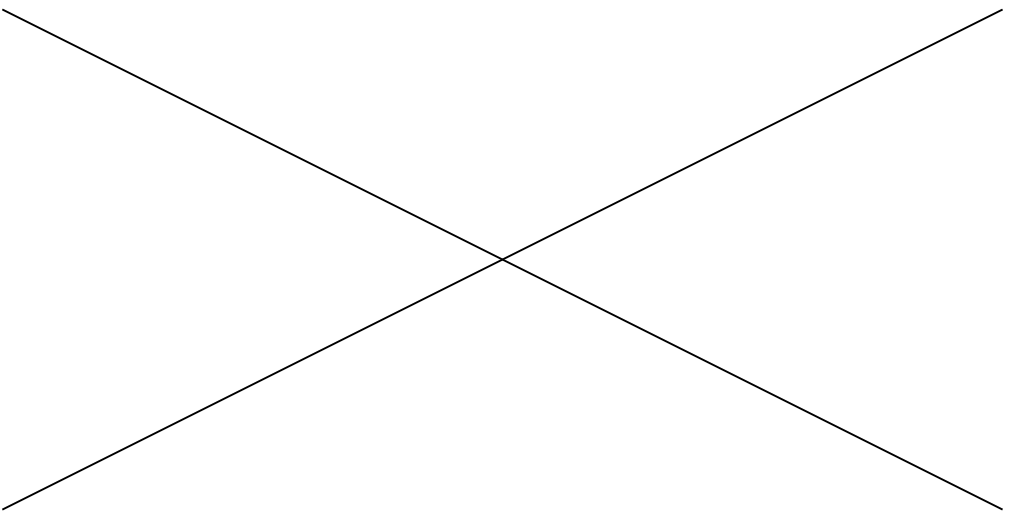}
\caption{ \label{Fig1} Diagrammatic representations of vector-enchange (on the left) \\and contact-interaction (on the right) contributions to the trispectrum.}
\end{figure}

\subsubsection{Vector-exchange diagrams\quad\quad\quad\quad\quad\quad\quad\quad\quad\quad\quad\quad\quad\quad\quad\quad\quad\quad\quad\quad\quad\quad}

Let us begin with the two vertex diagram. Using the language of Eq.~(\ref{compact}), it can be put in the form 
\bea\label{2v}\fl
\langle\Theta(\eta^{*})\rangle \supset \frac{(-i)^2}{2}\langle T \Big[\Theta \int_{-\infty}^{\eta^{*}}d\eta^{'}\left(H^{+}(\eta^{'})-H^{-}(\eta^{'})\right) \int_{-\infty}^{\eta^{*}}d\eta^{''}\left(H^{+}(\eta^{''})-H^{-}(\eta^{''})\right)\Big]\rangle
\eea
where now $H\equiv H^{(3)}_{int}$, $\Theta\equiv \de A_{\mu}^{a}\de A_{\nu}^{b} \de A_{\rho}^{c}\de A_{\sigma}^{d}$ and the inclusion symbol as usual points out that what stands on the right-hand side is only one of the contributions to $\langle\Theta(\eta^{*})\rangle$. The superscripts $+$ and $-$ refer to different rules of field-operator contractions, i.e.

\bean
\widehat{\delta B^{a,+}_{i}(\epr)\delta B^{b,+}_{j}(\eps)} = \tilde{\Pi}_{ij}^{ab}(\epr,\eps)\Theta(\epr-\eps)+\bar{\Pi}_{ij}^{ab}(\epr,\eps)\Theta(\eps-\epr),\\
\widehat{\delta B^{a,+}_{i}(\epr)\delta B^{b,-}_{j}(\eps)} = \bar{\Pi}_{ij}^{ab}(\epr,\eps),\\
\widehat{\delta B^{a,-}_{i}(\epr)\delta B^{b,+}_{j}(\eps)} = \tilde{\Pi}_{ij}^{ab}(\epr,\eps),\\
\widehat{\delta B^{a,-}_{i}(\epr)\delta B^{b,-}_{j}(\eps)} = \bar{\Pi}_{ij}^{ab}(\epr,\eps)\Theta(\epr-\eps)+\tilde{\Pi}_{ij}^{ab}(\epr,\eps)\Theta(\eps-\epr).
\eean
In Fourier space we have
\bea
\tilde{\Pi}_{ij}^{ab}(\vec{k})\equiv T_{ij}^{even}(\hat{k})\tilde{P}^{ab}_{+}+i T_{ij}^{odd}(\hat{k})\tilde{P}^{ab}_{ij}+T_{ij}^{long}(\hat{k})\tilde{P}^{ab}_{ij}\\
\bar{\Pi}_{ij}^{ab}(\vec{k})\equiv T_{ij}^{even}(\hat{k})\bar{P}^{ab}_{+}+i T_{ij}^{odd}(\hat{k})\bar{P}^{ab}_{ij}+T_{ij}^{long}(\hat{k})\bar{P}^{ab}_{ij}
\eea
where $\tilde{P}^{ab}_{\pm}\equiv (1/2)(\tilde{P}^{ab}_{R}\pm\tilde{P}^{ab}_{L})$, $\tilde{P}^{ab}_{R}$ being equal to the product of the two eigenfunctions $\delta_{ab}\delta B^{ab}_{R}(k,\eta)\delta B^{*ab}_{R}(k,\eta^{'})$ and $\bar{P}^{ab}_{\pm}=\left(\tilde{P}^{ab}_{\pm}\right)^{*}$ (similar definitions hold for $\tilde{P}^{ab}_{L}$ and $\tilde{P}^{ab}_{long}$). We remind the reader that in our model $T_{ij}^{odd}=0$.\\

\noindent Eq.~(\ref{2v}) can be rewritten as follows
\bea
\langle\Theta(\eta^{*})\rangle \supset\frac{(-i)^2}{2}\langle T\left[\Theta \left(\textit{A}+\textit{B}+\textit{C}+\textit{D}\right)\right]\rangle
\eea
where
\bea
\textit{A}\equiv \int_{-\infty}^{\eta^{*}}d\eta^{'}H^{+}(\eta^{'}) \int_{-\infty}^{\eta^{*}}d\eta^{''}H^{+}(\eta^{''})\nonumber\\
\textit{B}\equiv \int_{-\infty}^{\eta^{*}}d\eta^{'}H^{-}(\eta^{'}) \int_{-\infty}^{\eta^{*}}d\eta^{''}H^{-}(\eta^{''}) \nonumber\\
\textit{C}\equiv -\int_{-\infty}^{\eta^{*}}d\eta^{'}H^{+}(\eta^{'}) \int_{-\infty}^{\eta^{*}}d\eta^{''}H^{-}(\eta^{''}) \nonumber\\
\textit{D}\equiv -\int_{-\infty}^{\eta^{*}}d\eta^{'}H^{-}(\eta^{'}) \int_{-\infty}^{\eta^{*}}d\eta^{''}H^{+}(\eta^{''}) \nonumber
\eea
For each one of the integrals listed above, due to the presence of both the fields and their spatial derivatives in $H^{(3)}_{int}$, there are three different sets of contractions of the external with the vertex field-operators: for the first set, the field-operators with derivatives in the vertices are both contracted with external fields; for the second one, only one of the two field-operators with derivatives contracts with an external field (the other contracts with another internal field); for the third set, the field-operators with derivatives contract with each other.\\
A sample set of contractions of the first type is provided in the following equation
\bea\label{sample}\fl
T^{abcd}_{ijkl} &\supset&\frac{g_{c}^{2}}{2a^4(\eta^{*})}\ep^{a^{'}b^{'}c^{'}}\ep^{a^{''}b^{''}c^{''}}k_{m}k_{m^{'}}\nonumber\\\fl&\times&\int d\eta^{'}a^{4}(\eta^{'})\int d\eta^{''}a^{4}(\eta^{''})g^{mp}g^{m^{'}p^{'}}g^{nq}g^{n^{'}q^{'}}\tilde{\Pi}^{aa^{'}}_{in}\tilde{\Pi}^{bb^{'}}_{jp}\tilde{\Pi}_{kn^{'}}^{ca^{''}}\tilde{\Pi}^{db^{''}}_{lp^{'}}\tilde{\Pi}^{c^{'}c^{''}}_{qq^{'}}
\eea
where the first four $\tilde{\Pi}$s correspond to contractions between external and internal fields whereas the last one indicates the contraction between the two remaining internal fields. The $a^{-4}$ factor comes from expressing the external (physical) fields in terms of the comoving ones. As a reminder, we define $ (2\pi)^3\delta^{(3)}(\vec{k}_{1}+\vec{k}_{2}+\vec{k}_{3}+\vec{k}_{4})T^{abcd}_{ijkl}\equiv\langle \de A^{a}_{i} \de A^{b}_{j} \de A^{c}_{k} \de A^{d}_{l}\rangle$. The expression in (\ref{sample}) can be rewritten as follows

\bea\label{sec9}
T^{abcd}_{ijkl}&\supset&\frac{g_{c}^{2}}{2a^4(\eta^{*})}\ep^{a^{'}b^{'}c^{'}}\ep^{a^{''}b^{''}c^{''}}k_{m}k_{m^{'}}\delta^{aa^{'}}\delta^{bb^{'}}\delta^{ca^{''}}\delta^{db^{''}}\delta^{c^{'}c^{''}}\nonumber\\&\times&\sum_{\alpha,\beta,\gamma,\delta,\sigma}\left(\int d\eta^{'}\int d\eta^{''}\right)_{\alpha\beta\gamma\delta\sigma}T^{\alpha}_{in}T^{\beta}_{jm}T^{\gamma}_{kn^{'}}T^{\delta}_{lm^{'}}T^{\sigma}_{nn^{'}}
\eea

\noindent where, again, the greek indices of the sum indicate either the transverse (E) or longitudinal (l) modes, the $\left(\int d\eta^{'}\int d\eta^{''}\right)$ stand for the integrals over the wave functions, the chosen time variable being $x\equiv -k\eta$ ($k\equiv \sum_{i=1,...,4}k_{i}$, $k_{i}\equiv |\vec{k}_{i}|$).\\
Let us define the coefficients $T_{ijkl}^{\alpha\beta\gamma\delta\sigma}\equiv k_{m}k_{m^{'}} T^{\alpha}_{in}T^{\beta}_{jm}T^{\gamma}_{kn^{'}}T^{\delta}_{lm^{'}}T^{\sigma}_{nn^{'}}$. They should be calculated for each one of the three different sets of contractions and for each permutation within the specific set. This is a straghtforward but rather lengthy and not particularly interesting calculation. A convenient way to proceed could be the following: we first compute the time integrals in order to find out which one among the combinations of longitudinal and transverse mode functions in the string $[\alpha,\beta,\gamma,\delta,\sigma]$ provides the highest amplitude for the trispectrum (in order to be able to perform this comparison we work, as it is usually done when trying to quantify the amplitude of a three or of a four-point function, in the so called ``equilateral configuration'', which for the trispectrum means taking $k_{1}=k_{2}=k_{3}=k_{4}=k_{\hat{12}}=k_{\hat{14}}$); for the combination with the highest amplitude, we then calculate the coefficients $T_{ijkl}^{\alpha\beta\gamma\delta\sigma}$ for all the different sets of contractions and sum over all the permutations.\footnote{Notice that this procedure provides an approximate result, but it is certainly allowed since, when working in the equilateral configuration (as we do when we evaluate the amplitude of the trispectrum), it is easy to realize that the integrals $\left(\int d\eta^{'}\int d\eta^{''}\right)$ are independent of the specific set of operator contractions, which only differ by the $T_{ijkl}^{\alpha\beta\gamma\delta\sigma}$ and by the Levi-Civita coefficients.}\\
Let us now perform our calculations. The wavefunctions we are going to adopt were introduced in Sec.4 (see Eqs.~(\ref{T}) and (\ref{L})). It is possible to verify that 
$\textit{B}=\textit{A}^{*}$ and $\textit{D}=\textit{C}^{*}$ and that integrals of type \textit{A} are consistently smaller in amplitude than integrals of type \textit{C}. We therefore report the combined contribution $\textit{C}+\textit{D}=2 Re[\textit{C}]$ for one of the permutations

\bea
\fl\label{1}
\left(\int d\eta^{'}\int d\eta^{''}\right)_{EEEEE}&=&\frac{1}{8k^{3}_{1}k_{2}^{3}k_{3}^{3}k_{4}^{3}k_{\hat{12}}^{3}(k_{\hat{12}}+k_{1}+k_{2})(k_{\hat{12}}+k_{3}+k_{4})x^{*8}}\nonumber\\\fl&\times&\big[\big(M-2E\big)\big[\big(N-2F\big)\big(AB+CD\big)+\big(2H+L\big)\big(CB-AD\big)\big]\nonumber\\\fl&+&\big(2G+P\big)\big[\big(N-2F\big)\big(AD-CB\big)+\big(2H+L\big)\big(AB+CD\big)\big]\big]\\\fl
\left(\int d\eta^{'}\int d\eta^{''}\right)_{EEEEl}&=&n^{2}(x^{*})\left(\int d\eta^{'}\int d\eta^{''}\right)_{EEEEE}\\\fl
\left(\int d\eta^{'}\int d\eta^{''}\right)_{EEEll}&=&n^{4}(x^{*})\left(\int d\eta^{'}\int d\eta^{''}\right)_{EEEEE}\\\fl
\left(\int d\eta^{'}\int d\eta^{''}\right)_{EElll}&=&n^{6}(x^{*})\left(\int d\eta^{'}\int d\eta^{''}\right)_{EEEEE}\\\fl
\left(\int d\eta^{'}\int d\eta^{''}\right)_{Ellll}&=&n^{8}(x^{*})\left(\int d\eta^{'}\int d\eta^{''}\right)_{EEEEE}\\\fl\label{2}
\left(\int d\eta^{'}\int d\eta^{''}\right)_{lllll}&=&n^{10}(x^{*})\left(\int d\eta^{'}\int d\eta^{''}\right)_{EEEEE}
\eea
where $A$, $B$, $C$, $D$, $E$, $F$, $G$, $H$, $L$, $M$, $N$ and $P$ are functions of $x^{*}$ and of the momenta moduli to be provided in Appendix B (Eqs.~(\ref{eeeee1}) through (\ref{eeeee2})). Obviously the value of the integrals does not change when permuting its labels $\alpha\beta\gamma\delta\sigma$, besides having a different power of the coefficient $n(x^{*})$ multiplying them. We need now to find out if there is one, among the integrals in Eqs.~(\ref{1}) through (\ref{2}), that has the largest amplitude, i.e. understand if something can be said about the order of magnitude of $n(x^{*})$. We could try to extrapolate some information about $n(x^{*})$ from what happens at very late times. In the models discussed in 
Ref.~\cite{Dimopoulos:2006ms,Dimopoulos:2008yv} it turns out that the longitudinal mode is $\delta B^{||}=\sqrt{2}\delta B^{T}$.\footnote{
As another example, in models with varying kinetic function and mass \cite{Dimopoulos:2009am,Dimopoulos:2009vu}, we have verified that $n(x)\gg 1$ at late times and for a vector field that is light until the end of inflation.}If this is the correct asymptotic behaviour and we find it reasonable to extrapolate back until the horizon crossing epoch, it is then correct to conclude that in this case the amplitude is the largest for the integral among the ones listed in Eqs.~(\ref{1}) through (\ref{2}) containing the highest powers of $n$, i.e. for $\left(\int d\eta^{'}\int d\eta^{''}\right)_{lllll}$. The coefficients we intend to calculate are then of the kind $T_{ijkl}^{lllll}$ only. We list them below for the three different sets of contractions we introduced above and for one particular permutation 

\bea\label{tt}
T^{lllll(1)}_{ijkl}=k_{1}k_{3}k_{1234}\left(\hat{k}_{1}\cdot\hat{k}_{\hat{12}}\right)\left(\hat{k}_{1}\cdot\hat{k}_{2}\right)\left(\hat{k}_{3}\cdot\hat{k}_{4}\right)\left(\hat{k}_{3}\cdot\hat{k}_{\hat{12}}\right),\\
T^{lllll(2)}_{ijkl}=k_{3}k_{\hat{12}}k_{1234}\left(\hat{k}_{1}\cdot\hat{k}_{\hat{12}}\right)\left(\hat{k}_{2}\cdot\hat{k}_{\hat{12}}\right)\left(\hat{k}_{3}\cdot\hat{k}_{4}\right)\left(\hat{k}_{3}\cdot\hat{k}_{\hat{12}}\right),\\
T^{lllll(3)}_{ijkl}=k_{\hat{12}}k_{\hat{12}}k_{1234}\left(\hat{k}_{1}\cdot\hat{k}_{\hat{12}}\right)\left(\hat{k}_{2}\cdot\hat{k}_{\hat{12}}\right)\left(\hat{k}_{3}\cdot\hat{k}_{\hat{12}}\right)\left(\hat{k}_{4}\cdot\hat{k}_{\hat{12}}\right).
\eea 
We adopted the following notation: $k_{s}\equiv |\vec{k}_{s}|$, $\hat{k}_{s}\equiv\vec{k}_{s}/k_{s}$, the index $s$ running over the four external momenta; $k_{ss's''s'''}\equiv\hat{k}_{si}\hat{k}_{s'j}\hat{k}_{s''k}\hat{k}_{s'''l}$, with $s,s',s'',s'''=1,2,3,4$ and with the indices $i,j,k,l$ indicating the spatial components of the vectors; $\vec{k}_{\hat{ss'}}\equiv\vec{k}_{s}+\vec{k}_{s'}$, $k_{\hat{ss'}}\equiv|\vec{k}_{s}+\vec{k}_{s'}|$ and so $\hat{k}_{\hat{ss'}}\equiv\vec{k}_{\hat{ss'}}/k_{\hat{ss'}}$.\\

\noindent It is possible to prove that, once the Levi-Civita coefficients and the sum over the permutations are taken into account, only the first set of contractions is left (see again Appendix A for detailed calculations). The final result after these cancellations can be written in the following form

\bea\label{final-2v}
\fl
\langle \de A^{a}_{i} \de A^{b}_{j} \de A^{c}_{k} \de A^{d}_{l}\rangle_{*}  &\supset&(2 \pi)^3\delta^{(3)}(\vec{k}_{1}+\vec{k}_{2}+\vec{k}_{3}+\vec{k}_{4}) g_{c}^{2}\left(\frac{H_{*}x^{*}}{k}\right)^4\ep^{abc^{''}}\ep^{cdc^{''}}\Big[I \times k_{1234}\times \left(\sum_{i=1}^{4}t_{i}\right)\nonumber\fl\\&+&II \times k_{1324}\times \left(\sum_{i=5}^{8}t_{i}\right)+III \times k_{1432}\times \left(\sum_{i=9}^{12}t_{i}\right)\Big].
\eea
All the possible permutations have been included in the previous equation and, as a reminder, the indices $i,j,k,l$ are hidden in $k_{ss's''s'''}$ on the right-hand side. We define
\bea\fl
I&\equiv& n^{10}\times\left(\frac{1}{8k^{3}_{1}k_{2}^{3}k_{3}^{3}k_{4}^{3}k_{\hat{12}}^{3}(k_{\hat{12}}+k_{1}+k_{2})(k_{\hat{12}}+k_{3}+k_{4})x^{*8}}\right)\nonumber\\\fl&\times&\big[\big(M-2E\big)\big[\big(N-2F\big)\big(AB+CD\big)+\big(2H+L\big)\big(CB-AD\big)\big]\nonumber\\\fl&+&\big(2G+P\big)\big[\big(N-2F\big)\big(AD-CB\big)+\big(2H+L\big)\big(AB+CD\big)\big]\big]
\eea
(from Eqs.~(\ref{1})-(\ref{2})). The function $II$ is defined from $I$ by exchanging $k_{2}$ with $k_{3}$ and $k_{\hat{12}}$ with $k_{\hat{13}}$; similarly, $III$ is defined from $I$ by exchanging $k_{2}$ with $k_{4}$ and $k_{\hat{12}}$ with $k_{\hat{14}}$, so they are all functions of the horizon crossing time $x^{*}\equiv -k \eta^{*}$ and of the moduli of the external momenta and of their sums. This amounts to seven independent variables, $x^{*}$, $k_{1}$, $k_{2}$, $k_{3}$, $k_{4}$, $k_{\hat{12}}$ and $k_{\hat{14}}$. The coefficients $t_{i}$ ($i=1,...,12$) come from $T_{ijkl}^{\alpha\beta\gamma\delta\sigma}$ and so they are also functions of the momenta moduli (see Eqs.~(\ref{t1}) through (\ref{t12}) for their expressions). Finally, the anisotropic part of Eq.~(\ref{final-2v}) is represented by the $k_{ss's''s'''}$ terms, which, in the final expression for the curvature perturbation trispectrum, have their spatial indices contracted with the derivatives $N_{i}^{a}$ of the number of e-foldings w.r.t. the vector fields as follows

\bea\label{IA}\fl
\langle\zeta_{\vec{k}_{1}}\zeta_{\vec{k}_{2}}\zeta_{\vec{k}_{3}}\zeta_{\vec{k}_{4}} \rangle \supset N_{i}^{a}N_{j}^{b}N_{k}^{c}N_{l}^{d}\langle \de A^{a}_{i} \de A^{b}_{j} \de A^{c}_{k} \de A^{d}_{l}\rangle_{*} &\supset& (2 \pi)^3\delta^{(3)}(\vec{k}_{1}+\vec{k}_{2}+\vec{k}_{3}+\vec{k}_{4}) g_{c}^{2}\left(\frac{H_{*}x^{*}}{k}\right)^4\nonumber\\\fl&\times&I\times\left(\sum_{i=1}^{4}t_{i}\right)\times \Delta_{I}+perms.
\eea
where the anisotropic term in the first permutation is
\bea
\Delta_{I}\equiv \ep^{abc^{''}}\ep^{cdc^{''}}N_{i}^{a}N_{j}^{b}N_{k}^{c}N_{l}^{d}k_{1234}.
\eea
It can be interesting to rewrite $\Delta_{I}$ in terms of all its variables
\bea\fl
\Delta_{I}=\sum_{(a<b)a,b=1}^{3}{\left[\left(N^{a}\right)^2\left(N^{b}\right)^2\times\prod_{[i,j]=[1,2],[3,4]}{det\left(M^{i,j,a,b}_{I}\right)}\right]}
\eea
The $M_{I}$'s are $2\times 2$ matrices whose entries are represented by the cosines of the angles between the wavevectors and the $\vec{N}^{a}$ 

\[
M_{I}^{i,j,a,b}\equiv \left| \begin{array}{cc}
\cos\theta_{ia} & \cos\theta_{ja}  \\
\cos\theta_{ib} & \cos\theta_{jb} 
\end{array} \right|.\]
i.e. $\cos\theta_{ia}\equiv\hat{k}_{i}\cdot\hat{N}^{a}$ and so on.\\ 
The two permutations in Eq.~(\ref{IA}) can be written in a similar fashion with anisotropic coefficients 

\bea\fl
\Delta_{II}\equiv \ep^{abc^{''}}\ep^{cdc^{''}}N_{i}^{a}N_{j}^{b}N_{k}^{c}N_{l}^{d}k_{1324}=\sum_{(a<b)a,b=1}^{3}{\left[\left(N^{a}\right)^2\left(N^{b}\right)^2\times\prod_{[i,j]=[1,3],[2,4]}{det\left(M^{i,j,a,b}_{II}\right)}\right]},\nonumber\\
\fl
\Delta_{III}\equiv \ep^{abc^{''}}\ep^{cdc^{''}}N_{i}^{a}N_{j}^{b}N_{k}^{c}N_{l}^{d}k_{1432}=\sum_{(a<b)a,b=1}^{3}{\left[\left(N^{a}\right)^2\left(N^{b}\right)^2\times\prod_{[i,j]=[1,4],[3,2]}{det\left(M^{i,j,a,b}_{III}\right)}\right]}\nonumber.
\eea
The number of angular variables is equal to $12$. These are to be added to the six scalar variables from the isotropic part of (\ref{IA}) ($k_{1}$, $k_{2}$, $k_{3}$, $k_{4}$, $k_{\hat{12}}$ and $k_{\hat{14}}$) and to three parameters represented by the lengths of the vectors $\vec{N}^{a}$ in $\Delta$.\\
Like in (\ref{final-V}), the anisotropy coefficients here become equal to zero in the event of an alignment of the gauge vectors along a unique direction.\\
Notice that the isotropization we found out to occur in the $n(x)=1$ case for the trispectrum contribution in Eq.~(\ref{final-V}) (and that will be as well proven for the contact interaction in the next subsection), does not hold for the vector-exchange diagram. In fact, even without having to calculate the $T_{ijkl}^{\alpha\beta\gamma\delta\sigma}$ coefficients explicitly, it is easy to verify that, summing all of them up, we are left with a term proportional to $\epsilon^{a^{'}bc}\epsilon^{a^{'}de}\left(\vec{N}_{b}\cdot\vec{N}_{d}\right)\left(\hat{k}_{1}\cdot\vec{N}_{c}\right)\left(\hat{k}_{3}\cdot\vec{N}_{e}\right)$, coming from $T_{ijkl}^{EEEEE}$, plus its permutations. The anisotropy therefore survives because of the presence of spatial derivatives in the interaction Hamiltonian that are contracted with the external field operators.


\subsubsection{Point-interaction diagrams\quad\quad\quad\quad\quad\quad\quad\quad\quad\quad\quad\quad\quad\quad\quad\quad\quad\quad\quad\quad\quad\quad\quad\quad\quad\quad\quad\quad\quad\quad\quad\quad\quad\quad\quad\quad\quad\quad\quad\quad\quad\quad}Let us now move to the one-vertex diagrams
\bea
\langle\Theta(\eta^{*})\rangle &\supset& i\langle T \Big[\Theta \int_{-\infty}^{\eta^{*}}d\eta^{'}\left(H^{+}(\eta^{'})-H^{-}(\eta^{'})\right)\Big]\rangle, 
\eea
where now $H\equiv H_{int}^{(4)}$ and again $\Theta\equiv \de A_{\mu}^{a}\de A_{\nu}^{b} \de A_{\rho}^{c}\de A_{\sigma}^{d}$. After working out the Wick contractions, this becomes
\bea\label{trispec}\fl
N_{a}^{\mu}N_{b}^{\nu}N_{c}^{\rho}N_{d}^{\sigma}T_{\mu\nu\rho\sigma}^{abcd}(\vec{k}_{1},\vec{k}_{2},\vec{k}_{3},\vec{k}_{4}) &\supset&g_{c}^{2}\left(\frac{H_{*} x^{*}}{k}\right)^4\epsilon^{a'bc}\epsilon^{a'da}N^{m}_{a}N^{n}_{b}N^{o}_{c}N^{p}_{d}\nonumber\\\fl&\times&\sum_{\alpha\beta\gamma\delta}\left(\int dx\right)_{\alpha\beta\gamma\delta}T_{mi}^{\alpha}T_{nj}^{\beta}T_{oi}^{\gamma}T_{pj}^{\delta}+permutations.
\eea

\noindent Let us list the coefficients $T_{mnop}^{\alpha\beta\gamma\delta}\equiv  T_{mi}^{\alpha}T_{nj}^{\beta}T_{oi}^{\gamma}T_{pj}^{\delta}$ for one of the permutations

\bea\label{coeff1}\fl
T_{mnop}^{EEEE}&=& \de_{mo}\de_{np}-\de_{mo}\hat{k}_{p4}\hat{k}_{n4}-\de_{mo}\hat{k}_{n2}\hat{k}_{p2}+\de_{mo}\hat{k}_{n2}\hat{k}_{p4}\hat{k}_{2}\cdot\hat{k}_{4}-\de_{np}\hat{k}_{o3}\hat{k}_{m3}\nonumber\\\fl&+&k_{3434}+k_{3232}-\hat{k}_{2}\cdot\hat{k}_{4}\left(k_{3234}+k_{1214}\right)-\de_{np}\hat{k}_{m1}\hat{k}_{o1}+k_{1414}+k_{1212}\nonumber\\\fl&+&\de_{np}\hat{k}_{m1}\hat{k}_{o3}\hat{k}_{1}\cdot\hat{k}_{3}-\hat{k}_{1}\cdot\hat{k}_{3}\left(k_{1232}+k_{1434}\right)+\hat{k}_{1}\cdot\hat{k}_{3}\hat{k}_{2}\cdot\hat{k}_{4},\\\fl
T_{mnop}^{EEEl}&=& \de_{mo}\hat{k}_{p4}\hat{k}_{n4}-\de_{mo}\hat{k}_{n2}\hat{k}_{p4}\hat{k}_{2}\cdot\hat{k}_{4}-k_{3434}+\hat{k}_{2}\cdot\hat{k}_{4}\left(k_{3234}+k_{1214}\right)-k_{1414}\nonumber\\\fl&+&k_{1434}\hat{k}_{1}\cdot\hat{k}_{3}-k_{1234}\hat{k}_{1}\cdot\hat{k}_{3}\hat{k}_{2}\cdot\hat{k}_{4},\\\fl
T_{mnop}^{EElE}&=& \de_{np}\hat{k}_{o3}\hat{k}_{m3}-\de_{np}\hat{k}_{m1}\hat{k}_{o3}\hat{k}_{1}\cdot\hat{k}_{3}-k_{3434}+\hat{k}_{1}\cdot\hat{k}_{3}\left(k_{1434}+k_{1232}\right)-k_{3232}\nonumber\\\fl&+&k_{3234}\hat{k}_{2}\cdot\hat{k}_{4}-k_{1234}\hat{k}_{1}\cdot\hat{k}_{3}\hat{k}_{2}\cdot\hat{k}_{4},\\\fl
T_{mnop}^{ElEE}&=& \de_{mo}\hat{k}_{n2}\hat{k}_{p2}-\de_{mo}\hat{k}_{n2}\hat{k}_{p4}\hat{k}_{2}\cdot\hat{k}_{4}-k_{3232}+\hat{k}_{2}\cdot\hat{k}_{4}\left(k_{3234}+k_{1214}\right)-k_{1212}\nonumber\\\fl&+&k_{1232}\hat{k}_{1}\cdot\hat{k}_{3}-k_{1234}\hat{k}_{1}\cdot\hat{k}_{3}\hat{k}_{2}\cdot\hat{k}_{4},\\\fl
T_{mnop}^{lEEE}&=& \hat{k}_{m1}\hat{k}_{o1}\de_{np}-\hat{k}_{m1}\hat{k}_{o3}k_{13}\de_{np}-k_{1414}+\hat{k}_{1}\cdot\hat{k}_{3}\left(k_{1434}+k_{1232}\right)-k_{1212}\nonumber\\\fl&+&k_{1214}\hat{k}_{2}\cdot\hat{k}_{4}-k_{1234}\hat{k}_{1}\cdot\hat{k}_{3}\hat{k}_{2}\cdot\hat{k}_{4},\\\fl
T_{mnop}^{EEll}&=& k_{3434}-k_{3234}\hat{k}_{2}\cdot\hat{k}_{4}-k_{1434}\hat{k}_{1}\cdot\hat{k}_{3}+k_{1234}\hat{k}_{1}\cdot\hat{k}_{3}\hat{k}_{2}\cdot\hat{k}_{4},\\\fl
T_{mnop}^{ElEl}&=& \de_{mo}\hat{k}_{n2}\hat{k}_{p4}\hat{k}_{2}\cdot\hat{k}_{4}-\hat{k}_{2}\cdot\hat{k}_{4}\left(k_{3234}+k_{1214}\right)+k_{1234}\hat{k}_{1}\cdot\hat{k}_{3}\hat{k}_{2}\cdot\hat{k}_{4},\\\fl
T_{mnop}^{EllE}&=& k_{3232}-k_{3234}\hat{k}_{2}\cdot\hat{k}_{4}-k_{1232}\hat{k}_{1}\cdot\hat{k}_{3}+k_{1234}\hat{k}_{1}\cdot\hat{k}_{3}\hat{k}_{2}\cdot\hat{k}_{4},\\\fl
T_{mnop}^{llEE}&=& k_{1212}-k_{1214}\hat{k}_{2}\cdot\hat{k}_{4}-k_{1232}\hat{k}_{1}\cdot\hat{k}_{3}+k_{1234}\hat{k}_{1}\cdot\hat{k}_{3}\hat{k}_{2}\cdot\hat{k}_{4},\\\fl
T_{mnop}^{lEEl}&=& k_{1414}-k_{1214}\hat{k}_{2}\cdot\hat{k}_{4}-k_{1434}\hat{k}_{1}\cdot\hat{k}_{3}+k_{1234}\hat{k}_{1}\cdot\hat{k}_{3}\hat{k}_{2}\cdot\hat{k}_{4},\\\fl
T_{mnop}^{lElE}&=& \de_{np}\hat{k}_{m1}\hat{k}_{o3}\hat{k}_{1}\cdot\hat{k}_{3}-\hat{k}_{1}\cdot\hat{k}_{3}\left(k_{1434}+k_{1232}\right)+k_{1234}\hat{k}_{1}\cdot\hat{k}_{3}\hat{k}_{2}\cdot\hat{k}_{4},\\\fl
T_{mnop}^{lllE}&=& k_{1232}\hat{k}_{1}\cdot\hat{k}_{3}-k_{1234}\hat{k}_{1}\cdot\hat{k}_{3}\hat{k}_{2}\cdot\hat{k}_{4},\\\fl
T_{mnop}^{llEl}&=& k_{1214}\hat{k}_{2}\cdot\hat{k}_{4}-k_{1234}\hat{k}_{1}\cdot\hat{k}_{3}\hat{k}_{2}\cdot\hat{k}_{4},\\\fl
T_{mnop}^{lEll}&=& k_{1434}\hat{k}_{1}\cdot\hat{k}_{3}-k_{1234}\hat{k}_{1}\cdot\hat{k}_{3}\hat{k}_{2}\cdot\hat{k}_{4},\\\fl
T_{mnop}^{Elll}&=& k_{3234}\hat{k}_{2}\cdot\hat{k}_{4}-k_{1234}\hat{k}_{1}\cdot\hat{k}_{3}\hat{k}_{2}\cdot\hat{k}_{4},\\\label{coeff2}\fl
T_{mnop}^{llll}&=& k_{1234}\hat{k}_{1}\cdot\hat{k}_{3}\hat{k}_{2}\cdot\hat{k}_{4}.
\eea
When evaluating the integrals $\left(\int dx\right)_{\alpha\beta\gamma\delta}$, we use again the wavefunctions previously introduced in Eqs.~(\ref{L}) and (\ref{T}). The final result is

\bea\label{p-i}\fl
\left(\int dx\right)_{EEEE}&=& \frac{1}{24k^5k_{1}^{2}k_{2}^{2}k_{3}^{2}k_{4}^{2}x^{*7}}\big[Q_{EEEE}+A_{EEEE}\,\,\, ci x^{*}\left(B_{EEEE}\cos x^{*}+C_{EEEE}\sin x^{*}\right)\nonumber\\\fl&&+D_{EEEE}\,\,\, si x^{*}\left(E_{EEEE}\cos x^{*}+F_{EEEE} \sin x^{*} \right)\big] ,\\\fl 
\left(\int dx\right)_{EEEl}&=& n^{2}(x^{*})\left(\int dx\right)_{EEEE},\\\fl 
\left(\int dx\right)_{EEll}&=& n^{4}(x^{*})\left(\int dx\right)_{EEEE}  ,\\\fl 
\left(\int dx\right)_{lllE}&=&  n^{6}(x^{*})\left(\int dx\right)_{EEEE} ,\\\fl 
\left(\int dx\right)_{llll}&=& n^{8}(x^{*})\left(\int dx\right)_{EEEE} ,
\eea 
where $Q_{EEEE}$, $A_{EEEE}$, $B_{EEEE}$, $C_{EEEE}$, $D_{EEEE}$, $E_{EEEE}$ and $F_{EEEE}$ are functions of $x^{*}$ and of the momenta $k_{i}\equiv |\vec{k}_{i}|$, $ci$ and $si$ stand respectively for the CosIntegral and the SinIntegral functions. The expressions of these functions can be found in Appendix C.\\
It is again important noticing that the anisotropy coefficients become zero if the gauge fields are all aligned.\\
Finally, summing up the coefficients in Eqs.~(\ref{coeff1}) through (\ref{coeff1}), one realizes that if the longitudinal and the transverse mode evolve in the same way, the total contribution from the point-interaction diagram is isotropic

\bea\label{isot-trisp}\fl
\langle\zeta_{\vec{k}_{1}}\zeta_{\vec{k}_{2}}\zeta_{\vec{k}_{3}}\zeta_{\vec{k}_{4}} \rangle &\supset&(2 \pi)^3\delta^{3}\left(\vec{k}_{1}+\vec{k}_{2}+\vec{k}_{3}+\vec{k}_{4}+\right)g_{c}^{2}\left(\frac{H_{*} x^{*}}{k}\right)^4\epsilon^{a'bc}\epsilon^{a'da}\left(\vec{N}^{c}\cdot\vec{N}^{a}\right)\left(\vec{N}^{b}\cdot\vec{N}^{d}\right)\nonumber\\\fl&\times&\frac{1}{24k^5 k^{2}_{1}k^{2}_{2}k^{2}_{3}k^{2}_{4}x^{*7}}\big[Q_{EEEE}+A_{EEEE}ci x^{*}\left(B_{EEEE}\cos x^{*}+C_{EEEE}\sin x^{*}\right)\nonumber\\\fl&&+D_{EEEE}si x^{*}\left(E_{EEEE}\cos x^{*}+F_{EEEE} \sin x^{*} \right)\big]+permutations.
\eea

\section{Calculation of $\tau_{NL}$}

The non-Gaussianity of a given theory of inflation and cosmological perturbations can be studied by looking at the expression of the two well-known parameters $f_{NL}$ and $\tau_{NL}$.\\
The parameter $\tau_{NL}$ is defined from the trispectrum normalized with the isotropic part of the $\zeta$ power spectrum \footnote{This normalization choice, which simplifies the expressions of $\tau_{NL}$ compared to normalizing with the complete $P_{\zeta}$ from Eq.~(\ref{power-zeta}), is motivated by the fact that the anisotropic part of the power spectrum cannot noticeably affect our results, being weighted by parameters such as $g^{ab}$, that have to be kept much smaller than one in order to avoid an anisotropic expansion.}
\bea
\tau_{NL}=\frac{2T_{\zeta}(\vec{k}_{1},\vec{k}_{2},\vec{k}_{3},\vec{k}_{4})}{P^{iso}(k_{1})P^{iso}(k_{2})P^{iso}(k_{\hat{14}})+23\,\,  {\rm perms.}},\eea
where $P^{iso}$ was provided in Eq.~(\ref{piso}), and we recall that $\vec{k}_{\hat{14}}\equiv\vec{k}_{1}+\vec{k}_{4}$.\\
The total $\tau_{NL}$ will then in general be a function the vectors $\vec{k}_{i}$, $i=1,...,4$ and can be written as the sum of scalar and vector contributions (we remind the reader that we are working under the conditions that the mixed contributions are equal to zero)
\bea
\tau_{NL}=\tau_{NL}^{(S)}+\tau_{NL}^{(V)}.
\eea 
More explicitly, we have
\bea
\tau_{NL}^{(S)}=\tau_{NL}^{S,1}+\tau_{NL}^{S,2}+\tau_{NL}^{S,3}+\tau_{NL}^{S,4},\\
\tau_{NL}^{(V)}=\tau_{NL}^{V,1}+\tau_{NL}^{V,2}+\tau_{NL}^{V,3}+\tau_{NL}^{V,4},
\eea
where we counted, both for the scalar and the vector parts, four different contributions which can be read from lines $1$ through $4$ of the right-hand sides of both Eqs.~(\ref{scalar}) and (\ref{vector}). Notice that $\tau_{NL}^{V,1}$ and $\tau_{NL}^{V,2}$ are the non-Abelian contributions, $\tau_{NL}^{V,3}$ and $\tau_{NL}^{V,4}$ the Abelian ones. More precisely, $\tau_{NL}^{V,2}$ refers to the contribution from the vector field bispectrum, Eq.~(\ref{final-V}), $\tau_{NL}^{V,1}$ refers to the vector-field trispectrum and therefore includes both vector-exchange and contact-interaction terms. Let us now carry out a quick estimate of the different contributions to $\tau_{NL}$, ignoring vector and gauge indices. The scalar contributions are
\bea
\tau_{NL}^{S,1}&=& \frac{\epsilon}{\left(1+\beta\right)^3},\\
\tau_{NL}^{S,2}&=&\tau_{NL}^{S,3}=\tau_{NL}^{S,4}= \frac{\epsilon^2}{\left(1+\beta\right)^3} .
\eea
where we define $\beta\equiv\left(N_{A}/N_{\phi}\right)^2$. The vector contributions are included in the Table 1 for the general case (without specifying the explicit expressions for the derivatives $N_{A}$ and $N_{AA}$), in vector inflation and in the vector curvaton model.

\begin{table}[h]\centering
\caption{Order of magnitude of the vector contributions to $\tau_{NL}$\\ in different scenarios.\\}
\begin{tabular}{|c||c|c|c|c|}\hline 
 $$ & \scriptsize{$\tau_{NL}^{V,1}$} & \scriptsize{$\tau_{NL}^{V,2}$} & \scriptsize{$\tau_{NL}^{V,3}$} & \scriptsize{$\tau_{NL}^{V,4}$}   \\
\hline 
\scriptsize{general case} & \scriptsize{$10^3\frac{\beta^2\epsilon g_{c}^{2}}{\left(1+\beta\right)^3}\left(\frac{m_{P}}{H}\right)^2$} & \scriptsize{$10^{-5}\frac{\beta^{3/2}\epsilon^{3/2}g_{c}^{2}}{\left(1+\beta\right)^3}\left(\frac{A}{H}\right)\left(\frac{m_{P}}{H}\right)m_{P}^2N_{AA}$} & \scriptsize{$\frac{\beta\epsilon^2}{\left(1+\beta\right)^3}m_{P}^4N_{AA}^2$} & \scriptsize{$\frac{\beta^{3/2}\epsilon^{3/2}}{\left(1+\beta\right)^3}m_{P}^3N_{AAA}$} \\
\hline
\scriptsize{v.inflation} & \scriptsize{same as above} & \scriptsize{$10^{-5}\frac{\beta^{3/2}\epsilon^{3/2}g_{c}^{2}}{\left(1+\beta\right)^3}\left(\frac{A}{H}\right)\left(\frac{m_{P}}{H}\right)$} & \scriptsize{$\frac{\beta\epsilon^2}{\left(1+\beta\right)^3}$} & $0$  \\
\hline
\scriptsize{v.curvaton} & \scriptsize{same as above} & \scriptsize{$10^{-5}\frac{r\beta^{3/2}\epsilon^{3/2}g_{c}^{2}}{\left(1+\beta\right)^3}\left(\frac{A}{H}\right)\left(\frac{m_{P}}{H}\right)\left(\frac{m_{P}}{A}\right)^2$} & \scriptsize{$\frac{r^2\beta\epsilon^2}{\left(1+\beta\right)^3}\left(\frac{m_{P}}{A}\right)^4$} & \scriptsize{$\frac{r\beta^{3/2}\epsilon^{3/2}}{\left(1+\beta\right)^3}\left(\frac{m_{P}}{A}\right)^3$} \\
\hline
\end{tabular}
\label{table1}
\end{table}
\noindent The previous expressions were derived using $N_{\phi}\simeq(m_{P}\sqrt{\epsilon})^{-1}$, $N_{\phi\phi}\simeq m_{P}^{-2}$, $N_{\phi\phi\phi}\simeq \sqrt{\epsilon}/m_{P}^{3}$, with $\epsilon=(\dot{\phi}^2)/\left(2m_{P}^{2} H^{2}\right)$. Numerical coefficients of order one in the amplitudes were not reported. The  $10^3$ factor in $\tau_{NL}^{V,1}$ is entirely due to the vector-exchange contribution, which turns out to have an amplitude that is several orders of magnitude larger than the one of the point-interaction term. For the derivatives of $N$ w.r.t. the vector fields, the results of \cite{Dimopoulos:2008yv,Bartolo:2009pa} were employed, which we schematically (neglecting all indices) report below
\bea
N_{A}\simeq \frac{A}{m_{P}^2},\quad\quad N_{AA}\simeq \frac{1}{m_{P}^{2}},\quad\quad N_{AAA}=0
\eea
in vector inflation, and
\bea
N_{A}\simeq \frac{r}{A},\quad\quad\quad N_{AA}\simeq \frac{r}{A^{2}},\quad\quad N_{AAA}\simeq\frac{r}{A^3}
\eea
for the vector curvaton model, where $r\equiv \left(3\rho_{A}\right)\left(3\rho_{A}+4\rho_{\phi}\right)$ ($\rho_{A}$ and $\rho_{\phi}$ indicating respectively the energy densities of vector fields and inflaton at the epoch of the curvaton decay). Here what we call ``vector inflation'' is a regular scalar-field-driven inflation occurring in the presence of vector fields whose contribution to the total universe energy density is assumed to be subdominant w.r.t. the inflaton energy density in such a way that the expansion remains approximately (and within experimental bounds) isotropic. The vector curvaton model, proposed in \cite{Dimopoulos:2006ms} for the first time, is similar to the classic curvaton mechanism \cite{Mollerach,Enqvist:2001zp,Lyth:2001nq,Lyth:2002my,Moroi:2001ct,Bartolo:2003jx}, except for vector fields, rather than scalars, playing the role of the curvaton field.\\

The parameters $\beta$, $\epsilon$, $g_{c}$ and $r$ appearing in the previous expressions are all assumed to be smaller than unity. Concerning $\beta$, it is well-known in the $U(1)$ case that it has to be kept small in order to prevent too much anisotropy from affecting the power spectrum (see \cite{Groeneboom:2008fz} where data analysis is provided in this sense; this is also discussed in \cite{Dimopoulos:2008yv}); no data analysis has up to now been performed considering more than one preferred spatial direction, so, unless all the gauge vectors are about aligned with one another, the upper bound on $\beta$ is not known. It sounds nevertheless safe and reasonable to assume that this quantity is quite small also for a generic orientation of the vector fields in our $SU(2)$ model. The parameters $\epsilon$ and $g_{c}$ are obviously supposed to be small respectively for ensuring slow-roll of the inflaton field and allowing perturbation theory to hold. Finally, in the curvaton model, $r$ has to remain small at least until the end of inflation for an almost isotropic exponential expansion to occur. \\
The ratio $m_{P}/H$ is of order $10^{5}$ (assuming $\epsilon\simeq 10^{-1}$ and from the observed power spectrum of fluctuations); as to the other ratios $A/H$ and $m_{P}/A$ appearing in Table~1, their value is not strictly constrained. In fact, if on one hand it is reasonable to assume that the expectation value of the background gauge fields does not exceed the Planck mass, in principle no constraint can be put on $A/H$, which can generically be a very large number, even though specific gauge field configurations exist where it is of order one. An example of such configurations is the one where all the gauge fields have the same magnitude and are not approximately aligned (see Section 6 and Appendix A of \cite{Bartolo:2009pa} for a complete discussion). To conclude our observations on Table~1, we can say that $\tau_{NL}^{V,1}$ and $\tau_{NL}^{V,2}$ can be either smaller or much larger than one both in vector inflation and in the vector curvaton model, depending on which region of parameter space is considered, $\tau_{NL}^{V,3}$ and $\tau_{NL}^{V,4}$ are certainly very small in vector inflation whereas there is room for them to be large in vector curvaton.\\

A comparison between scalar and vector contributions can be done by looking at the expressions reported in Table~2. Again, one can conclude that the vector contribution being dominant or subdominant compared to the scalar one very much depends on the parameter space region we want to consider (similarly to the disussion above and recalling that the contribution from the scalar field fluctuations in slow-roll inflation is at most of order $\epsilon$). \\

\begin{table}[h]\centering
\caption{Order of magnitude of ther ratios $\tau_{NL}^{V}/\tau_{NL}^{S,1}$ in different scenarios.\\}
\begin{tabular}{|c||c|c|c|c|}\hline 
 $$ & \scriptsize{$\tau_{NL}^{V,1}/\tau_{NL}^{S,1}$} & \scriptsize{$\tau_{NL}^{V,2}/\tau_{NL}^{S,1}$} & \scriptsize{$\tau_{NL}^{V,3}/\tau_{NL}^{S,1}$} & \scriptsize{$\tau_{NL}^{V,4}/\tau_{NL}^{S,1}$}   \\
\hline 
\scriptsize{general case} & \scriptsize{$10^3\beta^2 g_{c}^{2}\left(\frac{m_{P}}{H}\right)^2$} & \scriptsize{$10^{-5}\beta^{3/2}\epsilon^{1/2}g_{c}^{2}\left(\frac{A}{H}\right)\left(\frac{m_{P}}{H}\right)m_{P}^2N_{AA}$} & \scriptsize{$\beta\epsilon m_{P}^4N_{AA}^2$} & \scriptsize{$\beta^{3/2}\epsilon^{1/2}m_{P}^3N_{AAA}$} \\
\hline
\scriptsize{v.inflation} & \scriptsize{same as above} & \scriptsize{$10^{-5}\beta^{3/2}\epsilon^{1/2}g_{c}^{2}\left(\frac{A}{H}\right)\left(\frac{m_{P}}{H}\right)$} & \scriptsize{$\beta\epsilon$} & $0$  \\
\hline
\scriptsize{v.curvaton} & \scriptsize{same as above} & \scriptsize{$10^{-5}r\beta^{3/2}\epsilon^{1/2}g_{c}^{2}\left(\frac{A}{H}\right)\left(\frac{m_{P}}{H}\right)\left(\frac{m_{P}}{A}\right)^2$} & \scriptsize{$r^2\beta\epsilon\left(\frac{m_{P}}{A}\right)^4$} & \scriptsize{$r\beta^{3/2}\epsilon^{1/2}\left(\frac{m_{P}}{A}\right)^3$} \\
\hline
\end{tabular}
\label{table2}
\end{table}

It is possible to observe that $f_{NL}$, as calculated in \cite{Bartolo:2009pa} (see Table~1 therein), and $\tau_{NL}^{V}$ depend on the same set of parameters (especially if the non-Abelian contribution to the bispectrum non-linearity parameter, i.e. $f_{NL}^{NA}$, is taken into account); this can be interesting for different reasons: for example, in the event that the bispectrum and the trispectrum were independently known, we could then attempt to restrict the extent of the parameter space of the underlying theory; also, it is evident from Table~3 that there is enough room in the parameter space of the models we considered for $\tau_{NL}$ to be large in spite of a small unobservable $f_{NL}$. 
 
\begin{table}[h]\centering
\caption{Order of magnitude of the ratios $\tau_{NL}^{V}/\left(f_{NL}^{NA}\right)^2$ in different scenarios.\\}
\begin{tabular}{|c||c|c|c|c|}\hline 
 $$ & \scriptsize{$\tau_{NL}^{V,1}/\left(f_{NL}^{NA}\right)^2$} & \scriptsize{$\tau_{NL}^{V,2}/\left(f_{NL}^{NA}\right)^2$} & \scriptsize{$\tau_{NL}^{V,3}/\left(f_{NL}^{NA}\right)^2$} & \scriptsize{$\tau_{NL}^{V,4}/\left(f_{NL}^{NA}\right)^2$}   \\
\hline 
\scriptsize{v.i.} & \scriptsize{$10^9\frac{\epsilon\left(1+\beta\right)}{g_{c}^{2}\beta^2}\left(\frac{H}{m_{P}}\right)^2$} & \scriptsize{$10\frac{\epsilon^{3/2}\left(1+\beta\right)}{\beta^{5/2}g_{c}^{2}}\left(\frac{A}{H}\right)\left(\frac{H}{m_{P}}\right)^3$} & \scriptsize{$10^6\frac{\epsilon^2\left(1+\beta\right)}{\beta^3g_{c}^{4}}\left(\frac{H}{m_{P}}\right)^4$} & $0$  \\
\hline
\scriptsize{v.c.} & \scriptsize{$10^9\frac{r^2\epsilon\left(1+\beta\right)}{g_{c}^{2}\beta^2}\frac{m_{P}^{2}}{A^{2}}\frac{H^{2}}{A^{2}}$} & \scriptsize{$10\frac{r^5\epsilon^{3/2}\left(1+\beta\right)}{\beta^{5/2}g_{c}^{2}}\frac{H^{3}}{A^{3}}\frac{m_{P}}{H}\frac{m_{P}^{2}}{A^{2}}$} & \scriptsize{$10^6\frac{r^6\epsilon^2\left(1+\beta\right)}{\beta^3 g_{c}^{4}}\left(\frac{m_{P}}{A}\right)^4\left(\frac{H}{A}\right)^4$} & \scriptsize{$10^6\frac{r^3\epsilon^{3/2}\left(1+\beta\right)}{g_{c}^{2}\beta^{5/2}}\frac{m_{P}^{3}}{A^{3}}\frac{H^{4}}{A^{4}}$} \\
\hline
\end{tabular}
\label{table3}
\end{table}
\noindent 

In Table 3 $f_{NL}^{NA}$ represents the amplitude of the bispectrum coming from the non-Abelian interactions which, as explained in 
Ref.~\cite{{Bartolo:2009pa}}, can be the dominant contribution. It is given by 
$f_{NL}^{NA} \equiv 10^{-3}\left(\beta g_{c}/(1+\beta)\right)^2\left(m/H\right)^2$, $m\simeq m_{P}$ in vector inflation and $m\simeq A/\sqrt{r}$ in the vector curvaton model.

\section{Profile of the trispectrum}

The shape of a correlation function such as the bispectrum or the trispectrum is in general very peculiar to the specific model or class of models under investigation, this is why it is important to be able to carefully analyse these profiles and to find distinctive features that can make them easily recognizable.\\
As to the bispectrum and for theories where it is isotropic, this task is fairly simple given the small number of parameters it depends on (i.e. the external momenta moduli $k_{1}$, $k_{2}$ and $k_{3}$). In theories where it is anisotropic, the bispectrum profile becomes harder to analyse: the number of free parameters notably increases with the addition of the angles defining the existing preferred spatial directions. We provided an example of this in \cite{Bartolo:2009pa}, where we proceeded by first analyzing the isotropic part of the bispectrum in order to identify the preferred momentum configuration and then, in that specific configuration, by studying the effect of the anisotropic part on amplitudes and profiles. Producing a plot of the complete shape was not possible though, unless narrowing the number of angular variables down, by being more specific about the spatial orientation of the $\vec{N}_{a}$ vector w.r.t. the $\vec{k}_{i}$.\\
For the trispectrum, the situation is already complicated at the isotropic level. In fact, even in this case, the total number of parameters is too large for straighforwardly generating a final plot. It is then necessary to restrict the total number of parameters with the help of some assumptions about the relative spatial orientation or about the moduli of the wave vectors. A list of possible configurations for the tetrahedron made up by the four wave vectors, was provided in \cite{Chen:2009bc}, where the trispectrum is calculated for a general Lagrangian in single-field inflation. \\
We are going to proceed in a similar fashion and, in analogy with what was done for the bispectrum \cite{Bartolo:2009pa}: we analyse the isotropic part first and, in the next section, we plot the  anisotropic coefficients in a sample momentum configuration. The main purpose is to provide a hint of how the existence of preferred spatial directions is responsible for modulating the amplitude of the trispectrum, given a particular momentum dependence.\\  
Our study of the profile by no means intends to be exhaustive, in fact we are going to consider only two of the many possible tetrahedron categories (the so called ``equilateral'' and ``specialized planar''); also, we will not plot all of the contributions to the trispectrum we analysed in this paper, but only focus on the contributions to the vector-field trispectrum, arising from vector-exchange and contact-interaction diagrams. We leave the task of working out a more systematic and complete shape analysis for future work.

\begin{figure}\centering
\includegraphics[width=0.4\textwidth]{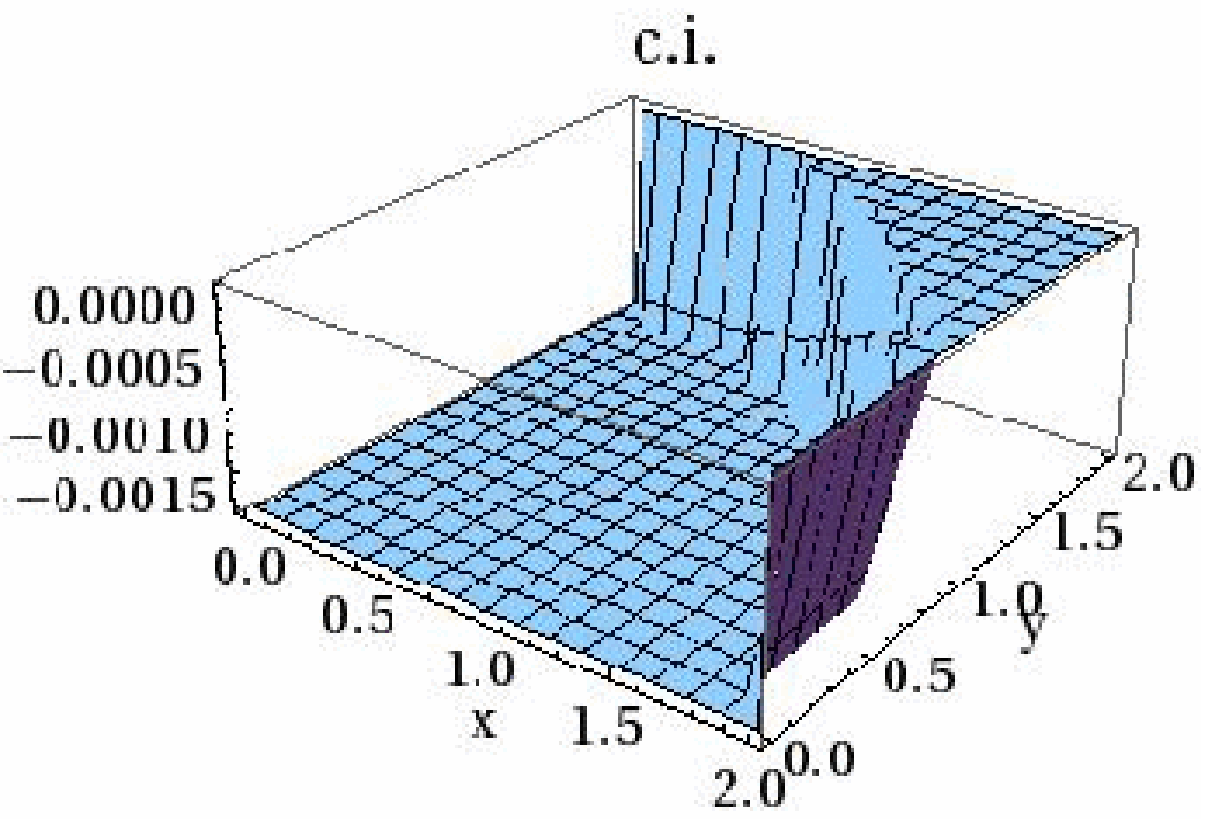}
\hspace{0.1\textwidth}
 \includegraphics[width=0.4\textwidth]{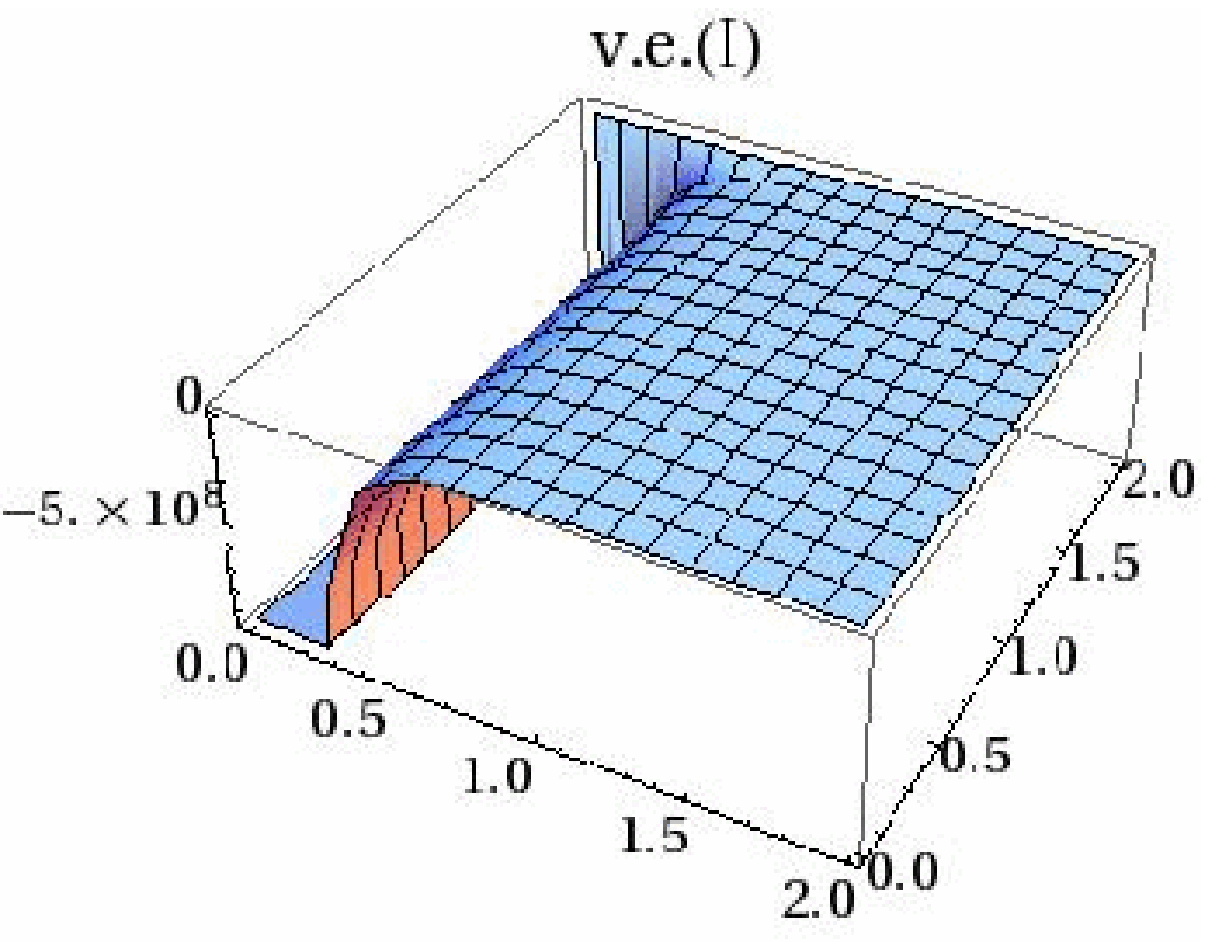}
\vspace{0.02\textwidth}
 \includegraphics[width=0.4\textwidth]{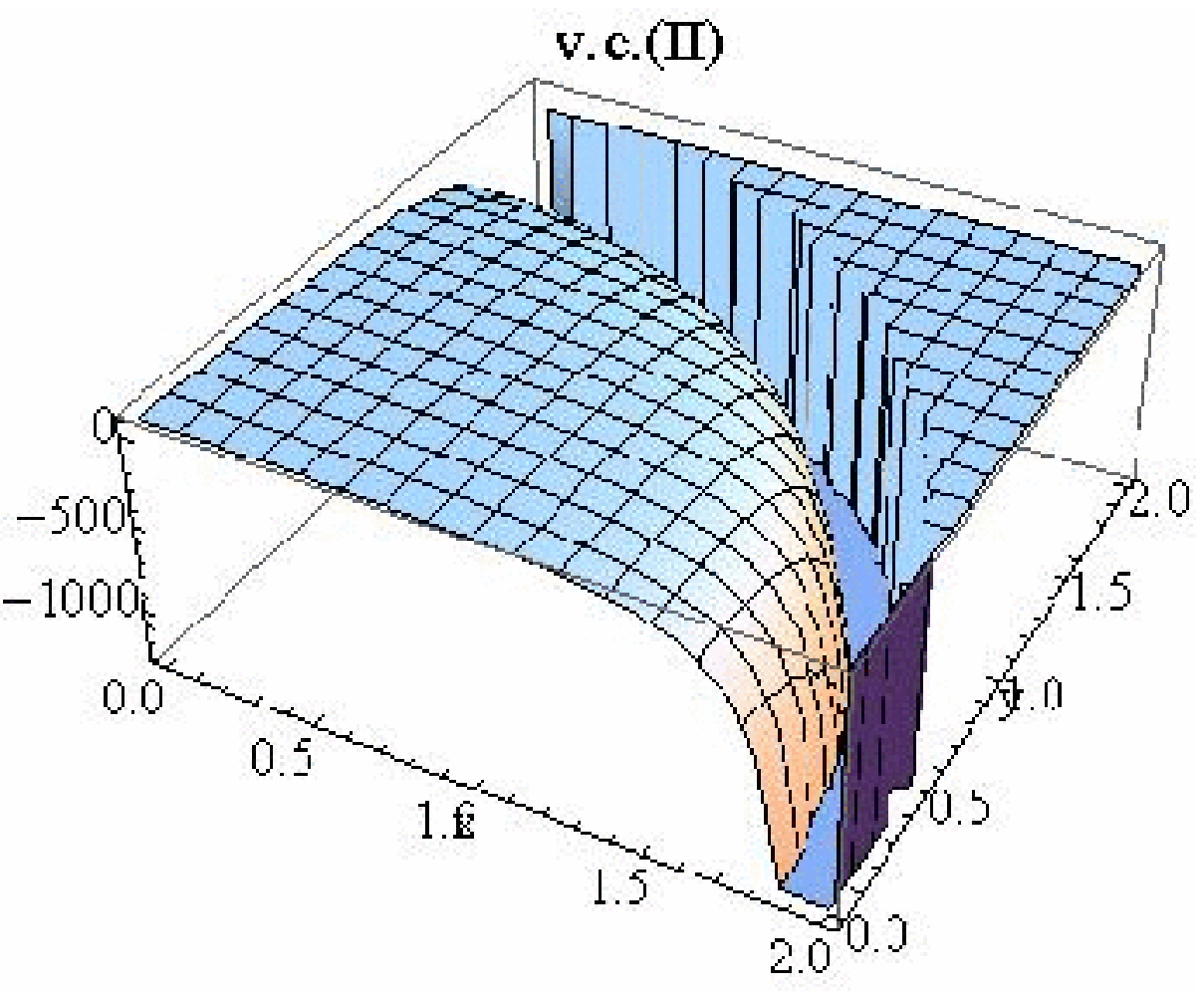}
\hspace{0.1\textwidth} \includegraphics[width=0.4\textwidth]{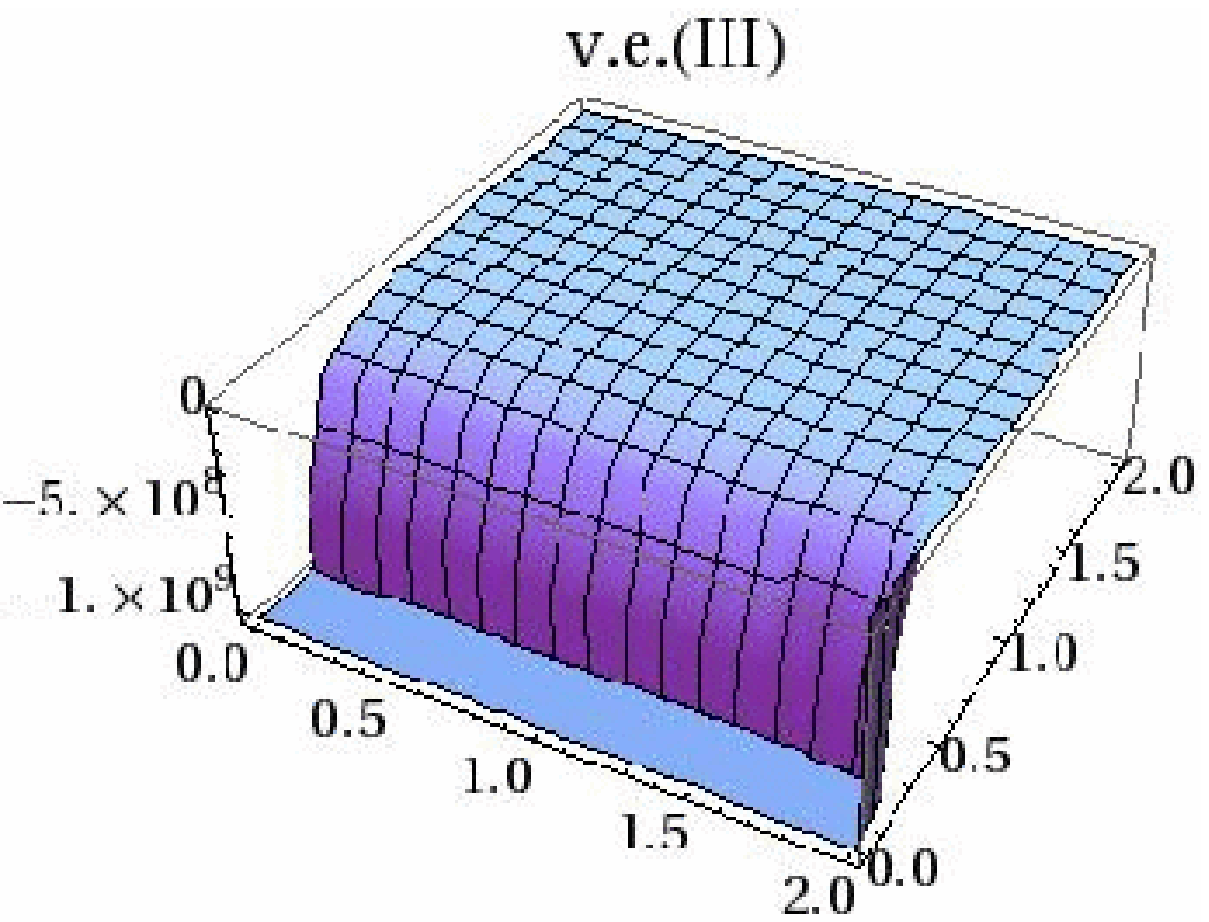}
\caption{ \label{Fig2}Plots of the contact interaction and of the vector-exchange contributions in the equilateral configuration. In this and in the next figures, we use the notation $c.i.\equiv k^{-4}\left(\int dx\right)_{EEEE}$, $v.e.(I)\equiv k^{-4}\times I \times \sum_{i=1}^{4}{t_{i}}$, $v.e.(II)\equiv k^{-4}\times II \times \sum_{i=5}^{8}{t_{i}}$ and $v.e.(III)\equiv k^{-4}\times III \times \sum_{i=9}^{12}{t_{i}}$ (the quantities $I$, $II$, $III$ and $t_{i}$ are introduced in Eq.~(\ref{final-2v})).}
\end{figure}

\begin{figure}\centering
\includegraphics[width=0.4\textwidth]{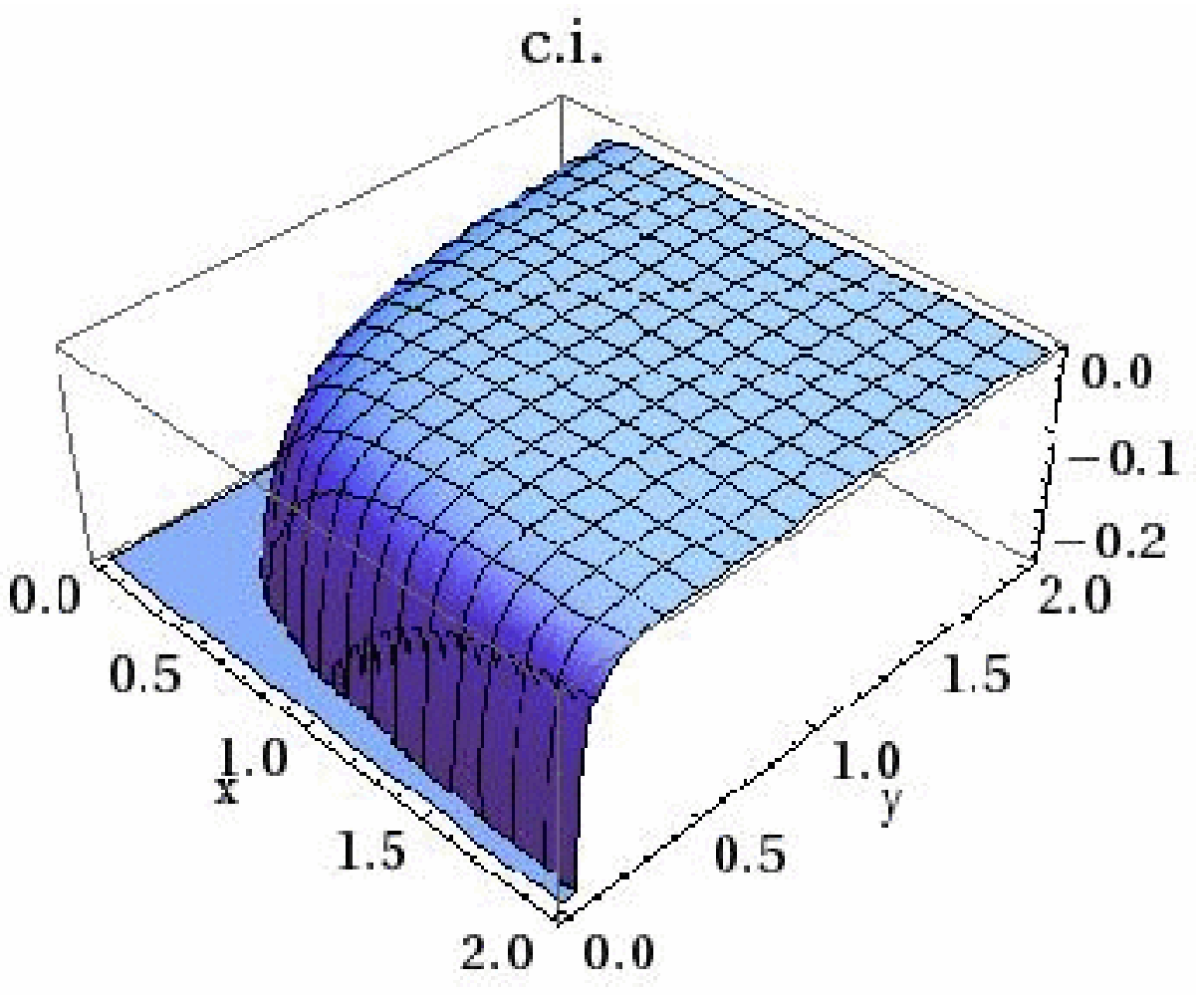}
\hspace{0.1\textwidth}
 \includegraphics[width=0.4\textwidth]{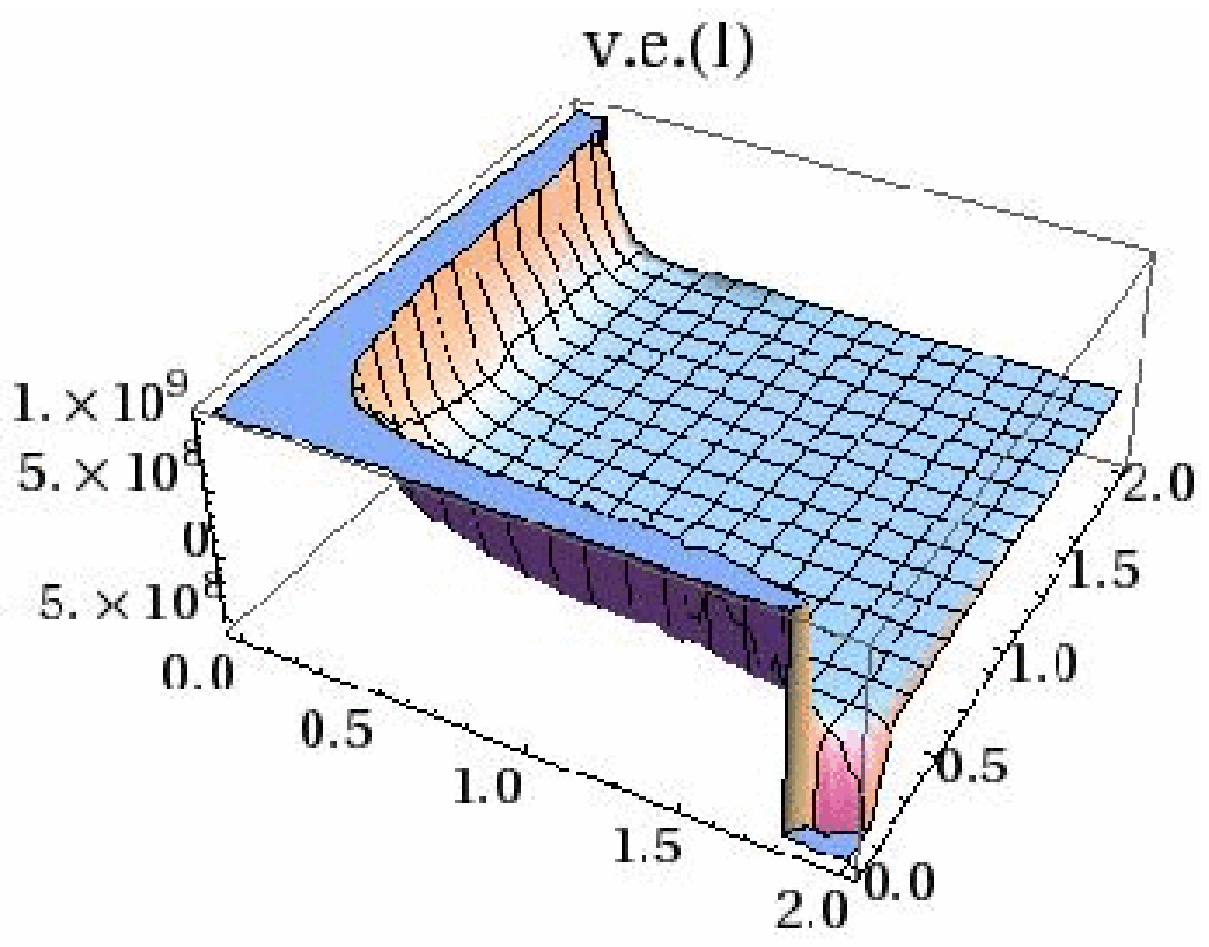}
\vspace{0.02\textwidth}
 \includegraphics[width=0.4\textwidth]{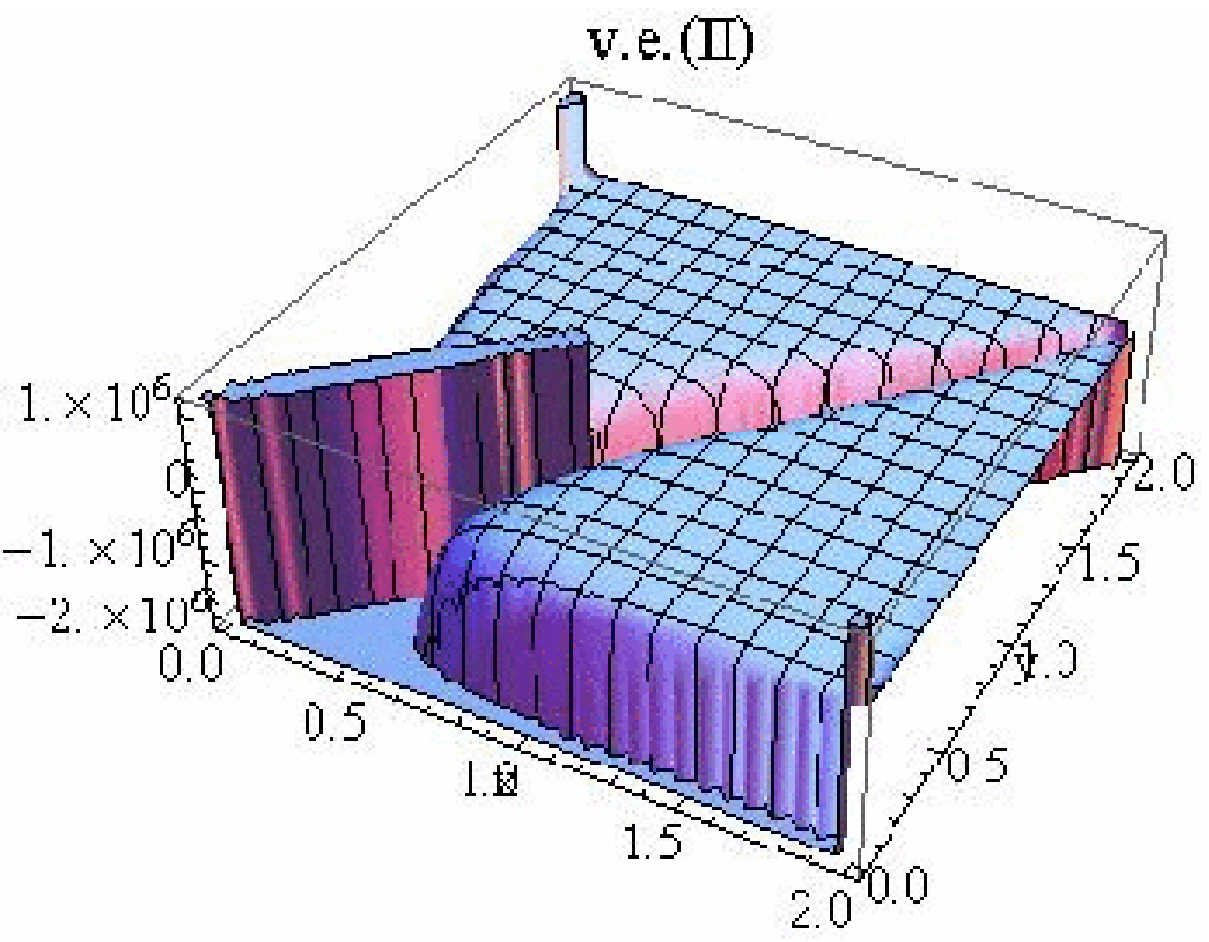}
\hspace{0.1\textwidth}
 \includegraphics[width=0.4\textwidth]{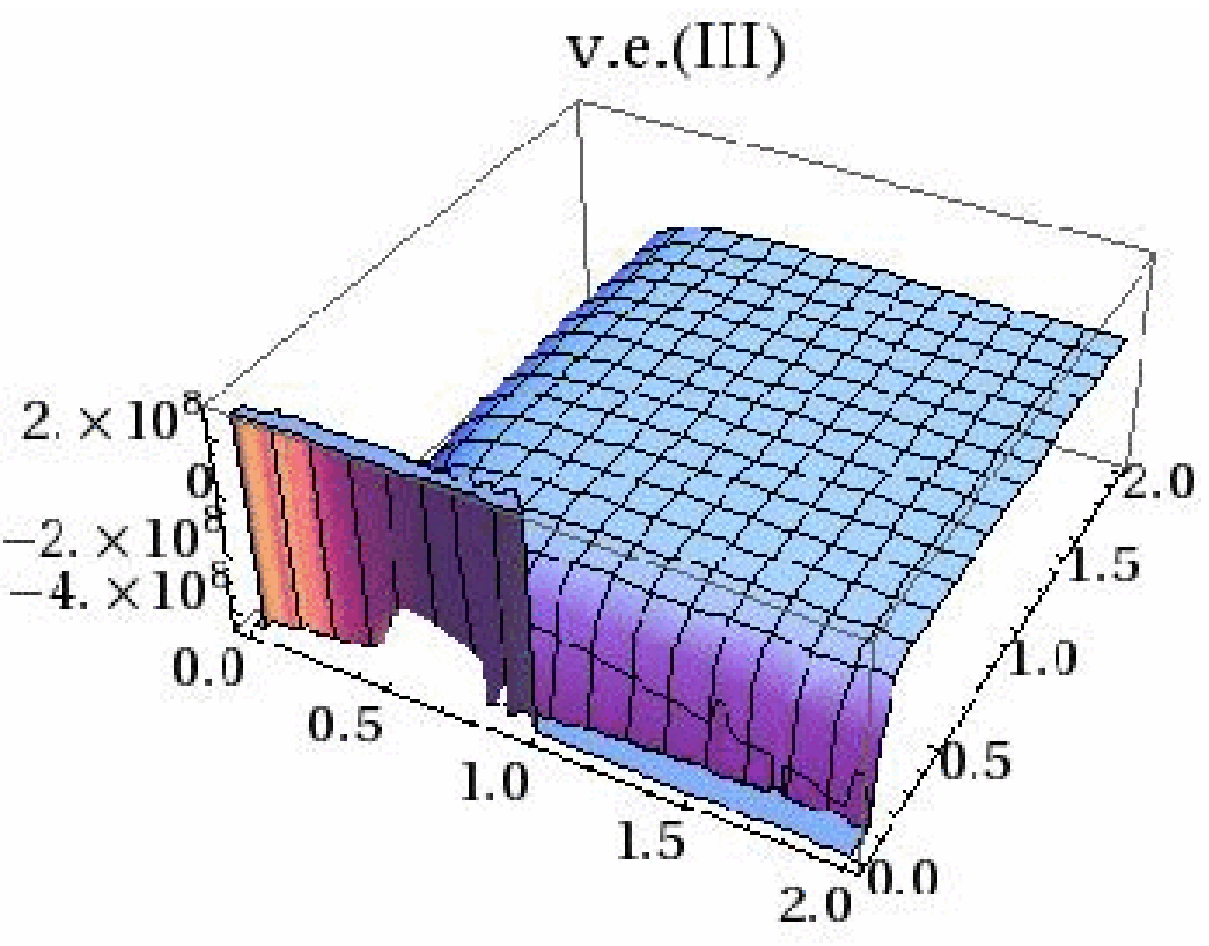}
\caption{ \label{Fig3} Plots of the contact interaction and of the vector-exchange contributions in the specialized planar configuration (plus sign).}
\end{figure}

\begin{figure}\centering
\includegraphics[width=0.4\textwidth]{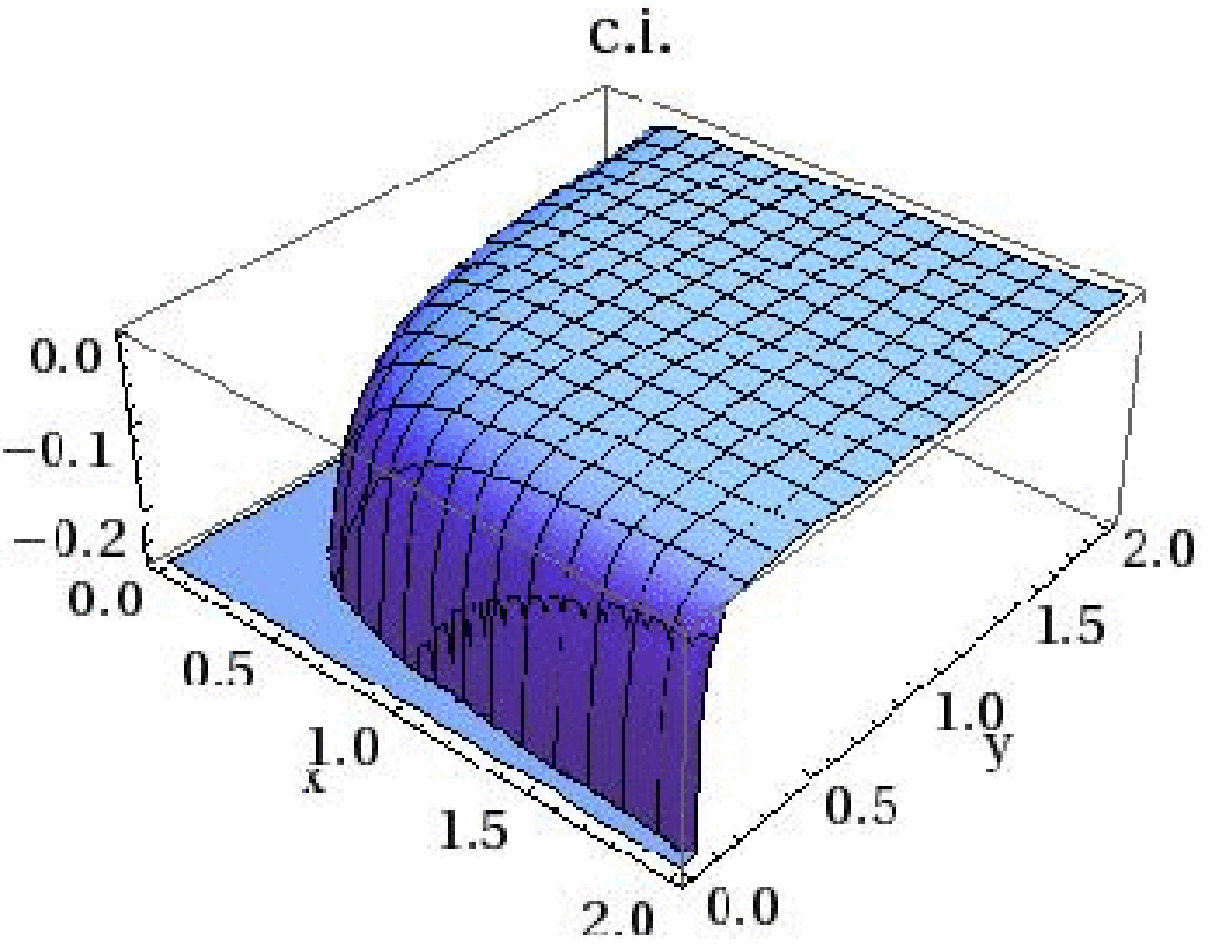}
\hspace{0.1\textwidth}
 \includegraphics[width=0.4\textwidth]{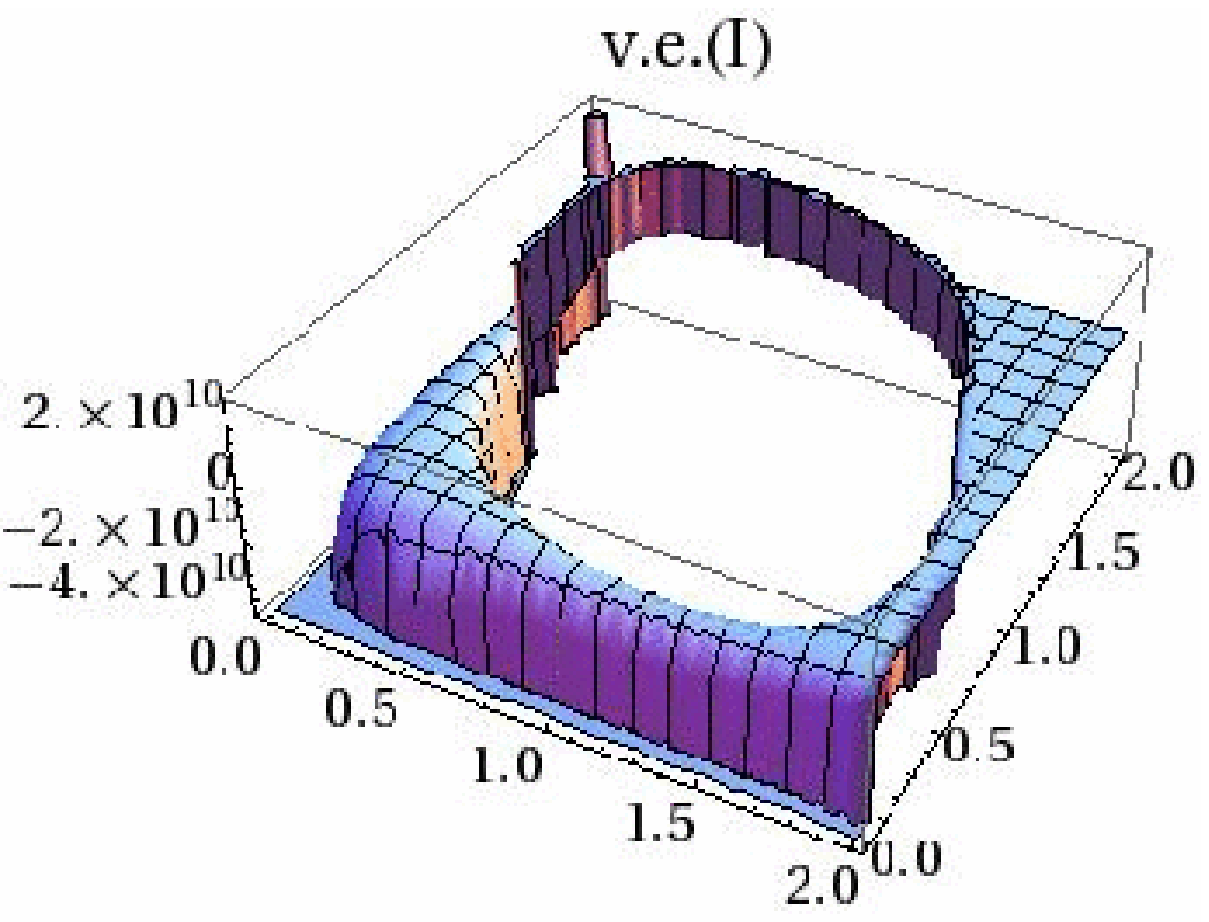}
\vspace{0.02\textwidth}
 \includegraphics[width=0.4\textwidth]{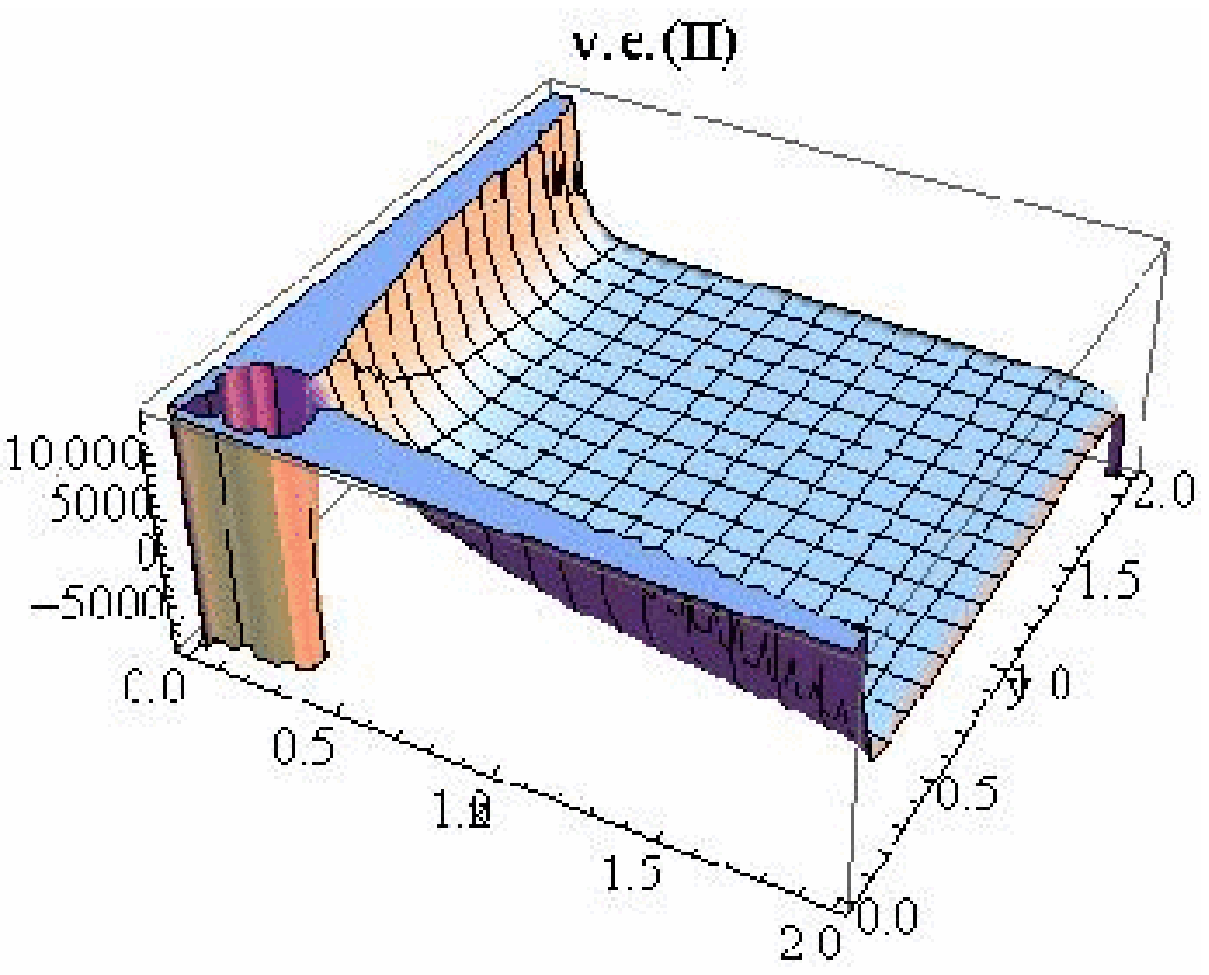}
\hspace{0.1\textwidth}
 \includegraphics[width=0.4\textwidth]{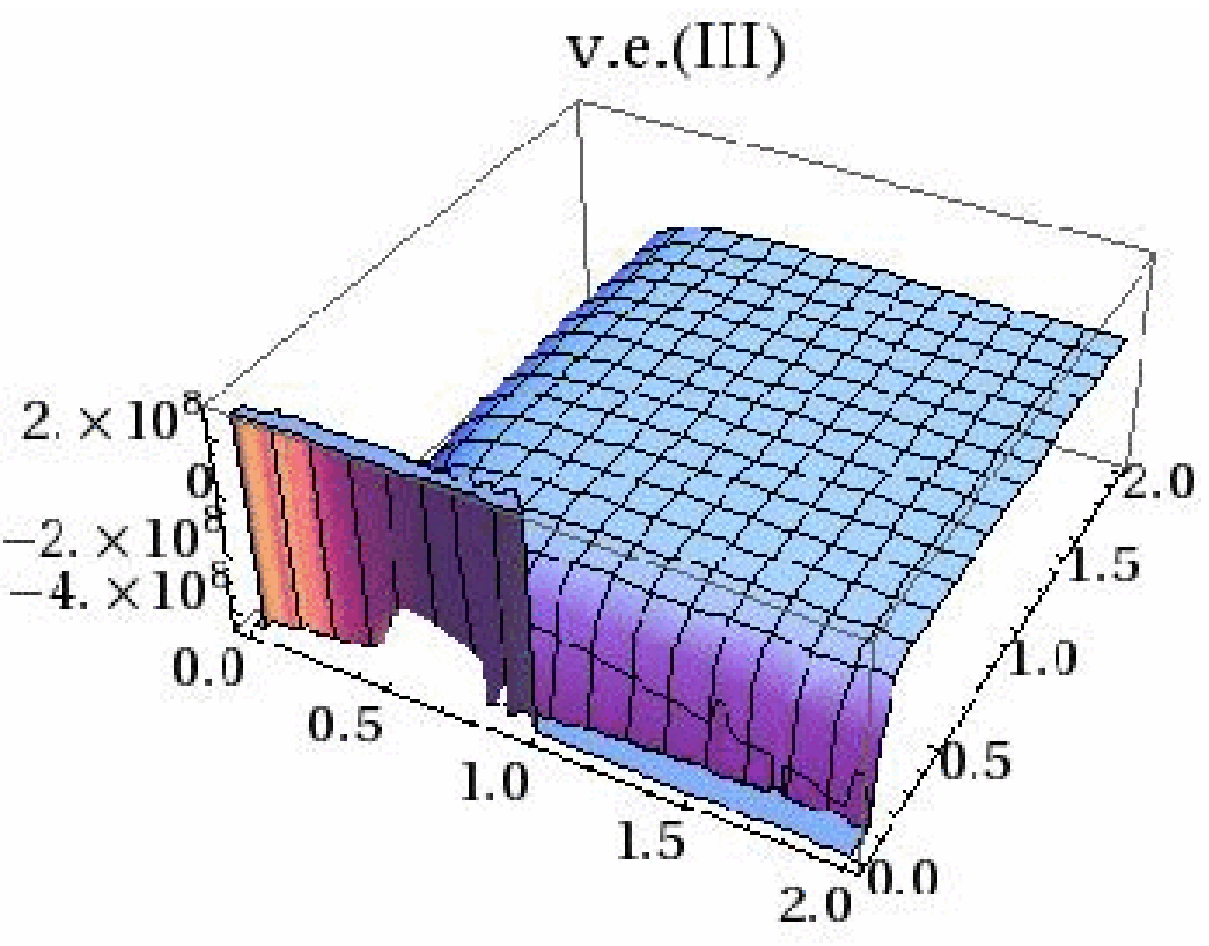}
\caption{ \label{Fig4} Plots of the contact interaction and of the vector-exchange contributions in the specialized planar configuration (minus sign).}
\end{figure}

\subsection{Equilateral configuration}

It is defined by setting all of the momenta moduli equal to each other ($k_{1}=k_{2}=k_{3}=k_{4}$). The variables of the plot can be chosen as $x\equiv k_{\hat{12}}/k_{1}$ and $y\equiv k_{\hat{14}}/k_{1}$, which vary in the interval $[0,2]$, with $k_{\hat{13}}/k_{1}=\sqrt{4-x^2-y^2}$. The plots of the isotropic parts of three vector-exchange (from Eq.~(\ref{IA}) and its permutations) and of the point-interation contributions to the trispectrum in the equilateral case are provided in Fig.~2.\\
The contact-interaction plot is, as expected, flat on the whole integration domain, since it does not depend on $k_{\hat{12}}$ and $k_{\hat{14}}$.This behaviour is very similar to the one of what was defined in \cite{Chen:2009bc} as $T_{loc2}$, i.e. the local trispetrum term that is proportional to the parameter $g_{NL}$ of local cubic non-linearities.\\
The vector-exchange diagrams instead, diverge as $k_{\hat{1i}}^{-3}$ as $(k_{\hat{1i}}/k_{1})\rightarrow 0$ (limit in which the tetrahedron becomes flat), $i=1,2,3$ respectively for $(v.e.I)$, $(v.e.II)$ and $(v.e.III)$. This behaviour is similar to what was defined in \cite{Chen:2009bc} as $T_{loc1}$, i.e. the local trispetrum term that is proportional ot $f_{NL}^{2}$. For the vector-exchange plots it is possible to check that the amplitudes run from a very large ($\sim 4\times 10^{6}$) and negative number at $x,y,z\simeq 1$ respectively for $(v.e.I,II,III)$, to a very large (of the same order) and positive number when $x,y$ and $z$ approach $2$ (we define $z\equiv k_{\hat{13}}/k_{1}$). This behaviour can be understood by looking at the analytic expressions which are characterized, on one hand, by a very smooth power law $x$ dependence (that is also responsible for the sign change when moving in the interval $x=1$ to $x=2$), on the other hand by very large numerical coefficients.\\ 
We can then conclude that the shapes of point-interaction and vector-exchange diagrams in the equilateral configuration are completely different.

\subsection{Specialized planar configuration}

This is the case defined by $k_{1}=k_{3}=k_{\hat{14}}$ and by requiring the tetrahedron to become flattened. The expressions for $k_{\hat{12}}$ and $k_{\hat{13}}$ are then given by 

\bea\label{k12}
\frac{k_{\hat{12}}}{k_{1}}=\sqrt{1+\frac{x^2 y^2}{2}\pm\frac{xy}{2}\sqrt{(4-x^2)(4-y^2)}}\\\label{k13}
\frac{k_{\hat{13}}}{k_{1}}=\sqrt{x^2+y^2-\frac{x^2 y^2}{2}\mp\frac{xy}{2}\sqrt{(4-x^2)(4-y^2)}}
\eea
where $x\equiv k_{2}/k_{1}$ and $y\equiv k_{3}/k_{1}$ and varying between $0$ and $2$. The plus and minus signs in Eqs.~(\ref{k12}) and (\ref{k13}) indicate a double degeneracy in the structure of the quadrangle to be expected in the specialized planar case: the plus sign corresponds to a quadrangle with internal angles $> \pi$, the minus sign to $\leq\pi$ \cite{Chen:2009bc}. The treatment of this case deserves some attention in the sense that, while in \cite{Chen:2009bc} the plus and minus cases are perfectly equivalent given the symmetry of the trispectrum w.r.t. the exchange of any pair of its $k_{i}$ ($i=1,...,4$) with each other, this symmetry does not a priori hold in our model because of its anisotropy. The two cases will therefore be considered separately in Figs.3 and 4. \\ 
Let us begin with the contact interaction diagram. This is independent of $k_{\hat{12}}$ and $k_{\hat{13}}$, it has therefore the same expression for both plus and minus sign cases. The graph diverges in the limit $x \rightarrow 0$ as $x^{-2}$ and similarly for $y$ (the expression is symmetric in $x$ and $y$) and it is roughly flat in the remaining part of the domain.\\
As to the vector-exchange plots, the situation is very similar. In the plus sign case, the diagram $(v.e.I)$ diverges as $x^{-3}$ and $y^{-3}$ respectively as $x \rightarrow 0$ and $y \rightarrow 0$ and is flat for the rest of the domain. The diagram $(v.e.II)$ shows similar divergences in the limit $x\rightarrow 0$ and $y\rightarrow 0$. As to $(v.e.III)$, besides for the regions $x \rightarrow 0$ and $y\rightarrow 0$, it also blows up for $x\rightarrow y$. This last divergence is due to the fact that ($v.eIII$) goes like $k_{\hat{13}}^{-3}$ and, when $x=y$, $k_{\hat{13}}=0$.\\
In conclusion, for this configuration, we can say that the contact-interaction together with the vector-exchange $(I)$ and $(II)$ graphs have similar features in terms of divergences and plateau locations. The graph $(III)$ has additional divergences for $x\sim y$, a feature in common with the local trispectra $T_{loc1}$ \cite{Chen:2009bc}. \\
Let us now come to the last set of graphs: specialized planar configuration with minus sign for the vector-exchange diagram (Fig.~4). First of all, we can notice that, while for $v.e.(I)$ and $v.e.(III)$ the plus and minus specialized planar cases are different, they conincide for $v.e.(II)$, being this independent of $k_{\hat{12}}$ and $k_{\hat{13}}$. \\
As to $v.e.(I)$, it is singular for $x=0$ and $y=0$ (the asymptotic behavious is in fact given by $\sim x^{-3}$ and $y^{-3}$ respectively as $x \rightarrow 0$ and $y \rightarrow 0$). The plot shows additional singularities in the central region of the plot due to $k_{\hat{12}}\rightarrow 0$. As to $v.e.(II)$, it has singularities for $x\rightarrow 0$ ($\sim x^{-3}$), $y\rightarrow 0$ ($\sim y^{-3}$) and $(x,y)\rightarrow (2,2)$ ($\sim (x-2)^{-3/2}$ and $\sim (y-2)^{-3/2}$).

\section{Features and typical level of anisotropy in the trispectrum}

\begin{figure}\centering
\includegraphics[width=0.4\textwidth]{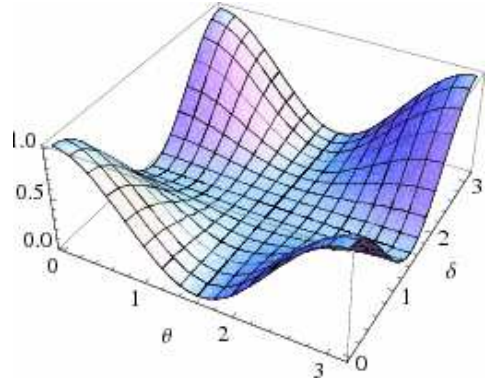}
\caption{ \label{Fig5} Plots of the anisotropic part of the trispectrum from the contribution due to vector-exchange diagrams in a sample angular configuration. The analytic expression for this plot is $\cos^{2}\theta\cos^{2}\delta$ and was derived from Eq.~(\ref{zzeta}) after normalizing to $g_{c}^{2}H^{4}_{*}N^{4}_{A}\left[ISO\right]$ and setting $x^{*}=1$.}
\end{figure}

Let us now consider the complete expression for the trispectrum, i.e. let us include its anisotropy coefficients. Again, we will focus on the non-Abelian trispectrum contributions and, in particular, on the vector-exchange one, given that, in general, its amplitude turns out to be several orders of magnitude larger than the one of the contact-interaction diagram. In order to discuss an example and provide a plot of the anisotropic part of the trispectrum, it is convenient to pick a well-defined configuration reducing the total number of variables to some $\leq 2$ number. One possible choice is to consider the case where the tetrahedron is flat and the wave vectors form a rectangle. We need to partially fix the orientation of the vectors $\vec{N}_{a}$ w.r.t. the wave vectors. Let us pick the following sample configuration
\bea
\hat{N}_{2}\cdot\hat{k}_{i}=0\,(i=1,...4)\nonumber\\
\hat{N}_{1}\cdot\hat{k}_{1}=\cos\delta,\quad\quad\quad\hat{N}_{1}\cdot\hat{k}_{2}=0\nonumber\\
\hat{N}_{3}\cdot\hat{k}_{2}=\cos\theta,\quad\quad\quad\hat{N}_{3}\cdot\hat{k}_{1}=0.
\eea 
In addition to that, let us assume that all the $\vec{N}_{a}$ have the same magnitude $N_{A}$. In this configuration, we have 
\bea
\Delta_{I}=\Delta_{III}=N^{A}_{4}\cos^{2}\theta\cos^{2}\delta,\quad\quad\quad\Delta_{II}=0,
\eea
therefore the the expression in Eq.~(\ref{IA}) becomes
\bea\label{zzeta}
T_{\zeta}\supset g_{c}^{2}H^{4}_{*}N_{A}^{4}\left[ISO\right]\cos^{2}\theta\cos^{2}\delta
\eea
where the expression in brackets includes an isotropic term (which is rotationally invariant) 
\bea
ISO\equiv\left(\frac{x^{*}}{k}\right)^4\left(I\sum_{i=1}^{4}t_{i}+III\sum_{i=9}^{12}t_{i}\right).
\eea
Fig. 5 plots the trispectrum contribution in Eq.~(\ref{zzeta}) normalized to its isotropic part. It shows that the trispectrum gets modulated by the preferred directions that break statistical isotropy.\\
The trispectrum amplitude $\tau_{NL}$ was calculated in the equilateral configuration in Section 6 (without taking into account the angles between the wavevectors and 
the $\vec{N}$ vectors). The amplitude is actually modulated by the anisotropy coefficients, as the previous example shows. Notice that in general the bispectrum and 
trispectrum can be expressed as the sum of an isotropic plus an anisotropic parts. This was for instance done in \cite{Yokoyama:2008xw} for the $f_{NL}$ parameter in the $U(1)$ case. For our case, the purely isotropic parts of the trispectrum are included in the last two lines of Eqs.(\ref{vector}), as far as the Abelian part is concerned, and are given by Eqs.(\ref{anisotropo}) and (\ref{isot-trisp}), for the non-Abelian terms. The level of the anisotropic and isotropic parts can be read from Tables 1 and 2. In particular $\tau^{V,2}_{\rm NL}$,$\tau^{V,3}_{\rm NL}$ and $\tau^{V,4}_{\rm NL}$ give the 
order of magnitude for the level of the corresponding isotropic and anisotropic contributions, showing that the two are comparable; on the other hand $\tau^{V,1}_{\rm NL}$ quantifies only a pure anisotropic contribution, and it turns out that it can be dominant w.r.t. the isotropic part of the full trispectrum.

\section{Trispectrum for $f(\tau)$ models of gauge interactions}

\begin{figure}\centering
\includegraphics[width=0.4\textwidth]{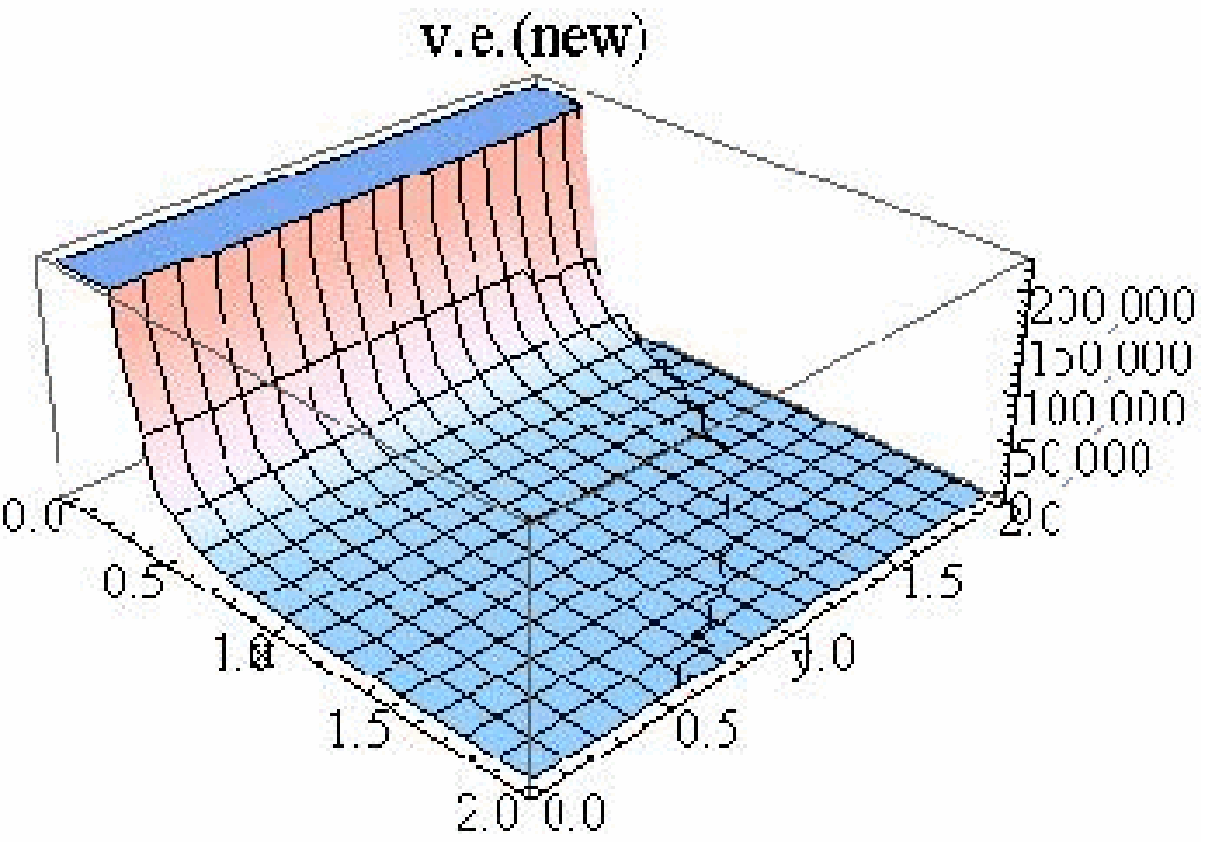}
\hspace{0.1\textwidth}
 \includegraphics[width=0.4\textwidth]{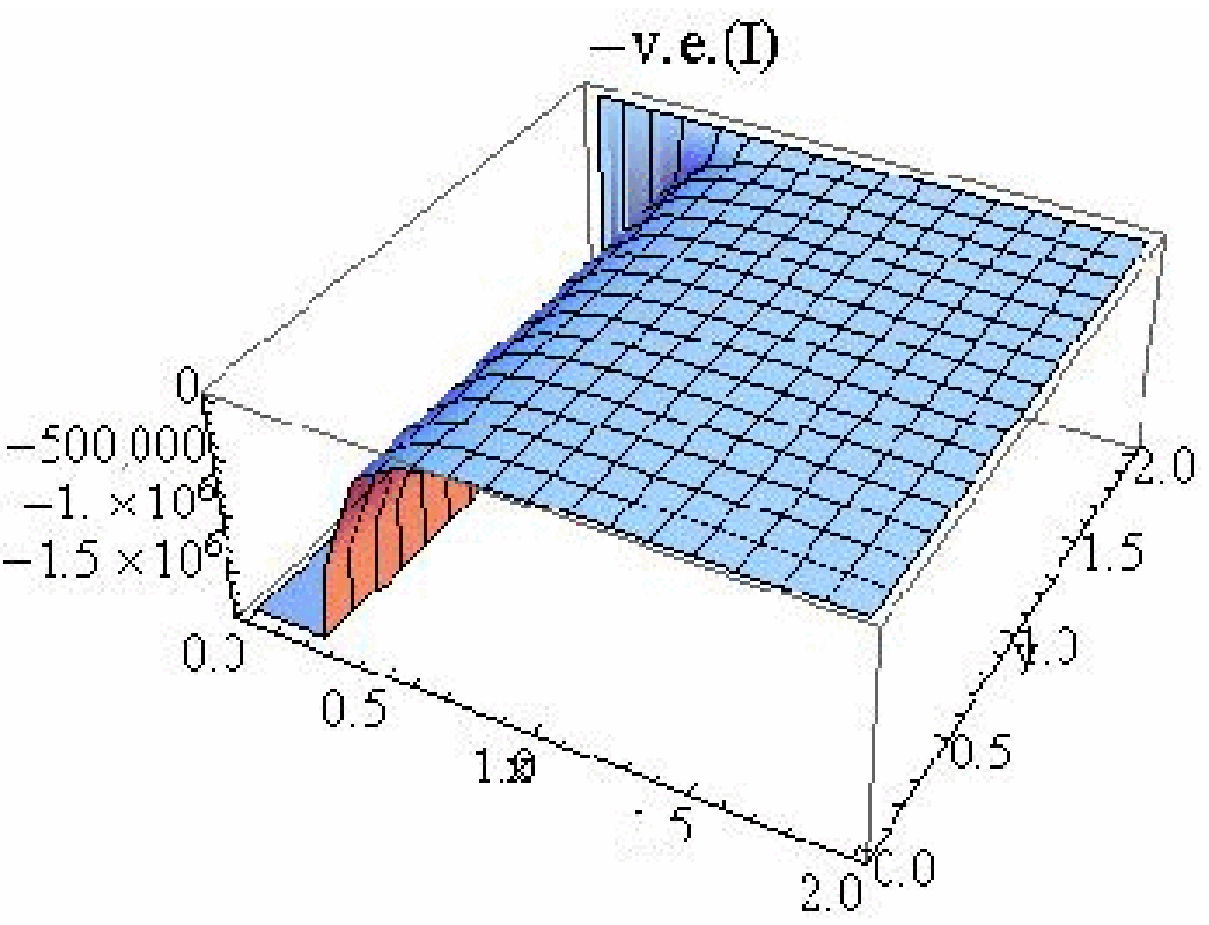}
\vspace{0.02\textwidth}
 \includegraphics[width=0.4\textwidth]{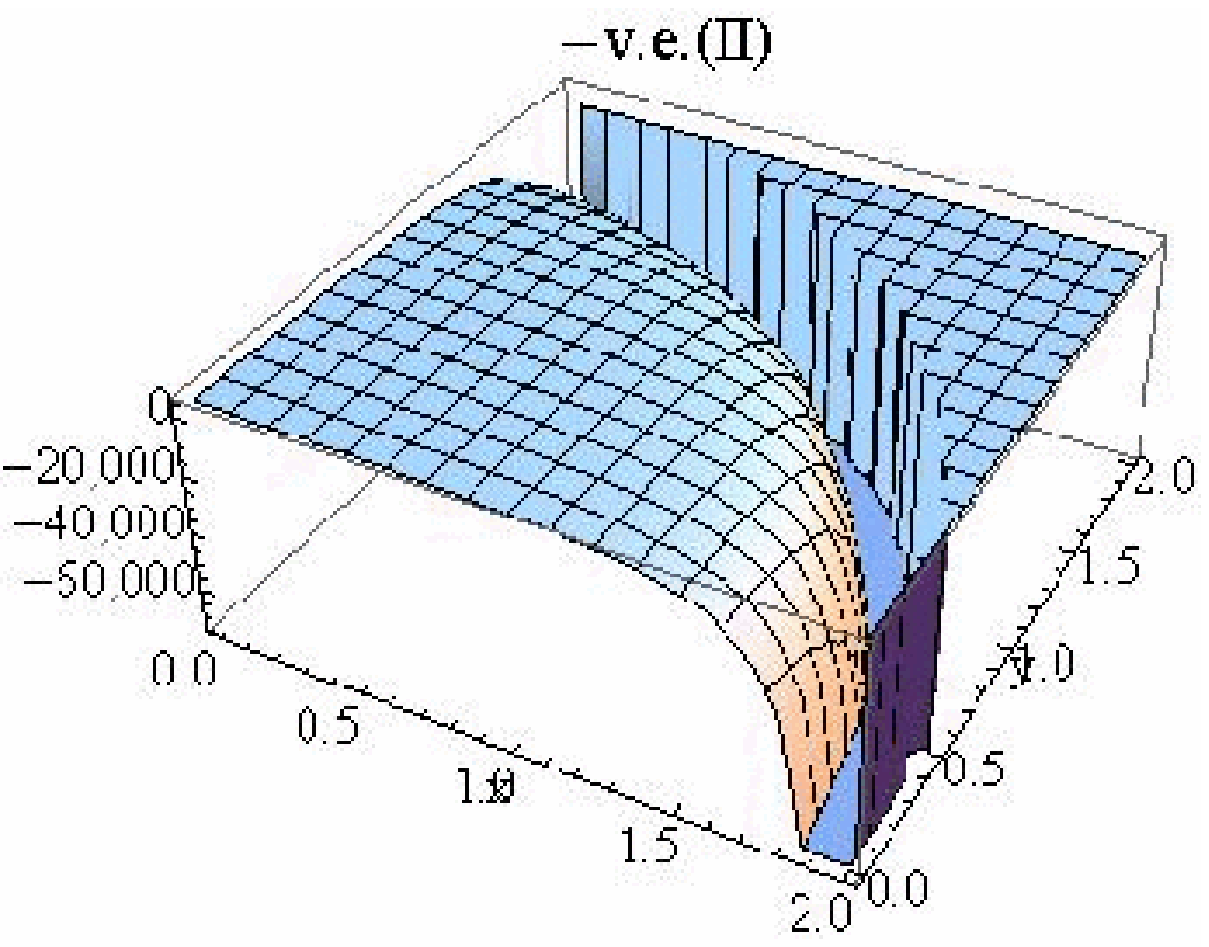}
\hspace{0.1\textwidth} \includegraphics[width=0.4\textwidth]{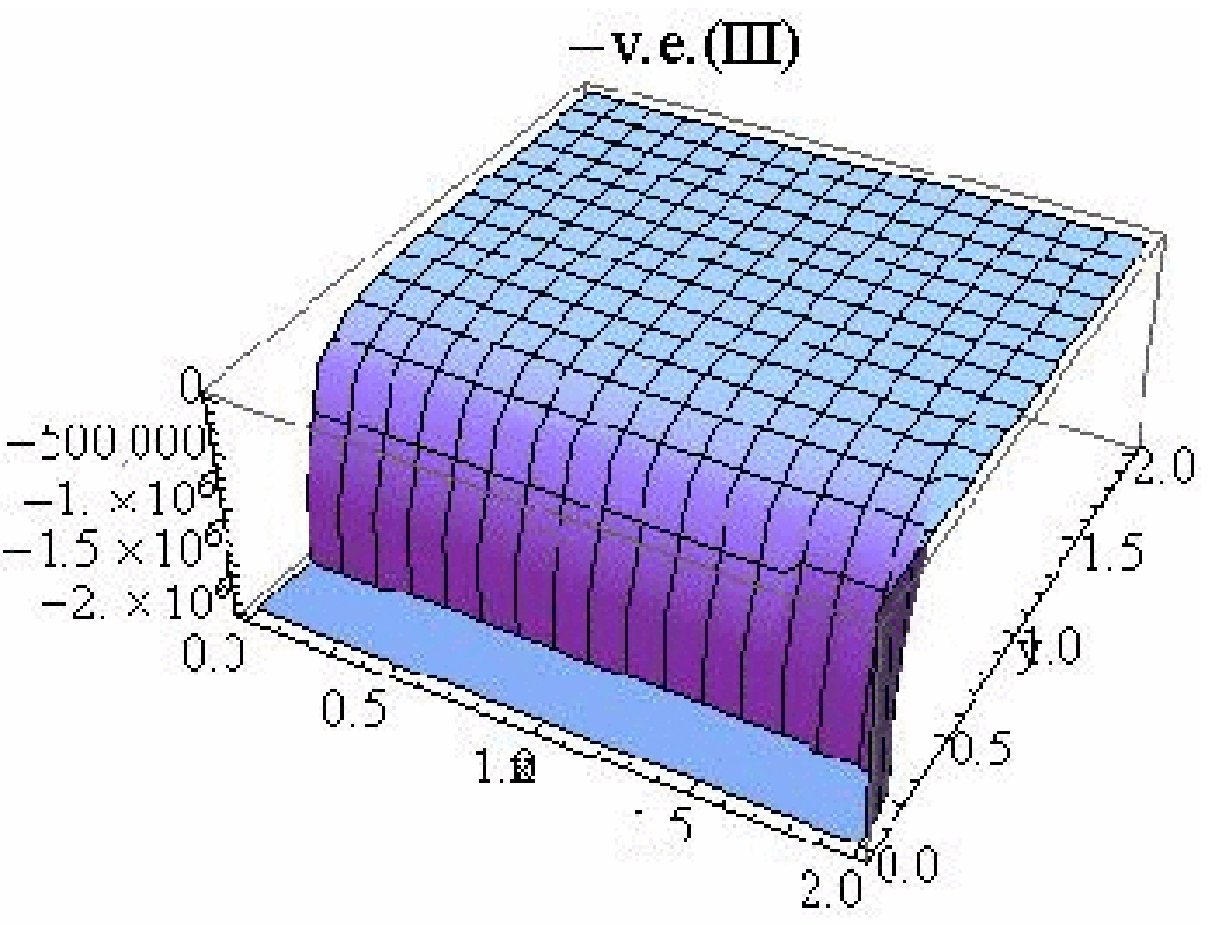}
\caption{ \label{Fig2}Plots of the isotropic functions of some of the vector-exchange contributions in the equilateral configuration, for the $f\simeq a^{-2}$ model. In this and in the next figures, ``v.e.(I,II,III)'' represent the isotropic functions associated with the very last term in square brackets in Eq.~(\ref{t}); ``v.e.(new)'' represents the isotropic function associated with the $k_{1144}$, $k_{2244}$, $k_{1133}$ and $k_{2233}$ terms in the second line of Eq.~(\ref{t}).}
\end{figure}

\begin{figure}\centering
\includegraphics[width=0.4\textwidth]{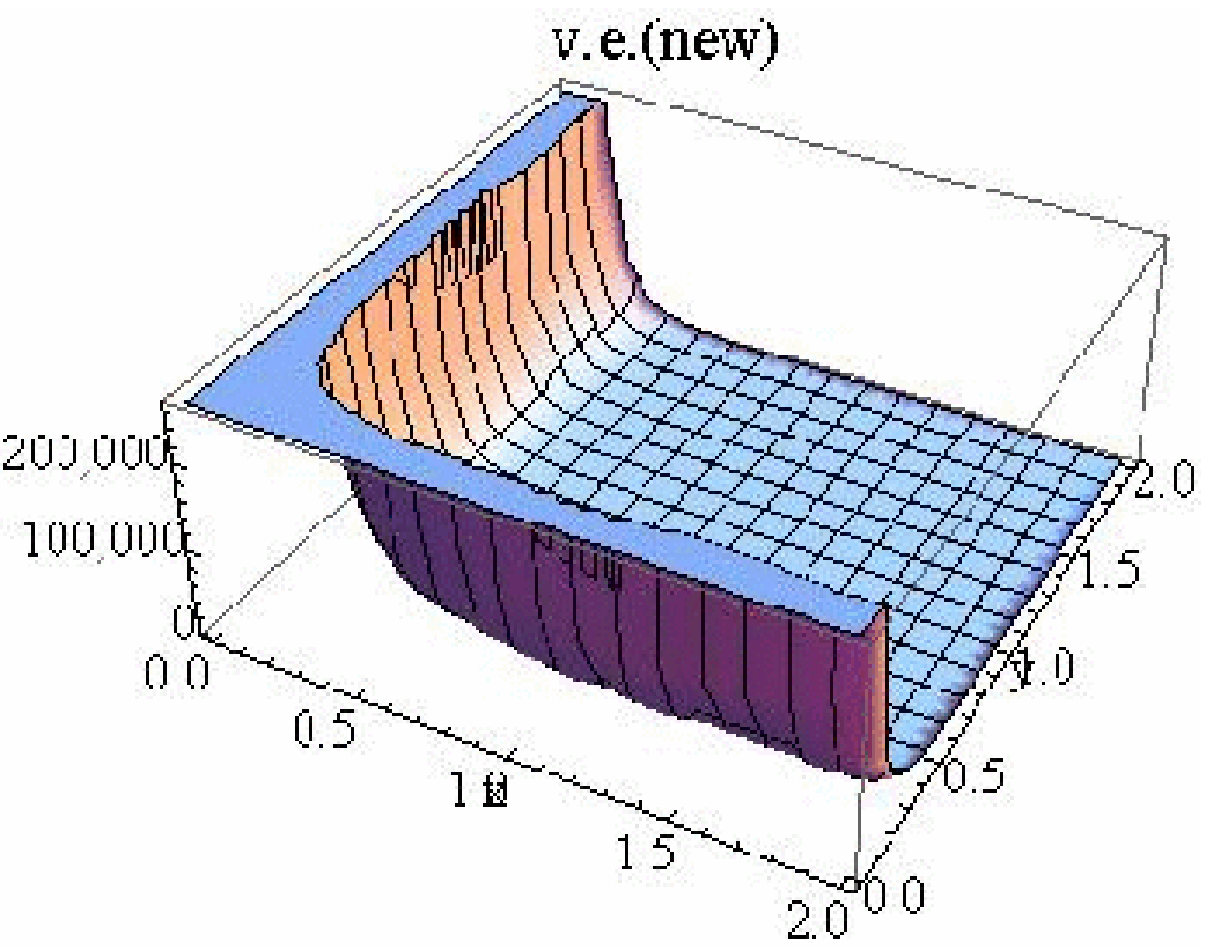}
\hspace{0.1\textwidth}
 \includegraphics[width=0.4\textwidth]{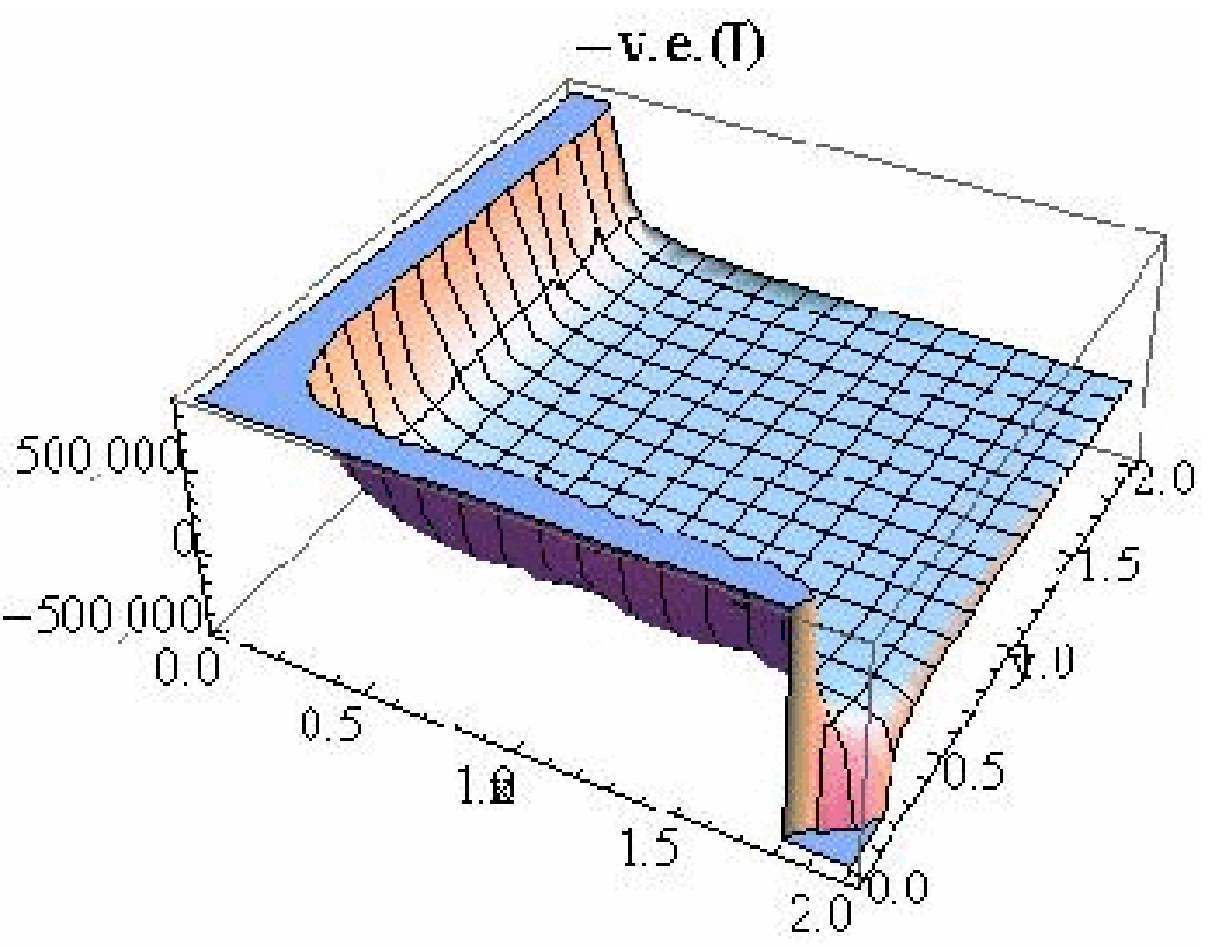}
\vspace{0.02\textwidth}
 \includegraphics[width=0.4\textwidth]{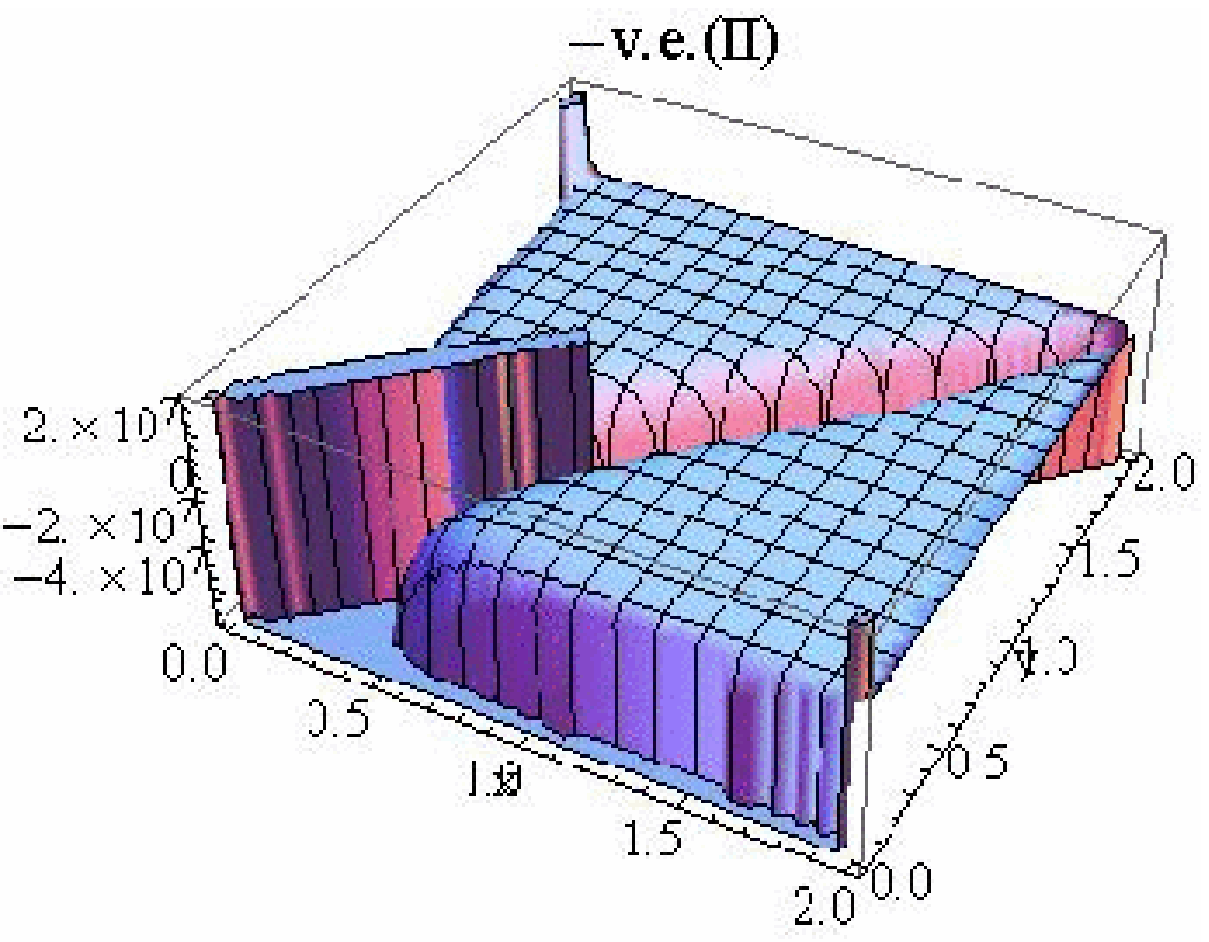}
\hspace{0.1\textwidth} \includegraphics[width=0.4\textwidth]{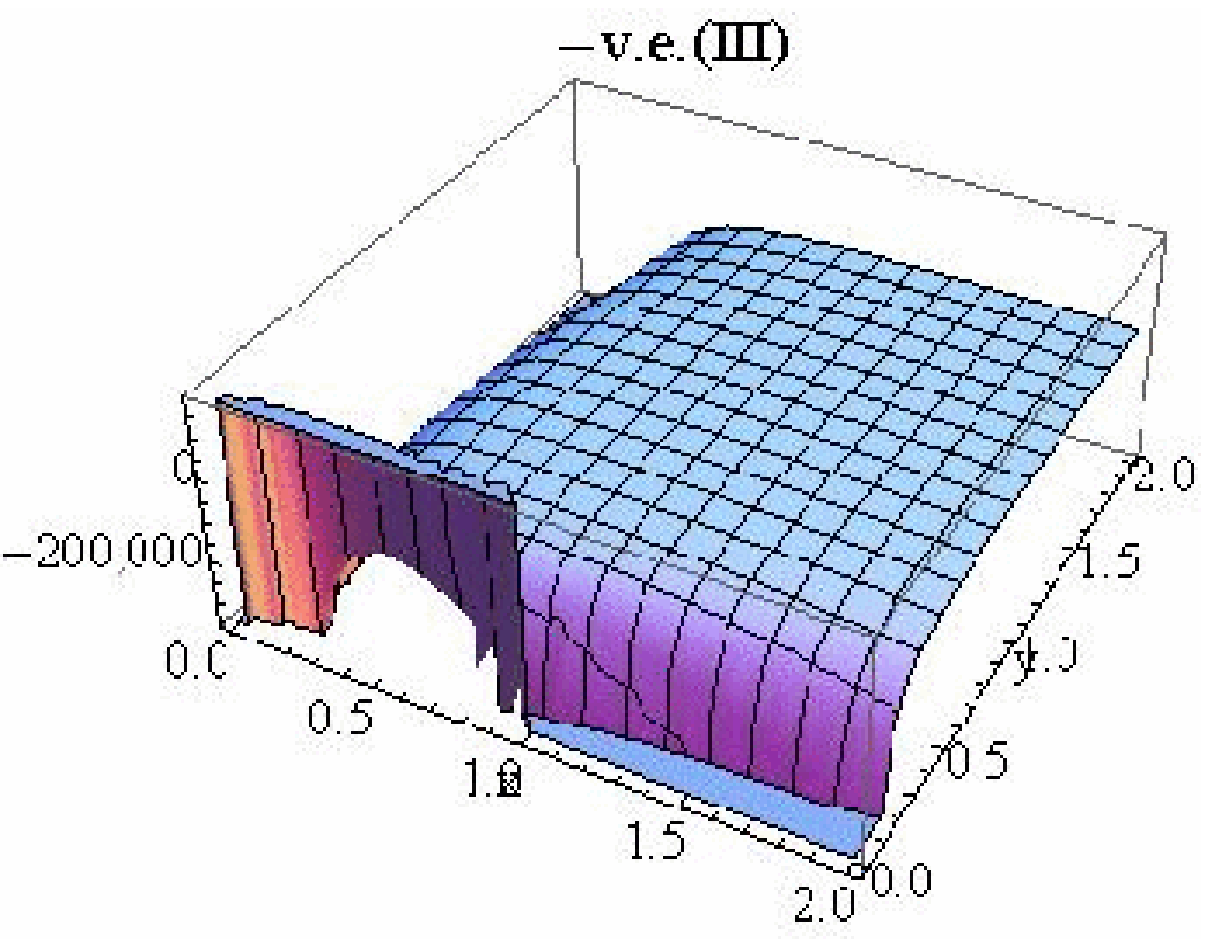}
\caption{ \label{Fig2}Plots of the isotropic functions of some of the vector-exchange contributions in the specialized planar configuration (plus sign), for the $f\simeq a^{-2}$ model.}
\end{figure}

\begin{figure}\centering
\includegraphics[width=0.4\textwidth]{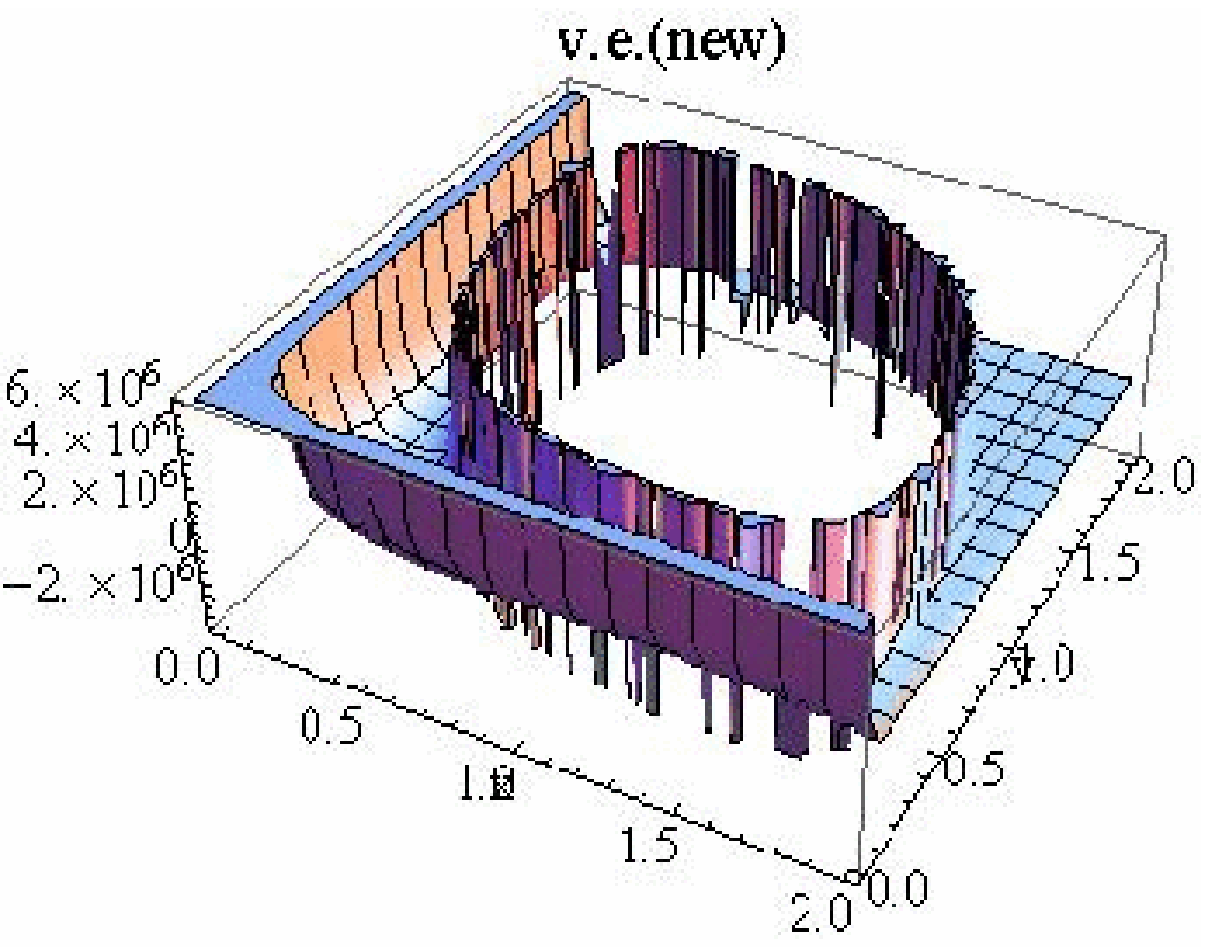}
\hspace{0.1\textwidth}
 \includegraphics[width=0.4\textwidth]{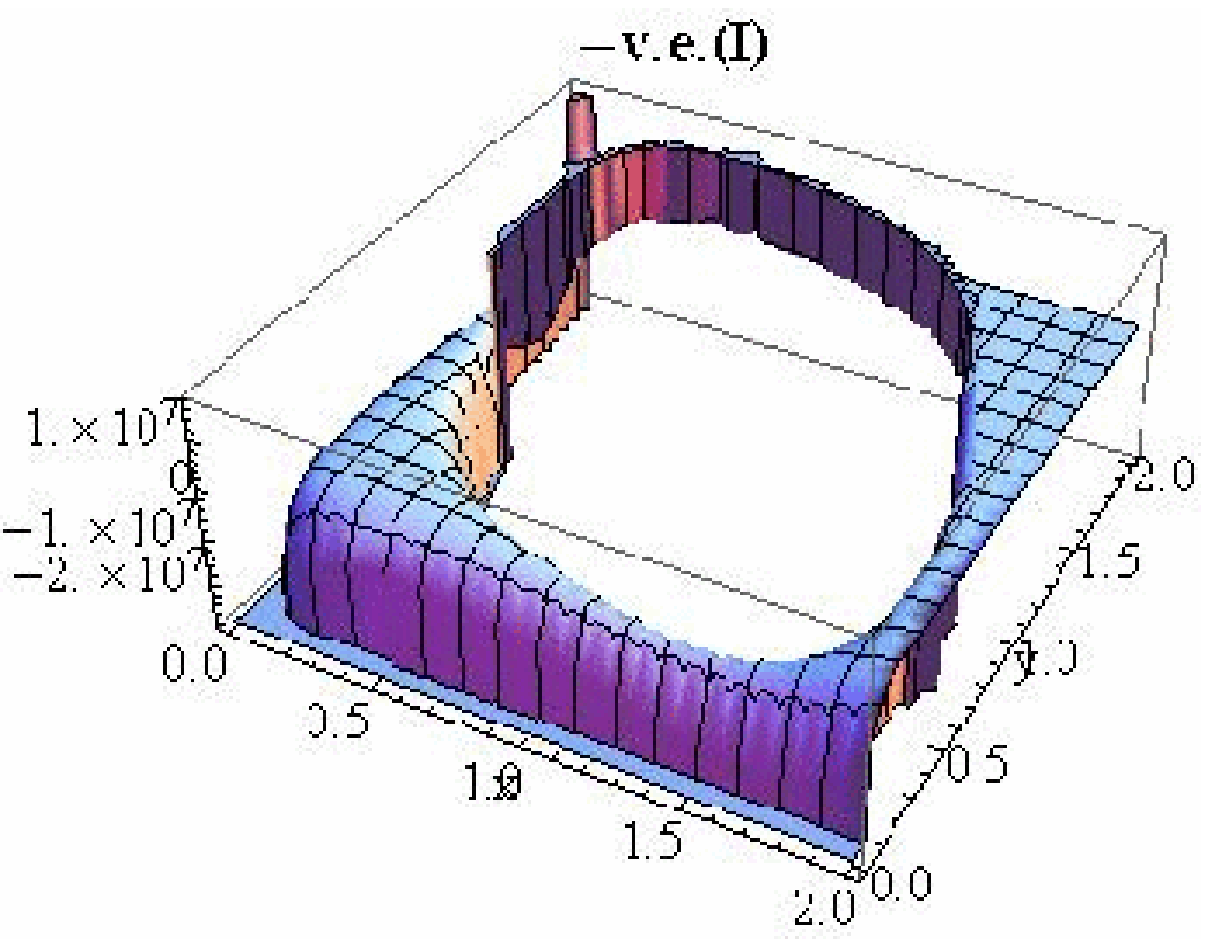}
\vspace{0.02\textwidth}
 \includegraphics[width=0.4\textwidth]{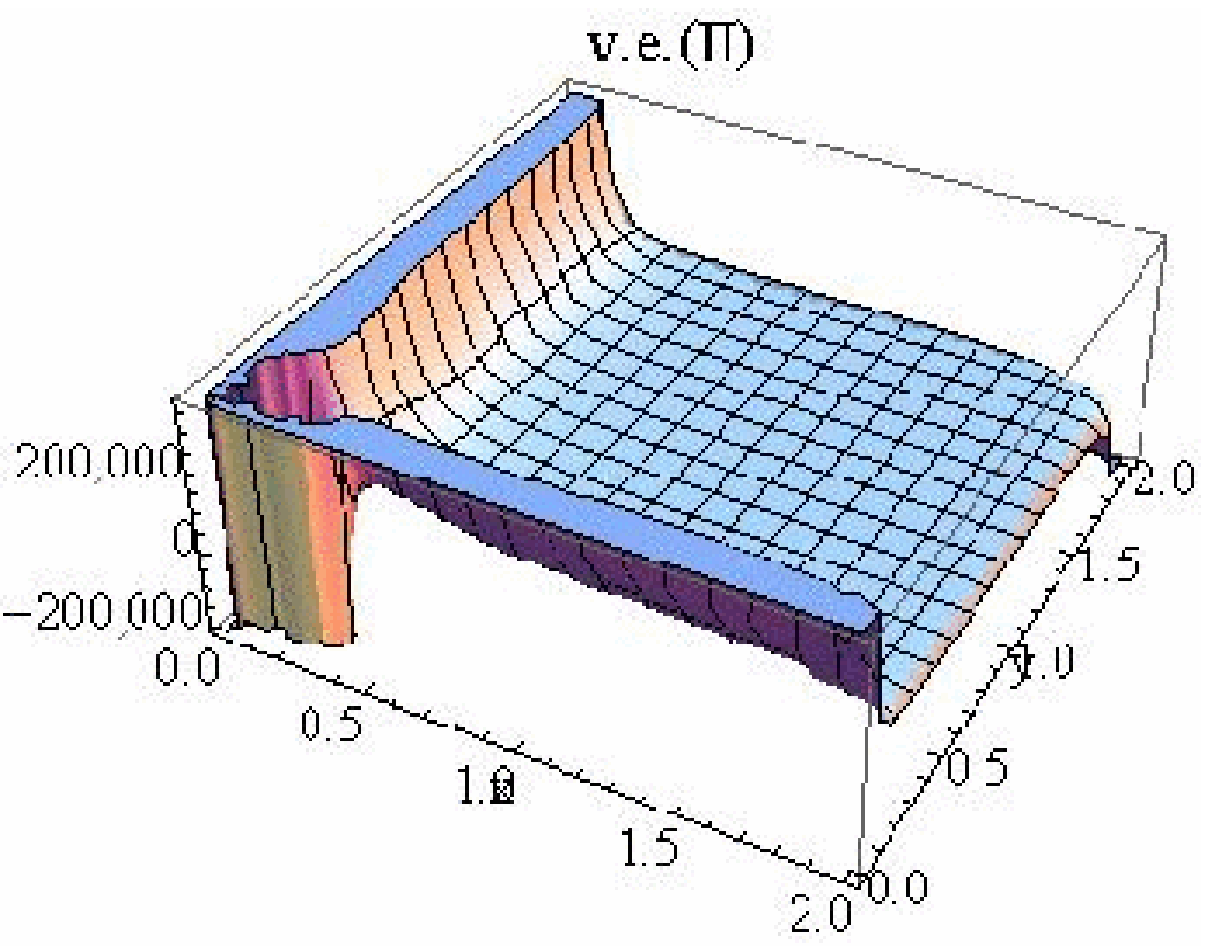}
\hspace{0.1\textwidth} \includegraphics[width=0.4\textwidth]{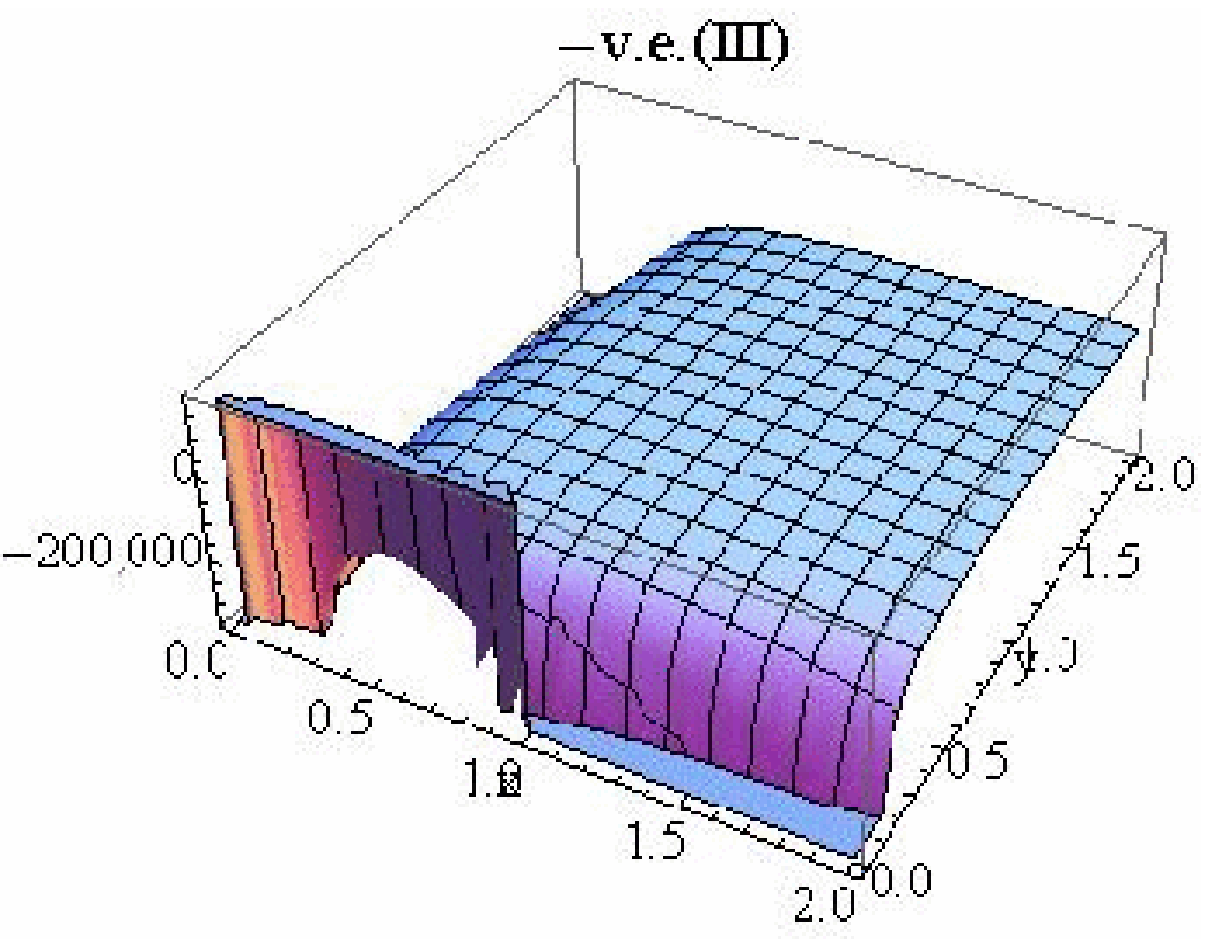}
\caption{ \label{Fig2}Plots of the isotropic functions of some of the vector-exchange contributions in the specialized planar configuration (minus sign), for the $f\simeq a^{-2}$ model.}
\end{figure}

As anticipated in Section 2, we will now show that it is quite straightforward to extend the calculations we performed for $f=1$ to cases where $f$ is not a constant. One interesting model is the one studied in \cite{Martin:2007ue} and also recently discussed in \cite{Yokoyama:2008xw}, where the field is effectively massless ($m_{0}=\xi=0$) so the action (\ref{ac}) for the gauge field becomes
\bea\label{newL}
S= \int d^4x \sqrt{-g}\left[-\frac{1}{4}f^{2}(\phi)g^{\mu\nu}g^{\rho\sigma}F_{\mu\nu}^{a}F_{\rho\sigma}^{a}+...\right],
\eea
where again $F_{\mu\nu}^{a}\equiv\p_{\mu}B^{a}_{\nu}-\p_{\nu}B^{a}_{\mu}+g_{c}\ep^{abc}B^{b}_{\mu}B^{c}_{\nu}$.\\
Let us introduce the fields $\tilde{A}_{i}^{a}$ and $A^{a}_{i}$, related by the equations $\tilde{A}_{i}^{a}\equiv f B_{i}^{a}=a A_{i}^{a}$. The $A^{a}_{i}$ are the physical fields.\\
We can expand the perturbations of $\tilde{A}^{a}_{i}$ in terms of creation and annihilation operators in the usual way
\bea
\delta\tilde{A}^{a}_{i}(\eta,\vec{x})=\int\frac{d^3q}{(2 \pi)^3}\sum_{\lambda=R,L}\Big[e_{i}^{\lambda}(\hat{q})a^{a\lambda}_{\vec{q}}\delta\tilde{A}^{a}_{\lambda}(\eta,q)+h.c.\Big].
\eea
If $f=f_{0} a^{\alpha}$, with $\alpha$ equal either to $1$ or $-2$ ($f_{0}$ is a constant), it is possible to prove \cite{Yokoyama:2008xw} that the equation of motion for $\delta\tilde{A}^{a}_{\lambda}$ is the same as the one for $\delta B^{T}$, where by $\delta B^{T}$ we mean the transverse mode function in Eq.~(\ref{T}). This is equivalent to saying that, under the assumption $\alpha=1,-2$, the physical gauge fields are governed by the same equation of motion as a light scalar field in a de Sitter space and so they generate a scale invariant power spectrum. \\
\noindent We are going to sketch the calculation of the trispectrum in this theory. \\
The general expressions in Eq.~(\ref{vector}) obviously still hold as well as the compact structure of the Abelian terms, Eq.~(\ref{VV}), except that the power spectrum in Eq.~(\ref{ps}) reduces to 
\bea
P^{ab}_{ij}=T^{ab}_{ij}P_{+},
\eea
having gauged the longitudinal modes away.\\
Let us now have a look at the non-Abelian part and see how Eqs.~(\ref{final-V}), (\ref{sec9}) and (\ref{trispec}) change when we switch to massless $f(\tau)$ Lagrangians. First of all, we need to set $n(x^{*})=0$. The interaction Hamiltonian to third and fourth order are the same as in Eqs.~(\ref{third}) and (\ref{fourth}), except for an extra $f^{2}(\tau)$ factor. The anisotropy coefficients $\tilde{I}_{EEE}$, $T^{EEEEE}_{ijkl}$ and $T^{EEEE}_{mnop}$ that survive after setting the longitudinal mode to zero do not change. On the other hand, the wavefunctions for the gauge fields are now given by $\delta B=\delta\tilde{A}/f=\delta B^{T}/f_{0}a^{\alpha}$. The new trispectrum therefore differs because of extra scale factors inside and outside the time integrals, which in general imply a different power of $H_{*}$ in the final results and a different momentum dependence in the isotropic part of the expressions.\\
The three non-Abelian vector contributions in the $f=1$ case can be schematically written as follows 
\bea\label{ftau1}\fl
\langle \delta B^4\rangle_{line2}&\simeq& \left(\delta B^{3}\right)\int d\eta \left(\delta B\right)^{3}=\left(\delta B^{T}\right)^{3}\int d\eta \left(\delta B^{T}\right)^{3},\\\fl
\langle\delta B^4\rangle_{v.e.}&\simeq& \left(\delta B\right)^{4}\int d\eta^{'} \left(\delta B\right)^{3}\int d\eta^{''} \left(\delta B\right)^{3}\nonumber\\\fl&=&\left(\delta B^{T}\right)^{4}\int d\eta^{'} \left(\delta B^{T}\right)^{3}\int d\eta^{''} \left(\delta B^{T}\right)^{3},\\\label{ftau2}\fl
\langle \delta B^4\rangle_{c.i.}&\simeq& \left(\delta B\right)^{4}\int d\eta \left(\delta B\right)^{4}=\left(\delta B^{T}\right)^{4}\int d\eta \left(\delta B^{T}\right)^{4},
\eea
respectively from Eqs.~(\ref{final-V}), (\ref{sec9}) and (\ref{trispec}), where $\langle \delta B^4\rangle\equiv \langle \delta B^{a}_{i}(\vec{k}_{1})\delta B^{b}_{j}(\vec{k}_{2})\delta B^{c}_{k}(\vec{k}_{3})\delta B^{d}_{l}(\vec{k}_{4})\rangle$ and we have omitted all the gauge and vector indices, as well as complex conjugate symbols. Let us see now how Eqs.~(\ref{ftau1})-(\ref{ftau2}) change if $f=f_{0}a^{\alpha}$ ($\alpha=1,-2$)
\bea\label{ftau3}\fl
\langle \delta B^4\rangle_{line2}&\simeq& \left(\delta B\right)^{3}\int d\eta f^{2}\left(\delta B\right)^{3}\simeq\left(\frac{\delta B^{T}}{a^{\alpha}}\right)^{3}\int d\eta a^{2\alpha}\left(\frac{\delta B^{T}}{a^{\alpha}}\right)^{3},\\\fl
\langle \delta B^4\rangle_{v.e.}&\simeq& \left(\delta B\right)^{4}\int d\eta^{'} f^{2}\left(\delta B\right)^{3}\int d\eta^{''}f^{2} \left(\delta B \right)^{3}\nonumber\\\fl&\simeq&\left(\frac{\delta B^{T}}{a^{\alpha}}\right)^{4}\int d\eta^{'} a^{2\alpha}\left(\frac{\delta B^{T}}{a^{\alpha}}\right)^{3}\int d\eta^{''}a^{2\alpha} \left(\frac{\delta B^{T}}{a^{\alpha}} \right)^{3},\\\label{ftau4}\fl
\langle \delta B^4\rangle_{c.i.}&\simeq& \left(\delta B\right)^{4}\int d\eta f^{2} \left(\delta B \right)^{4}\simeq\left(\frac{\delta B^{T}}{a^{\alpha}}\right)^{4}\int d\eta a^{2\alpha} \left(\frac{\delta B^{T}}{a^{\alpha}} \right)^{4}.
\eea
Using $a=(-H\eta)^{-1}$ in the previous equations, we get

\bea\label{ftau5}\fl
\langle \delta B^4\rangle_{line2}&\simeq& H^{4\alpha}_{*}\left(\delta B^{T}(-\eta^{*})^{\alpha}\right)^3\int d\eta \left(\delta B^{T}\right)^{3}(-\eta)^{\alpha},\\\fl
\langle \delta B^4\rangle_{v.e.}&\simeq& H^{6\alpha}_{*}\left(\delta B^{T}(-\eta^{*})^{\alpha}\right)^{4}\int d\eta^{'} \left(\delta B^{T}\right)^{3}(-\eta^{'})^{\alpha}\nonumber\\\fl&\times&\int d\eta^{''} \left(\delta B^{T} \right)^{3}(-\eta^{''})^{\alpha},\\\label{ftau6}\fl
\langle \delta B^4\rangle_{c.i.}&\simeq& H^{6\alpha}_{*}\left(\delta B^{T}(-\eta^{*})^{\alpha}\right)^{4}\int d\eta \left(\delta B^{T}\right)^{4}(-\eta)^{2\alpha} .
\eea

Let us now consider more in details the $\alpha=-2$ case for contact-interaction and vector-exchange contributions. The expressions for the anisotropy coefficients are respectively given by

\bea\label{t}\fl
T^{EEEEE}_{ijkl}&=&k_{1}k_{3}\big(\hat{k}_{1}\cdot\hat{k}_{3}-\hat{k}_{1}\cdot\hat{k}_{\hat{12}}\hat{k}_{3}\cdot\hat{k}_{\hat{12}}\big)\big[\delta_{ij}\delta_{kl}-\delta_{ij}\hat{k}_{k4}\hat{k}_{l4}-\delta_{ij}\hat{k}_{k3}\hat{k}_{l3}-\delta_{kl}\hat{k}_{i2}\hat{k}_{j2}-\delta_{kl}\hat{k}_{i1}\hat{k}_{j1}\nonumber\\\fl&+&\delta_{ij}\hat{k}_{k3}\hat{k}_{l4}\hat{k}_{3}\cdot\hat{k}_{4}+\delta_{kl}\hat{k}_{i1}\hat{k}_{j2}\hat{k}_{1}\cdot\hat{k}_{2}+k_{1144}+k_{2244}+k_{1133}+k_{2233}-k_{2234}\hat{k}_{3}\cdot\hat{k}_{4}\nonumber\\\fl&-&k_{1134}\hat{k}_{3}\cdot\hat{k}_{4}-k_{1244}\hat{k}_{1}\cdot\hat{k}_{2}+k_{1233}\hat{k}_{1}\cdot\hat{k}_{2}+k_{1234}\hat{k}_{1}\cdot\hat{k}_{2}\hat{k}_{3}\cdot\hat{k}_{4}\big]
\eea
and by Eq.~(\ref{coeff1}), for one of the possible permutations. These expressions are more complicated w.r.t $T_{ijkl}^{lllll}$ in Eq.~(\ref{tt}) and $T_{mnop}^{llll}$ in Eq.~(\ref{coeff2}) for the longitudinal modes. As a result, when studying the shape of the trispectrum, for the isotropic functions appearing in it, several diagrams need to be taken into account, one for each term in $T^{EEEE}_{ijkl}$ and $T^{EEEEE}_{ijkl}$. For comparison with the $f=1$ case, we plotted the isotropic functions associated with the very last term in square brackets in Eq.~(\ref{t}) (see ``$-v.e(I)$'', ``$-v.e(II)$'' and ``$-v.e(III)$'' in Figs.~6, 7 and 8). By comparing these plots with the ones in Figs.~2, 3 and 4, it is evident that they have very similar shapes. On the other hand, when we consider the isotropic functions associated with terms that are not present in the $f=1$ case, several differences arise in the plots; we provide a sample in Figs.~6, 7 and 8 with the ``$v.e.(new)$'' plots, which represent isotropic functions associated with the $k_{1144}$, $k_{2244}$, $k_{1133}$ and $k_{2233}$ terms in the second line of Eq.~(\ref{t}). We verified that similar observations can be made concerning the shapes of the contact-interaction contributions.

\section{Overview and conclusions}

We have calculated the trispectrum of curvature perturbation in an $SU(2)$ model of gauge bosons coupled to gravity during inflation, where the latter is driven by a slowly-rolling scalar field, Eq.(\ref{ac}). The $\delta$N formalism has been employed to relate the curvature perturbations to the perturbations of all the primordial fields. \\

The trispectrum is made up of contributions due both to the scalar, Eq.~(\ref{scalar}), and to the gauge fields, Eq.~(\ref{vector}). The latter can be distinguished in ``Abelian", i.e. retrievable in the Abelian limit of the theory, Eq.~(\ref{VV}), and ``non-Abelian'', i.e. specific to our model, Eqs.~(\ref{final-V}), (\ref{IA}) and (\ref{trispec}). It is worth to point out that as a by-product of our results, the trispectrum in the pure Abelian case has also been computed for the first time. All of the vector field contributions can be written as the product of an isotropic function of the momenta moduli ($k_{1}$, $k_{2}$, $k_{3}$, $k_{4}$, $k_{\hat{12}}$ and $k_{\hat{14}}$) and of the time labeling the initial hypersurface $\eta^{*}$, times anisotropy coefficients that depend on the angles defining the direction of the gauge fields w.r.t. each other and to the wave vectors. The non-Abelian contribution, calculated using the Schwinger-Keldysh formalism, is made up of a term that is proportional to the bispectrum of the gauge fields, Eqs.~(\ref{final-V}), and terms that originate from the trispectrum of the gauge fields, Eqs.~(\ref{IA}) and (\ref{trispec}). The latter can be diagrammatically represented by two categories, depending on the specific 
interaction Hamiltonian term involved: vector-exchange and contact-interaction. They present a similar analytic expression in terms of powers of the $SU(2)$ coupling constant and of the 
Hubble parameter $H_{*}$, they are though very different in terms of their anisotropy coefficients and of their momentum dependence. \\

We have verified that, in the case where the longitudinal and the trasverse components of the wave vectors have exactly the same evolution and the model does not violate parity, the trispectrum becomes isotropic in all its contributions, except for the vector-exchange one. We also showed that the non-Abelian part of the trispectrum vanishes in the case where all the gauge fields are aligned along a unique spatial direction.\\

We have studied the amplitude $\tau_{NL}$ of the trispectrum. We neglected all the gauge and vector indices (Table 1), in order to provide an approximate evaluation of $\tau_{NL}$ in terms of the parameters of the theory. We showed that the impact of these parameters on suppressing or enhancing the amplitude of the trispectrum very much depends on the specific subset we decide to pick in their parameter space (it depends also on the choice of a specific gauge field configuration, i.e. on the magnitude of their initial value and their spatial orientation). We also noticed that, when comparing the contact-interaction and the vector-exchange contributions (which present a similar parametric dependence from the $SU(2)$ coupling $g_{c}$ and from the Hubble parameter $H_{*}$), the latter are generally several order of magnitude larger than the former. This is of course true unless we are in a configuration for which the anisotropy coefficients somehow reverse this situation ending up by suppressing the vector-exchange contribution.  \\

We finally carried out some shape analysis, focusing on the contribution due to the trispectrum of the gauge fields. As specified above, each contribution to the trispectrum is written as the product of a term that is invariant under rotations of the tetrahedron times some anisotropy coefficients. The study of the shape was performed in two steps: first the rotationally invariant part was analysed for two distinct momentum configurations, the equilateral and the so called specialized planar configurations, Figs.~2, 3 and 4, then the anisotropic coefficients were evaluated in a sample angular configurations, Fig.~5. For the equilateral case we found many similarities between all our plots and the shape of a local trispectrum, whereas, for the specialized planar case, our plots turn out to be easily distinguishable from the local plots except for one of the vector-exchange graphs $(v.e.III)$. Also, the plots for point-interaction and vector-exchange have some common features in the specialized planar configuration whereas they turn out to be completely different in the equilateral case. In Fig.~5 we showed how the amplitude of the trispectrum is modulated by the presence of the anisotropy coefficients. We also verified that the amplitudes of the anisotropic and isotropic parts of the trispectrum can be of the same order of magnitude. This last analysis was not meant to be exhaustive but merely provide the reader with an idea of how the shapes can be studied and how anisotropy can leave a strong imprint on them. \\

Studying the generalization of our calculations to other gauge symmetries is an interesting possibility and it would translate into different results w.r.t. the $SU(2)$ case. In general, the expression of the interaction Hamiltonian would change and therefore the bispectrum and the trispectrum of
the vector bosons are expected to be different. Consider for example a generic $SU(N)$ group, with $N>2$; one realizes that, since the number of vector bosons is equal to the number of generators of the group, the number of preferred spatial directions would increase as we go to higher $N$. This is an important feature that allows to distinguish among different symmetry groups.

\section*{Acknowledgments}
We thank Marco Peloso for useful and interesting discussions and
correspondence.\\
This research has been partially supported by ASI contract I/016/07/0 ``COFIS''.

\vskip 1cm
\appendix
\setcounter{equation}{0}
\def\theequation{A.\arabic{equation}}
\vskip 0.2cm
\section{Anisotropy coefficients for the trispectrum mixed contributions}\label{Anisotropy coefficients for the trispectrum mixed contributions}

We list the coefficients appearing in Eq.~(\ref{final-M}) 

\bea\label{anisotropic-coefficients1}\fl
I_{EEE}^{(8)}&\equiv&\ep^{a'ab}\ep^{a'ce}\Big[6\left(\vec{N}_{a}\cdot\vec{N}_{c}\right)\left(\vec{N}_{\phi b}\cdot\vec{A}^{e}\right)\nonumber\\\fl&+&\left(\vec{N}_{\phi b}\cdot\vec{A}^{e}\right)\Big[\Big(-2\left(\hat{k}_{3}\cdot\vec{N}_{a}\right)\left(\hat{k}_{3}\cdot\vec{N}_{c}\right)-2\left(\hat{k}_{1}\cdot\vec{N}_{a}\right)\left(\hat{k}_{1}\cdot\vec{N}_{c}\right)\nonumber\\\fl&+&\left(\hat{k}_{1}\cdot\vec{N}_{a}\right)\left(\hat{k}_{3}\cdot\vec{N}_{c}\right)\hat{k}_{1}\cdot\hat{k}_{3}+\left(\hat{k}_{3}\cdot\vec{N}_{a}\right)\left(\hat{k}_{1}\cdot\vec{N}_{c}\right)\hat{k}_{1}\cdot\hat{k}_{3}\Big)+(1\rightarrow 2)+(3\rightarrow 2)\Big]\nonumber\\\fl&-&\left[\left(2\left(\vec{N}_{a}\cdot\vec{N}_{c}\right)\left(\hat{k}_{2}\cdot\vec{N}_{\phi b}\right)\left(\hat{k}_{2}\cdot\vec{A}^{e}\right)\right)+(2\rightarrow 1)+(2\rightarrow 3)\right]\nonumber\\\fl&+&\Big[\Big(\hat{k}_{2}\cdot\vec{A}^{e}\Big[2\left(\hat{k}_{3}\cdot\vec{N}_{a}\right)\left(\hat{k}_{2}\cdot\vec{N}_{\phi b}\right)\left(\hat{k}_{3}\cdot\vec{N}_{c}\right)+2\left(\hat{k}_{1}\cdot\vec{N}_{a}\right)\left(\hat{k}_{2}\cdot\vec{N}_{\phi b}\right)\left(\hat{k}_{1}\cdot\vec{N}_{c}\right)\nonumber\\\fl&-&\left(\hat{k}_{1}\cdot\vec{N}_{a}\right)\left(\hat{k}_{2}\cdot\vec{N}_{\phi b}\right)\left(\hat{k}_{3}\cdot\vec{N}_{c}\right)\hat{k}_{1}\cdot\hat{k}_{3}-\left(\hat{k}_{1}\cdot\vec{N}_{a}\right)\left(\hat{k}_{2}\cdot\vec{N}_{\phi b}\right)\left(\hat{k}_{3}\cdot\vec{N}_{c}\right)\hat{k}_{1}\cdot\hat{k}_{3}\Big]\Big)\nonumber\\\fl&+&(2\leftrightarrow 1)+(3\leftrightarrow 2)\Big]\Big], \\\fl
I_{lll}^{(8)}&\equiv&\ep^{a'ab}\ep^{a'ce}\Big[\Big(\left(\hat{k}_{1}\cdot\vec{N}_{a}\right)\left(\hat{k}_{3}\cdot\vec{N}_{\phi b}\right)\left(\hat{k}_{2}\cdot\vec{N}_{c}\right)\left(\hat{k}_{1}\cdot\hat{k}_{2}\right)\left(\hat{k}_{3}\cdot\vec{A}^{e}\right)\nonumber\\\fl&-&\left(\hat{k}_{3}\cdot\vec{N}_{a}\right)\left(\hat{k}_{2}\cdot\vec{N}_{\phi b}\right)\left(\hat{k}_{1}\cdot\vec{N}_{c}\right)\left(\hat{k}_{1}\cdot\hat{k}_{2}\right)\left(\hat{k}_{3}\cdot\vec{A}^{e}\right)\Big)+(1\leftrightarrow 3)+(2\leftrightarrow 3) \Big],\\\fl
I_{llE}^{(8)}&\equiv& \ep^{a'ab}\ep^{a'ce}\Big[\left(\vec{N}_{\phi b}\cdot\vec{A}^{e}\right)\Big(\left(\hat{k}_{1}\cdot\vec{N}_{a}\right)\left(\hat{k}_{2}\cdot\vec{N}_{c}\right)+\left(\hat{k}_{2}\cdot\vec{N}_{a}\right)\left(\hat{k}_{1}\cdot\vec{N}_{c}\right)\Big)\hat{k}_{1}\cdot\hat{k}_{2}\nonumber\\\fl&+&\left[\Big(2\left(\hat{k}_{2}\cdot\vec{A}^{e}\right)\left(\hat{k}_{1}\cdot\vec{N}_{a}\right)\left(\hat{k}_{2}\cdot\vec{N}_{\phi b}\right)\left(\hat{k}_{1}\cdot\vec{N}_{c}\right)\Big)+(1\leftrightarrow 2)\right]\nonumber\\\fl&-&\Big[\Big(\left(\left(\hat{k}_{1}\cdot\vec{N}_{a}\right)\left(\hat{k}_{2}\cdot\vec{N}_{c}\right)+\left(\hat{k}_{2}\cdot\vec{N}_{a}\right)\left(\hat{k}_{1}\cdot\vec{N}_{c}\right)\right)\left(\hat{k}_{3}\cdot\vec{N}_{\phi b}\right)\left(\hat{k}_{3}\cdot\vec{A}^{e}\right)\hat{k}_{1}\cdot\hat{k}_{2}\Big)\nonumber\\\fl&+&(1\leftrightarrow 3)+(2\leftrightarrow 3)\Big]\Big],\\\label{anisotropic-coefficients2}\fl
I_{EEl}^{(8)}&\equiv& \ep^{a'ab}\ep^{a'ce}\Big[4\left(\vec{N}_{\phi b}\cdot\vec{A}^{e}\right)\left(\hat{k}_{3}\cdot\vec{N}_{a}\right)\left(\hat{k}_{3}\cdot\vec{N}_{c}\right)\nonumber\\\fl&+&\Big[\Big(\left(\hat{k}_{2}\cdot\vec{N}_{\phi b}\right)\left(\hat{k}_{2}\cdot\vec{A}^{e}\right)\left(\left(\hat{k}_{1}\cdot\vec{N}_{a}\right)\left(\hat{k}_{3}\cdot\vec{N}_{c}\right)+\left(\hat{k}_{3}\cdot\vec{N}_{a}\right)\left(\hat{k}_{1}\cdot\vec{N}_{c}\right)\right)\hat{k}_{1}\cdot\hat{k}_{3}\Big)\nonumber\\\fl&+&(2\leftrightarrow 1)+(2\leftrightarrow 3)\Big]\nonumber\\\fl&-&\left[\left(2\left(\hat{k}_{2}\cdot\vec{A}^{e}\right)\left(\hat{k}_{2}\cdot\vec{N}_{a}\right)\left(\hat{k}_{3}\cdot\vec{N}_{\phi b}\right)\left(\hat{k}_{2}\cdot\vec{N}_{c}\right)\right)+(1\leftrightarrow 2)+(2\leftrightarrow 3)+(1\leftrightarrow 3)\right]\nonumber\\\fl&-&\left[\left(\left(\vec{N}_{\phi b}\cdot\vec{A}^{e}\right)\hat{k}_{1}\cdot\hat{k}_{3}\left(\left(\hat{k}_{1}\cdot\vec{N}_{a}\right)\left(\hat{k}_{3}\cdot\vec{N}_{c}\right)+\left(\hat{k}_{3}\cdot\vec{N}_{a}\right)\left(\hat{k}_{1}\cdot\vec{N}_{c}\right)\right)\right)+(1\leftrightarrow 2)\right] \nonumber\\\fl&+&\left[\left(\vec{N}_{a}\cdot\vec{N}_{\phi b}\right)\left(\hat{k}_{3}\cdot\vec{N}_{c}\right)\left(\hat{k}_{3}\cdot\vec{A}^{e}\right)+\left(\vec{N}_{c}\cdot\vec{N}_{\phi b}\right)\left(\hat{k}_{3}\cdot\vec{N}_{a}\right)\left(\hat{k}_{3}\cdot\vec{A}^{e}\right)\right]\Big],
\eea

\noindent where $\vec{N}_{\phi b}\equiv (N^{1}_{\phi b},N^{2}_{\phi b},N^{3}_{\phi b})$. \\
As to the coefficients $\tilde{I}_{\alpha\beta\gamma}$ appearing in Eq.~(\ref{final-V}), they can be easily derived by replacing $\vec{\tilde{N}}_{b}$ to $\vec{N}_{\phi b}$ in the expressions of ${I}_{\alpha\beta\gamma}^{(8)}$, where we defined $\tilde{N}_{b}^{i}\equiv M_{j}^{f}N^{ij}_{fb}$.

\def\theequation{B.\arabic{equation}}
\vskip 0.2cm

\section{More details on computing the analytic expressions of vector-exchange diagrams}
\label{More details on computing the analytic expressions of vector-exchange diagrams}

We report the expressions of the functions $A$, $B$...$P$ introduced in Eq.~(\ref{1})

\bea\label{eeeee1}\fl
A&\equiv& \left(-16k^{2}+k_{1}k_{2}x^{*2}\right)\cos\left[\frac{\left(k_{1}+k_{2}\right)x^{*}}{4k}\right]-4k\left(k_{1}+k_{2}\right)x^{*}\sin\left[\frac{\left(k_{1}+k_{2}\right)x^{*}}{4k}\right],\\\fl
B&\equiv& A\left[k_{1}\rightarrow k_{3},k_{2}\rightarrow k_{4}\right],\\\fl
C&\equiv& 4k\left(k_{1}+k_{2}\right)x^{*}\cos\left[\frac{\left(k_{1}+k_{2}\right)x^{*}}{4k}\right]+\left(-16k^{2}+k_{1}k_{2}x^{*2}\right)\sin\left[\frac{\left(k_{1}+k_{2}\right)x^{*}}{4k}\right],\\\fl
D&\equiv& C\left[k_{1}\rightarrow k_{3},k_{2}\rightarrow k_{4}\right],\\\fl
E&\equiv& \left(8k^{2}\left(k_{\hat{12}}+k_{3}+k_{4}\right)-k_{\hat{12}}k_{3}k_{4}x^{*2}\right)\cos\left[\frac{\left(k_{\hat{12}}+k_{3}+k_{4}\right)x^{*}}{4k}\right]\nonumber\\\fl&+&2k\left(k_{\hat{12}}+k_{3}+k_{4}\right)^2 x^{*}\sin\left[\frac{\left(k_{\hat{12}}+k_{3}+k_{4}\right)x^{*}}{4k}\right],\\\fl
F&\equiv& E\left[k_{3}\rightarrow k_{1},k_{4}\rightarrow k_{2}\right],\\\fl
G&\equiv&2k\left(k_{\hat{12}}+k_{1}+k_{2}\right)^2 x^{*}\cos\left[\frac{\left(k_{\hat{12}}+k_{1}+k_{2}\right)x^{*}}{4k}\right] \nonumber\\\fl&+&\left(-8k^{2}\left(k_{\hat{12}}+k_{1}+k_{2}\right)-k_{\hat{12}}k_{1}k_{2}x^{*2}\right)\sin\left[\frac{\left(k_{\hat{12}}+k_{1}+k_{2}\right)x^{*}}{4k}\right],\\\fl
L&\equiv& \left(k_{1}^{3}+k_{\hat{12}}^{3}+k_{\hat{12}}^{2}k_{2}+k_{\hat{12}}k_{2}^{2}+k_{2}^{3}(k_{\hat{12}}+k_{2})+k_{1}(k_{\hat{12}}^{2}+k_{2}^{2})\right)x^{*2}si\left[\frac{\left(k_{\hat{12}}+k_{1}+k_{2}\right)x^{*}}{4k}\right],\nonumber
\\\fl
M&\equiv& \left(k_{1}^{3}+k_{\hat{12}}^{3}+k_{\hat{12}}^{2}k_{2}+k_{\hat{12}}k_{2}^{2}+k_{2}^{3}(k_{\hat{12}}+k_{2})+k_{1}(k_{\hat{12}}^{2}+k_{2}^{2})\right)x^{*2}ci\left[\frac{\left(k_{\hat{12}}+k_{1}+k_{2}\right)x^{*}}{4k}\right],\nonumber\\\fl
N&\equiv& M\left[k_{3}\rightarrow k_{1},k_{4}\rightarrow k_{2}\right],\\\fl\label{eeeee2}
P&\equiv& L\left[k_{1}\rightarrow k_{3},k_{2}\rightarrow k_{4}\right].
\eea

\noindent The anisotropy coefficients introduced in Eq.~(\ref{final-2v}) have the following expressions

\bea\label{t1}\fl
t_{1}\equiv k_{1}k_{3}\left(\hat{k}_{1}\cdot\hat{k}_{\hat{12}}\right)\left(\hat{k}_{1}\cdot\hat{k}_{2}\right)\left(\hat{k}_{3}\cdot\hat{k}_{4}\right)\left(\hat{k}_{3}\cdot\hat{k}_{\hat{12}}\right)\\\fl
t_{2}\equiv  k_{1}k_{4}\left(\hat{k}_{1}\cdot\hat{k}_{\hat{12}}\right)\left(\hat{k}_{1}\cdot\hat{k}_{2}\right)\left(\hat{k}_{3}\cdot\hat{k}_{4}\right)\left(\hat{k}_{4}\cdot\hat{k}_{\hat{12}}\right),\\\fl
t_{3}\equiv  k_{2}k_{3}\left(\hat{k}_{2}\cdot\hat{k}_{\hat{12}}\right)\left(\hat{k}_{1}\cdot\hat{k}_{2}\right)\left(\hat{k}_{3}\cdot\hat{k}_{4}\right)\left(\hat{k}_{3}\cdot\hat{k}_{\hat{12}}\right),\\\fl
t_{4}\equiv  k_{2}k_{4}\left(\hat{k}_{2}\cdot\hat{k}_{\hat{12}}\right)\left(\hat{k}_{1}\cdot\hat{k}_{2}\right)\left(\hat{k}_{3}\cdot\hat{k}_{4}\right)\left(\hat{k}_{4}\cdot\hat{k}_{\hat{12}}\right),\\\fl
t_{5}\equiv  k_{1}k_{2}\left(\hat{k}_{1}\cdot\hat{k}_{\hat{13}}\right)\left(\hat{k}_{1}\cdot\hat{k}_{3}\right)\left(\hat{k}_{2}\cdot\hat{k}_{4}\right)\left(\hat{k}_{2}\cdot\hat{k}_{\hat{13}}\right),\\\fl
t_{6}\equiv  k_{1}k_{4}\left(\hat{k}_{1}\cdot\hat{k}_{\hat{13}}\right)\left(\hat{k}_{1}\cdot\hat{k}_{3}\right)\left(\hat{k}_{2}\cdot\hat{k}_{4}\right)\left(\hat{k}_{4}\cdot\hat{k}_{\hat{13}}\right),\\\fl
t_{7}\equiv  k_{2}k_{3}\left(\hat{k}_{3}\cdot\hat{k}_{\hat{13}}\right)\left(\hat{k}_{1}\cdot\hat{k}_{3}\right)\left(\hat{k}_{2}\cdot\hat{k}_{4}\right)\left(\hat{k}_{2}\cdot\hat{k}_{\hat{13}}\right),\\\fl
t_{8}\equiv  k_{3}k_{4}\left(\hat{k}_{3}\cdot\hat{k}_{\hat{13}}\right)\left(\hat{k}_{1}\cdot\hat{k}_{3}\right)\left(\hat{k}_{2}\cdot\hat{k}_{4}\right)\left(\hat{k}_{4}\cdot\hat{k}_{\hat{13}}\right),\\\fl
t_{9}\equiv  k_{1}k_{2}\left(\hat{k}_{1}\cdot\hat{k}_{\hat{14}}\right)\left(\hat{k}_{1}\cdot\hat{k}_{4}\right)\left(\hat{k}_{2}\cdot\hat{k}_{3}\right)\left(\hat{k}_{2}\cdot\hat{k}_{\hat{14}}\right),\\\fl
t_{10}\equiv k_{1}k_{3}\left(\hat{k}_{1}\cdot\hat{k}_{\hat{14}}\right)\left(\hat{k}_{1}\cdot\hat{k}_{4}\right)\left(\hat{k}_{2}\cdot\hat{k}_{3}\right)\left(\hat{k}_{3}\cdot\hat{k}_{\hat{14}}\right),\\\fl
t_{11}\equiv k_{1}k_{3}\left(\hat{k}_{1}\cdot\hat{k}_{\hat{12}}\right)\left(\hat{k}_{1}\cdot\hat{k}_{2}\right)\left(\hat{k}_{3}\cdot\hat{k}_{4}\right)\left(\hat{k}_{3}\cdot\hat{k}_{\hat{12}}\right),\\\label{t12}\fl
t_{12}\equiv k_{1}k_{3}\left(\hat{k}_{1}\cdot\hat{k}_{\hat{12}}\right)\left(\hat{k}_{1}\cdot\hat{k}_{2}\right)\left(\hat{k}_{3}\cdot\hat{k}_{4}\right)\left(\hat{k}_{3}\cdot\hat{k}_{\hat{12}}\right).
\eea

\noindent Let us now list all the scalar products appearing in the equations above\\

\begin{tabular}{|cc|c|cc|}\hline
$\hat{k}_{1}\cdot\hat{k}_{\hat{12}}=\frac{k_{\hat{12}}^2+k^{2}_{1}+k_{2}^{2}}{2k_{1}k_{\hat{12}}}$ & $$ & $\hat{k}_{1}\cdot\hat{k}_{\hat{13}}=\frac{2k_{1}^2+k^{2}_{4}-k_{\hat{12}}^{2}-k_{\hat{14}}^{2}}{2k_{1}k_{\hat{13}}}$ &$$ & $\hat{k}_{1}\cdot\hat{k}_{\hat{14}}=\frac{k_{\hat{14}}^2+k^{2}_{1}-k_{4}^{2}}{2k_{1}k_{\hat{14}}}$  \\
\hline
$\hat{k}_{2}\cdot\hat{k}_{\hat{12}}=\frac{k_{\hat{12}}^2+k^{2}_{1}-k_{1}^{2}}{2k_{2}k_{\hat{12}}}$ & $$ & $\hat{k}_{3}\cdot\hat{k}_{\hat{13}}=\frac{2k_{3}^{2}+k_{4}^2-k_{\hat{12}}^{2}-k_{\hat{14}}^{2}+k_{2}^{2}}{2k_{3}k_{\hat{13}}}$ & $$ & $\hat{k}_{4}\cdot\hat{k}_{\hat{14}}=\frac{k_{\hat{14}}^2-k^{2}_{1}+k_{4}^{2}}{2k_{4}k_{\hat{14}}}$ \\
\hline
$\hat{k}_{3}\cdot\hat{k}_{\hat{12}}=\frac{k_{\hat{12}}^2+k_{\hat{14}}^{2}-2k_{2}^{2}-k^{2}_{1}-k_{3}^{2}}{2k_{2}k_{\hat{13}}}$ & $$ & $\hat{k}_{2}\cdot\hat{k}_{\hat{13}}=\frac{k_{\hat{12}}^2+k_{\hat{14}}^{2}-k_{1}^{2}-2k
_{2}^{2}-k_{3}^{2}}{2k_{1}k_{3}}$ & $$ & $\hat{k}_{2}\cdot\hat{k}_{4}=\frac{k_{1}^{2}+k_{3}^{2}-k_{\hat{12}}^{2}-k_{\hat{14}}^{2}}{2k_{2}k_{4}}$\\
\hline
$\hat{k}_{4}\cdot\hat{k}_{\hat{12}}=\frac{k_{\hat{14}}^2+k_{\hat{12}}^2-k^{2}_{1}-k_{3}^{2}-2k_{4}^{2}}{2k_{4}k_{\hat{13}}}$ & $$ & $\hat{k}_{4}\cdot\hat{k}_{\hat{13}}=\frac{k_{\hat{14}}^{2}+k_{\hat{12}}^2-k^{2}_{1}-k^{2}_{3}-2k^{2}_{4}}{2k_{4}k_{\hat{13}}}$ & $$ &$\hat{k}_{2}\cdot\hat{k}_{\hat{14}}=\frac{k_{3}^{2}-k_{\hat{14}}^2-k^{2}_{2}}{2k_{2}k_{\hat{14}}}$\\
\hline
$\hat{k}_{1}\cdot\hat{k}_{2}=\frac{k_{1}^{2}+k_{2}^{2}-k_{\hat{12}}^{2}}{2k_{1}k_{2}}$ & $$ & $\hat{k}_{1}\cdot\hat{k}_{3}=\frac{k_{4}^{2}+k_{2}^{2}-k_{\hat{12}}^{2}-k_{\hat{14}}^{2}}{2k_{1}k_{3}}$ & $$ & $\hat{k}_{1}\cdot\hat{k}_{4}=\frac{k_{\hat{14}}^{2}-k_{1}^{2}-k_{4}^{2}}{2k_{1}k_{4}}$ \\
\hline
$\hat{k}_{3}\cdot\hat{k}_{4}=\frac{k_{\hat{12}}^{2}-k_{4}^{2}-k_{3}^{2}}{2 k_{4}k_{3}}$ & $$  & $\hat{k}_{3}\cdot\hat{k}_{\hat{14}}=\frac{k_{2}^2-k^{2}_{3}-k_{\hat{14}}^{2}}{2k_{3}k_{\hat{14}}}$ & $$ & $\hat{k}_{2}\cdot\hat{k}_{3}=\frac{k_{\hat{14}}^{2}-k_{2}^{2}-k_{3}^{2}}{2k_{2}k_{3}}$ \\
\hline
\end{tabular}
\\

\noindent where $k_{i}\equiv |\vec{k}_{i}|$, $\hat{k}_{i}\equiv\vec{k}_{i}/k_{i}$, $\vec{k}_{\hat{ij}}\equiv \vec{k}_{i}+\vec{k}_{j}$, $k_{ij}\equiv |\vec{k}_{i}+\vec{k}_{j}|$, $i$ and $j$ running over the four external wave vectors.

\def\theequation{C.\arabic{equation}}
\vskip 0.2cm

\section{Complete expressions for functions appearing in point-interation diagrams}
\label{Complete expressions for functions appearing in point-interaction diagrams}

We provide here the expressions for the coefficients appearing in Eq.~(\ref{p-i})
\bea\fl
Q_{EEEE}&\equiv& x^{*3}\big[-k\big(k^{3}k^{4}_{1}+k^{3}k^{4}_{2}-k^{3}_{3}+k^{3}k_{3}^{4}+k^{5}k_{3}k_{4}-3k^{3}k^{2}_{2}k_{3}k_{4}+k^{3}k^{3}_{3}k_{4}\nonumber\\\fl&-&k^{3}_{4}+k^{3}k_{3}k_{4}^{3}+k^{3}k_{4}^{4}+k^{3}k_{2}\big(k_{3}^{3}-kk_{3}k_{4}-3k_{3}^{2}k_{4}-3k_{3}k_{4}^{2}+k_{4}^{3}\nonumber\\\fl&+&k^{2}\big(k_{3}+k_{4}\big)\big)+k_{2}^{3}\big(-1+k^{3}\big(k_{3}+k_{4}\big)\big)-3k^{3}k^{2}_{1}\big(k_{3}k_{4}+k_{2}\big(k_{3}+k_{4}\big)\big)\nonumber\\\fl&+&k^{3}_{1}\big(-1+k^{3}\big(k_{2}+k_{3}+k_{4}\big)\big)+k^{3}k_{1}\big(k_{2}^{3}+k_{3}^{3}-3k_{3}^{2}k_{4}-3k_{3}k_{4}^{2}+k_{4}^{3}\nonumber\\\fl&-&3k^{2}_{2}\big(k_{3}+k_{4}\big)+k^{2}\big(k_{2}+k_{3}+k_{4}\big)-3k_{2}\big(k_{3}^{2}+3k_{3}k_{4}+k_{4}^{2}\big)-k\big(k_{3}k_{4}\nonumber\\\fl&+&k_{2}\big(k_{3}+k_{4}\big)\big)\big)\big)\big]+x^{*5}\big[k^2\big(k^{3}_{1}k_{2}k_{3}k_{4}+k_{2}k_{3}k_{4}\big(k_{2}^{3}+k_{3}^{3}-3k_{2}k_{3}k_{4}+k_{4}^{3}\big)\nonumber\\\fl&+&k^{4}_{1}\big(k_{3}k_{4}+k_{2}\big(k_{3}+k_{4}\big)\big)-3k_{1}^{2}\big(k_{3}^{2}k_{4}^{2}+k_{2}k_{3}k_{4}\big(k_{3}+k_{4}\big)+k^{2}_{2}\big(k_{3}^{2}+k_{3}k_{4}\nonumber\\\fl&+&k^{2}_{4}\big)\big)+k_{1}\big(k_{2}^{3}k_{3}k_{4}+k_{2}^{4}\big(k_{3}+k_{4}\big)-3k_{2}^{2}k_{3}k_{4}\big(k_{3}+k_{4}\big)+k_{3}k_{4}\big(k_{3}^{3}+k_{4}^{3}\big)\nonumber\\\fl&+&k_{2}\big(k_{3}^{4}+k_{3}^{3}k_{4}-3k_{3}^{2}k_{4}^{2}+k_{4}^{4}+k_{3}k_{4}\big(k^{2}+k_{4}^{2}\big)\big)\big)\big)\big]-3x^{*7}k_{1}^{2}k_{2}^{2}k_{3}^{2}k_{4}^{2} ,\\ \fl
A_{EEEE}&\equiv&  k\left(k_{1}^{3}+k_{2}^{3}+k_{3}^{3}+k_{4}^{3}\right)x^{*3},\\ \fl
B_{EEEE}&\equiv&  k^{4}-k^{2}\left(k_{3}k_{4}+k_{2}\left(k_{3}+k_{4}+k_{1}\left(k_{2}+k_{3}+k_{4}\right)x^{*2}+k_{1}k_{2}k_{3}k_{4}x^{*4}\right)\right),\\ \fl
C_{EEEE}&\equiv& kx^{*}\left(k^{3}-\left(k_{2}k_{3}k_{4}+k_{1}\left(k_{3}k_{4}+k_{2}\left(k_{3}+k_{4}\right)\right)\right)x^{*2}\right) ,\\ \fl
D_{EEEE}&\equiv&  -kx^{*}\left(k^{3}-\left(k_{2}k_{3}k_{4}+k_{1}\left(k_{3}k_{4}+k_{2}\left(k_{3}k_{4}\right)\right)\right)x^{*2}\right),\\ \fl
E_{EEEE}&\equiv& k^{4}-k^{2}\left(k_{3}k_{4}+k_{2}\left(k_{3}+k_{4}\right)+k_{1}\left(k_{2}+k_{3}+k_{4}\right)\right)x^{*2}+k_{1}k_{2}k_{3}k_{4}x^{*4} ,\\ \fl
F_{EEEE}&\equiv&  k\left(k_{1}^{3}+k_{2}^{3}+k_{3}^{3}+k_{4}^{3}\right)x^{*3}.
\eea


\section*{References}

\begin{thebibliography}{99}

\bibitem{Lyth:1998xn}
  D.~H.~Lyth and A.~Riotto,
  Phys.\ Rept.\  {\bf 314}, 1 (1999)
  [arXiv:hep-ph/9807278].

\bibitem{http://map.gsfc.nasa.gov/}
See http://map.gsfc.nasa.gov/.

\bibitem{Spergel:2006hy}
  D.~N.~Spergel {\it et al.}  [WMAP Collaboration],
  Astrophys.\ J.\ Suppl.\  {\bf 170}, 377 (2007)
  [arXiv:astro-ph/0603449].

\bibitem{Komatsu:2008hk}
  E.~Komatsu {\it et al.}  [WMAP Collaboration],
  Astrophys.\ J.\ Suppl.\  {\bf 180}, 330 (2009)
  [arXiv:0803.0547 [astro-ph]].

\bibitem{http://planck.esa.int/}
See http://planck.esa.int/.


\bibitem{Bartolo:2004if}
  N.~Bartolo, E.~Komatsu, S.~Matarrese and A.~Riotto,
  Phys.\ Rept.\  {\bf 402}, 103 (2004)
  [arXiv:astro-ph/0406398].


\bibitem{Komatsu:2009kd}
  E.~Komatsu {\it et al.},
  arXiv:0902.4759 [astro-ph.CO].



\bibitem{Acqua}
  V.~Acquaviva, N.~Bartolo, S.~Matarrese and A.~Riotto,
  Nucl.\ Phys.\  B {\bf 667}, 119 (2003)
  [arXiv:astro-ph/0209156].

\bibitem{Maldacena:2002vr}
  J.~M.~Maldacena,
  JHEP {\bf 0305}, 013 (2003)
  [arXiv:astro-ph/0210603].



\bibitem{Bmulti}
  N.~Bartolo, S.~Matarrese and A.~Riotto,
  Phys.\ Rev.\  D {\bf 65}, 103505 (2002)
  [arXiv:hep-ph/0112261].


\bibitem{SL1}
  D.~Seery and J.~E.~Lidsey,
  JCAP {\bf 0506}, 003 (2005)
  [arXiv:astro-ph/0503692].


\bibitem{SL2}
  D.~Seery and J.~E.~Lidsey,
  JCAP {\bf 0509}, 011 (2005)
  [arXiv:astro-ph/0506056].


\bibitem{Dimopoulos:2006ms}
  K.~Dimopoulos,
  Phys.\ Rev.\  D {\bf 74}, 083502 (2006)
  [arXiv:hep-ph/0607229].

\bibitem{Mollerach}
  S.~Mollerach,
  Phys.\ Rev.\  D {\bf 42}, 313 (1990).


\bibitem{Enqvist:2001zp}
  K.~Enqvist and M.~S.~Sloth,
  Nucl.\ Phys.\  B {\bf 626}, 395 (2002)
  [arXiv:hep-ph/0109214].




\bibitem{Lyth:2001nq}
  D.~H.~Lyth and D.~Wands,
  Phys.\ Lett.\  B {\bf 524}, 5 (2002)
  [arXiv:hep-ph/0110002].




\bibitem{Lyth:2002my}
  D.~H.~Lyth, C.~Ungarelli and D.~Wands,
  Phys.\ Rev.\  D {\bf 67}, 023503 (2003)
  [arXiv:astro-ph/0208055].



\bibitem{Moroi:2001ct}
  T.~Moroi and T.~Takahashi,
  Phys.\ Lett.\  B {\bf 522}, 215 (2001)
  [Erratum-ibid.\  B {\bf 539}, 303 (2002)]
  [arXiv:hep-ph/0110096].



\bibitem{Bartolo:2003jx}
  N.~Bartolo, S.~Matarrese and A.~Riotto,
  Phys.\ Rev.\  D {\bf 69}, 043503 (2004)
  [arXiv:hep-ph/0309033].





\bibitem{Ali}
  M.~Alishahiha, E.~Silverstein and D.~Tong,
  Phys.\ Rev.\  D {\bf 70} (2004) 123505
  [arXiv:hep-th/0404084].

\bibitem{Chen:2006nt}
  X.~Chen, M.~x.~Huang, S.~Kachru and G.~Shiu,
  JCAP {\bf 0701}, 002 (2007)
  [arXiv:hep-th/0605045].



\bibitem{ghostinfl}
  N.~Arkani-Hamed, P.~Creminelli, S.~Mukohyama and M.~Zaldarriaga,
  JCAP {\bf 0404}, 001 (2004)
  [arXiv:hep-th/0312100].

\bibitem{Okamoto:2002ik}
  T.~Okamoto and W.~Hu,
  Phys.\ Rev.\  D {\bf 66}, 063008 (2002)
  [arXiv:astro-ph/0206155].

\bibitem{Kogo:2006kh}
  N.~Kogo and E.~Komatsu,
  Phys.\ Rev.\  D {\bf 73}, 083007 (2006)
  [arXiv:astro-ph/0602099].

\bibitem{Barttrisp}
  N.~Bartolo, S.~Matarrese and A.~Riotto,
  JCAP {\bf 0508}, 010 (2005)
  [arXiv:astro-ph/0506410].



\bibitem{Seery:2006vu}
  D.~Seery, J.~E.~Lidsey and M.~S.~Sloth,
  JCAP {\bf 0701}, 027 (2007)
  [arXiv:astro-ph/0610210].

\bibitem{SL3}
  D.~Seery and J.~E.~Lidsey,
  JCAP {\bf 0701}, 008 (2007)
  [arXiv:astro-ph/0611034].





\bibitem{Byrnes}
  C.~T.~Byrnes, M.~Sasaki and D.~Wands,
  Phys.\ Rev.\  D {\bf 74}, 123519 (2006)
  [arXiv:astro-ph/0611075].




\bibitem{Seery:2008ax}
  D.~Seery, M.~S.~Sloth and F.~Vernizzi,
  JCAP {\bf 0903}, 018 (2009)
  [arXiv:0811.3934 [astro-ph]].


\bibitem{SVW}
  M.~Sasaki, J.~Valiviita and D.~Wands,
  Phys.\ Rev.\  D {\bf 74}, 103003 (2006)
  [arXiv:astro-ph/0607627].




\bibitem{Huang:2006eha}
  X.~Chen, M.~x.~Huang and G.~Shiu,
  Phys.\ Rev.\  D {\bf 74}, 121301 (2006)
  [arXiv:hep-th/0610235].

\bibitem{Arroja:2008ga}
  F.~Arroja and K.~Koyama,
  Phys.\ Rev.\  D {\bf 77}, 083517 (2008)
  [arXiv:0802.1167 [hep-th]].

\bibitem{Arroja:2009pd}
  F.~Arroja, S.~Mizuno, K.~Koyama and T.~Tanaka,
  Phys.\ Rev.\  D {\bf 80}, 043527 (2009)
  [arXiv:0905.3641 [hep-th]].

\bibitem{Chen:2009bc}
  X.~Chen, B.~Hu, M.~x.~Huang, G.~Shiu and Y.~Wang,
  JCAP {\bf 0908}, 008 (2009)
  [arXiv:0905.3494 [astro-ph.CO]].


\bibitem{Martin:2007ue}
  J.~Martin and J.~Yokoyama,
  JCAP {\bf 0801}, 025 (2008)
  [arXiv:0711.4307 [astro-ph]].

\bibitem{Gao:2009gd}
  X.~Gao and B.~Hu,
  JCAP {\bf 0908}, 012 (2009)
  [arXiv:0903.1920 [astro-ph.CO]].


\bibitem{Mizuno:2009mv}
  S.~Mizuno, F.~Arroja and K.~Koyama,
  arXiv:0907.2439 [hep-th].

\bibitem{Kunzetal}
  M.~Kunz, A.~J.~Banday, P.~G.~Castro, P.~G.~Ferreira and K.~M.~Gorski,
  arXiv:astro-ph/0111250.


\bibitem{Komatsuthesis}
  E.~Komatsu,
  arXiv:astro-ph/0206039.



\bibitem{DS}
  V.~Desjacques and U.~Seljak,
  arXiv:0907.2257 [astro-ph.CO].






\bibitem{Ack}
  L.~Ackerman, S.~M.~Carroll and M.~B.~Wise,
  Phys.\ Rev.\  D {\bf 75}, 083502 (2007)
  [arXiv:astro-ph/0701357]

\bibitem{Pullen:2007tu}
  A.~R.~Pullen and M.~Kamionkowski,
  Phys.\ Rev.\  D {\bf 76}, 103529 (2007)
  [arXiv:0709.1144 [astro-ph]].



\bibitem{Golovnev:2008cf}
  A.~Golovnev, V.~Mukhanov and V.~Vanchurin,
  JCAP {\bf 0806}, 009 (2008)
  [arXiv:0802.2068 [astro-ph]].



\bibitem{Dimopoulos:2008rf}
  K.~Dimopoulos and M.~Karciauskas,
  JHEP {\bf 0807}, 119 (2008)
  [arXiv:0803.3041 [hep-th]].


\bibitem{Yokoyama:2008xw}
  S.~Yokoyama and J.~Soda,
  JCAP {\bf 0808}, 005 (2008)
  [arXiv:0805.4265 [astro-ph]].



\bibitem{Dimopoulos:2008yv}
  K.~Dimopoulos, M.~Karciauskas, D.~H.~Lyth and Y.~Rodriguez,
  arXiv:0809.1055 [astro-ph].





\bibitem{Karciauskas:2008bc}
  M.~Karciauskas, K.~Dimopoulos and D.~H.~Lyth,
  arXiv:0812.0264 [astro-ph].


\bibitem{ValenzuelaToledo:2009af}
  C.~A.~Valenzuela-Toledo, Y.~Rodriguez and D.~H.~Lyth,
  arXiv:0909.4064 [astro-ph.CO].







\bibitem{Golovnev:2009ks}
  A.~Golovnev and V.~Vanchurin,
  arXiv:0903.2977 [astro-ph.CO].



\bibitem{Eriksen:2003db}
  H.~K.~Eriksen, F.~K.~Hansen, A.~J.~Banday, K.~M.~Gorski and P.~B.~Lilje,
  Astrophys.\ J.\  {\bf 605}, 14 (2004)
  [Erratum-ibid.\  {\bf 609}, 1198 (2004)]
  [arXiv:astro-ph/0307507].







\bibitem{Hansen:2004vq}
  F.~K.~Hansen, A.~J.~Banday and K.~M.~Gorski,
  Mon.\ Not.\ Roy.\ Astron.\ Soc.\  {\bf 354}, 641 (2004)
  [arXiv:astro-ph/0404206].


\bibitem{Jaffe:2005pw}
  T.~R.~Jaffe, A.~J.~Banday, H.~K.~Eriksen, K.~M.~Gorski and F.~K.~Hansen,
  Astrophys.\ J.\  {\bf 629}, L1 (2005)
  [arXiv:astro-ph/0503213].


\bibitem{Eriksen:2007pc}
  H.~K.~Eriksen, A.~J.~Banday, K.~M.~Gorski, F.~K.~Hansen and P.~B.~Lilje,
  Astrophys.\ J.\  {\bf 660}, L81 (2007)
  [arXiv:astro-ph/0701089].

\bibitem{Groeneboom:2008fz}
  N.~E.~Groeneboom and H.~K.~Eriksen,
  Astrophys.\ J.\  {\bf 690}, 1807 (2009)
  [arXiv:0807.2242 [astro-ph]].




\bibitem{Bartolo:2009pa}
  N.~Bartolo, E.~Dimastrogiovanni, S.~Matarrese and A.~Riotto,
  arXiv:0906.4944 [astro-ph.CO](JCAP, in press).


\bibitem{deltaN1}
A.~A.~Starobinsky,
  JETP Lett.\  {\bf 42} (1985) 152
  [Pisma Zh.\ Eksp.\ Teor.\ Fiz.\  {\bf 42} (1985) 124].



\bibitem{deltaN2}
M.~Sasaki and E.~D.~Stewart,
  Prog.\ Theor.\ Phys.\  {\bf 95}, 71 (1996)
  [arXiv:astro-ph/9507001].



\bibitem{deltaN3}
D.~H.~Lyth, K.~A.~Malik and M.~Sasaki,
  JCAP {\bf 0505}, 004 (2005)
  [arXiv:astro-ph/0411220].



\bibitem{deltaN4}
D.~H.~Lyth and Y.~Rodriguez,
  Phys.\ Rev.\ Lett.\  {\bf 95}, 121302 (2005)
  [arXiv:astro-ph/0504045].




\bibitem{in-in1}
 J.~S.~Schwinger,
  J.\ Math.\ Phys.\  {\bf 2}, 407 (1961).




\bibitem{in-in2}
E.~Calzetta and B.~L.~Hu,
  Phys.\ Rev.\  D {\bf 35}, 495 (1987).



\bibitem{in-in3}
R.~D.~Jordan,
  Phys.\ Rev.\  D {\bf 33}, 444 (1986).



\bibitem{Dimopoulos:2009am}
  K.~Dimopoulos, M.~Karciauskas and J.~M.~Wagstaff,
  arXiv:0907.1838 [hep-ph].


\bibitem{Dimopoulos:2009vu}
  K.~Dimopoulos, M.~Karciauskas and J.~M.~Wagstaff,
  arXiv:0909.0475 [hep-ph].

\bibitem{Himmetoglu:2008zp}
  B.~Himmetoglu, C.~R.~Contaldi and M.~Peloso,
  Phys.\ Rev.\ Lett.\  {\bf 102}, 111301 (2009)
  [arXiv:0809.2779 [astro-ph]].

\bibitem{Himmetoglu:2008hx}
  B.~Himmetoglu, C.~R.~Contaldi and M.~Peloso,
  Phys.\ Rev.\  D {\bf 79}, 063517 (2009)
  [arXiv:0812.1231 [astro-ph]].

\bibitem{Koivisto:2009sd}
  T.~S.~Koivisto, D.~F.~Mota and C.~Pitrou,
  JHEP {\bf 0909}, 092 (2009)
  [arXiv:0903.4158 [astro-ph.CO]].



\bibitem{Himmetoglu:2009qi}
  B.~Himmetoglu, C.~R.~Contaldi and M.~Peloso,
  arXiv:0909.3524 [astro-ph.CO].





\bibitem{Vernizzi:2006ve}
  F.~Vernizzi and D.~Wands,
  JCAP {\bf 0605}, 019 (2006)
  [arXiv:astro-ph/0603799].




\end{thebibliography}
\end{document}